\documentclass[12pt, a4paper]{article}
\usepackage{styleBuding}
\usepackage{amsmath}
\usepackage{bm}
\usepackage{amssymb}
\usepackage{listings}
\usepackage{multirow}
\usepackage{makecell}
\usepackage{array}
\usepackage{subfigure}
\usepackage{siunitx}
\usepackage{booktabs}

\usepackage{csquotes}
\newcommand{\tabincell}[2]{\begin{tabular}{@{}#1@{}}#2\end{tabular}}

\title{\boldmath Improved $(g-2)_\mu$ measurement and singlino dark matter in $\mu$-term extended $\mathbb{Z}_3$-NMSSM}

\author[]{Junjie Cao,}
\author[]{Jingwei Lian,}
\author[]{Yusi Pan,}
\author[]{Di Zhang}
\author[]{and Pengxuan Zhu}

\affiliation[]{Department of Physics, Henan Normal University, 453007, China}
\emailAdd{junjiec@alumni.itp.ac.cn}
\emailAdd{ljwfly@hotmail.com}
\emailAdd{panyusi0406@foxmail.com}
\emailAdd{dz481655@gmail.com}
\emailAdd{zhupx99@icloud.com}

\abstract{
Very recently, a Fermilab report of muon $g-2$ showed a $4.2\sigma$ discrepancy between it and the standard model (SM) prediction. Motivated by this inspiring result and the increasing tension in supersymmetric interpretation of the anomalous magnetic moment, it is argued that in the general next-to-minimal supersymmetric standard model (GNMSSM), a singlino-dominated neutralino can act as a feasible dark matter (DM) candidate in explaining the discrepancy naturally. In this case, the singlino-dominated DM and singlet-dominated Higgs bosons can form a secluded DM sector with $\tilde{\chi}_1^0\tilde{\chi}_1^0 \to h_s A_s$ responsible for the measured DM relic abundance
when $m_{\tilde{\chi}_1^0} \gtrsim 150~{\rm GeV}$ and the Yukawa coupling $\kappa$ is around $0.2$. This sector communicates with the SM sector by weak singlet-doublet Higgs mixing, so the scatterings of the singlino-dominated DM with nucleons are suppressed. Furthermore, due to the singlet nature of the DM and the complex mass hierarchy, sparticle decay chains in the GNMSSM are lengthened in comparison with the prediction of the minimal supersymmetric standard model. These characteristics lead to sparticle detection at the Large Hadron Collider (LHC) being rather tricky. This study surveys a specific scenario of the GNMSSM, which extends the $\mathbb{Z}_3$-NMSSM by adding an explicit $\mu$-term, to reveal the features. It indicates that the theory can readily explain the discrepancy of the muon anomalous magnetic moment without conflicting with the experimental results in DM and Higgs physics, and the LHC searches for sparticles.
}

\begin{document}
\maketitle
\section{Introduction}
\par Recently, experimentalists at Fermilab National Accelerator Laboratory (FNAL) reported the most accurate measurement of the muon anomalous magnetic moment $a_\mu^{\rm exp}({\rm FNAL})$~\cite{PhysRevLett.126.141801}. Combined with the previous Brookhaven National Laboratory E821 result $a_{\mu}^{\rm exp}({\rm BNL})$~\cite{Bennett:2006fi}, the statistical average $a_\mu^{\rm exp}$ reads
\begin{equation}\begin{split}
		a_{\mu}^{\rm exp}({\rm FNAL}) &= 116592040(54) \times 10^{-11}, \\
		a_{\mu}^{\rm exp}({\rm BNL}) &= 116592080(63) \times 10^{-11}, \\
		a_\mu^{\rm exp} &= 116592061(41) \times 10^{-11},
\end{split}\end{equation}
which reveals a $4.2\sigma$ discrepancy with the SM prediction $a_\mu^{\rm SM} = 116 591 810(43) \times 10^{-11}$~\cite{Aoyama:2020ynm,Aoyama:2012wk,Aoyama:2019ryr,Czarnecki:2002nt,Gnendiger:2013pva,Davier:2017zfy,Keshavarzi:2018mgv,Colangelo:2018mtw,Hoferichter:2019gzf,Davier:2019can,
Keshavarzi:2019abf,Kurz:2014wya,Melnikov:2003xd,Masjuan:2017tvw,Colangelo:2017fiz,Hoferichter:2018kwz,Gerardin:2019vio,Bijnens:2019ghy,Colangelo:2019uex,
Blum:2019ugy,Colangelo:2014qya}:
\begin{equation}\label{eq:intro-damu}
	\Delta a_\mu = a_{\mu}^{\rm exp} - a_{\mu}^{\rm SM} = \left( 251 \pm 59 \right) \times 10^{-11}.
\end{equation}
Besides, the Run-I result indicates that the future complete results of Fermilab and/or Japan Proton Accelerator Research Complex (J-PARC) experiments are very likely to confirm the excess of $a_\mu$ at $5\sigma$ discovery level. This expectation implies that the long-standing discrepancy of the muon anomalous magnetic moment between the SM prediction and experimental measurements, $\Delta a_\mu$, may be the most promising hint of the new physics beyond the SM. As the best candidate theory for new physics, supersymmetric models predict the scalar partners of muon and $\mu$-type neutrino. It was speculated about twenty years ago that the source of the observed $\Delta a_\mu$ might be the quantum effect contributed by these supersymmetric particles (sparticles)~\cite{Martin:2001st,Czarnecki:2001pv,Stockinger:2006zn}. Along this direction, numerous studies have been carried out in the minimal supersymmetric standard model (MSSM) and its extensions (see, e.g.,~\cite{Cao:2011sn,Kang:2016iok,Zhu:2016ncq,Yanagida:2017dao, Hagiwara:2017lse, Cox:2018qyi,Tran:2018kxv,Padley:2015uma,
Choudhury:2017fuu,Okada:2016wlm,Du:2017str, Ning:2017dng, Wang:2018vxp,Yang:2018guw,Liu:2020nsm,Cao:2019evo,Cao:2021lmj}).

At present, low-energy supersymmetric theories are subjected to the increasingly tight constraints from LHC experiments and DM search experiments. In the MSSM, the lightest neutralino $\tilde{\chi}_1^0$, if it is the lightest supersymmetric particle (LSP), can act as a DM candidate accounting for the Planck measured relic density~\cite{Aghanim:2018eyx}. The likelihood analysis of the phenomenological MSSM in 11 free-parameter space showed that the DM $\tilde{\chi}_1^0$ must be  bino-dominated within the $1\sigma$ confidence level (see Figure 12 and Table 6 of Ref.~\cite{Bagnaschi:2017tru}). Aiming at the proper recasting of the LHC Run-II data for sparticle search and assuming that $\tilde{\chi}_1^0$ provides the full DM relic density, recent studies investigated the phenomenology of the bino DM co-annihilating with wino-dominated $\tilde{\chi}_1^\pm$ or with sleptons $\tilde{\ell}_{L/R}$~\cite{Chakraborti:2020vjp, Chakraborti:2021dli}. The result revealed that it is difficult to obtain the correct DM relic density and experimentally compatible DM-nucleon scatterings in the natural parameter space to explain $\Delta a_\mu$\footnote{When we mention the term ``natural" in this work, it means that higgsinos lighter than about $500~{\rm GeV}$ are preferred to predict $Z$ boson mass without causing serious offsets between the different contributions of $m_Z$. }. The more recent research opened up the possibility of wino and higgsino DM by giving up $\tilde{\chi}_1^0$ to provide the full DM relic density~\cite{Chakraborti:2021kkr}.
In the next-to minimal supersymmetric standard model with a $\mathbb{Z}_3$ symmetry ($\mathbb{Z}_3$-NMSSM), which is another economic realization of supersymmetry, the DM $\tilde{\chi}_1^0$ may be either bino-dominated or singlino-dominated~\cite{Cao:2016nix,Ellwanger:2016sur,Xiang:2016ndq,Baum:2017enm,Ellwanger:2018zxt,Domingo:2018ykx, Baum:2019uzg,vanBeekveld:2019tqp, Abdallah:2019znp,Cao:2019qng,Guchait:2020wqn}. The situation of the theory is similar to that of the MSSM, i.e., the parameter space to explain $\Delta a_\mu$ naturally have been
constrained tightly by the LHC and DM experiments~\cite{Cao:2016cnv,Cao:2018rix,Zhou:2021pit}.

\par Considering the increasing tension between natural interpretations of $a_\mu$ and the experimental constraints, the present paper aims to study the combined constraints on the singlino-dominated DM scenario in the GNMSSM from the DM relic density, spin-dependent (SD) and spin-independent (SI) direct detection experiments, sparticle direct searches at the LHC, and the existing muon $(g-2)$ measurement in Eq.~(\ref{eq:intro-damu}). The rest of this paper is organized as follows. In the next section, the basics of the GNMSSM and the annihilation mechanism of singlino-dominated DM is reviewed. In Section \ref{sec:ana}, the relevant parameter space is scanned in a specific scenario of the GNMSSM, which extends the $\mathbb{Z}_3$-NMSSM by adding an explicit $\mu$-term and is called $\mu$NMSSM hereafter. It is found that $(g-2)_\mu$ can be properly interpreted without conflicting with any experimental observations. In Section \ref{sec:LHC}, the constraints from the LHC searches for supersymmetry on the interpretation are studied comprehensively by specific Monte Carlo simulations to reveal their features. The theory's capabilities to relax the experimental constraints are emphasized. Conclusions about the $\mu$NMSSM interpretation of $\Delta a_\mu$ are drawn in Section \ref{sec:summary}.

\section{Theoretical preliminaries}
\subsection{The basics of GNMSSM}
Compared with the MSSM, the NMSSM introduces a singlet Higgs superfield $\hat{S}$. Given the superfield composition, the general form of the NMSSM superpotential is~\cite{Maniatis:2009re, Ellwanger:2009dp}
\begin{equation}\label{eq:gnmssm-sp}
	W_{\rm GNMSSM} = W_{\rm Yukawa} + \left( \mu + \lambda \hat{S} \right) \hat{H_u} \cdot \hat{H_d}+\frac{1}{3} \kappa \hat{S}^3 + \frac{1}{2} \nu \hat{S}^2,
\end{equation}
where the Yukawa terms $W_{\rm Yukawa}$ are the same as those in MSSM. The scenario of $\mu = \nu = 0$ possess an accidental $\mathbb{Z}_3$-symmetry, and the theory is defined in a scale-invariant form. Clearly, if the dimensional parameters $\mu$ and $\nu$ are non-vanishing, they should be at the weak or supersymmetry-breaking scale to break the electroweak symmetry without fine-tuning. These $\mathbb{Z}_3$ broken terms $\mu$ and $\nu$ are introduced in some works to solve the tadpole problem~\cite{Ellwanger:1983mg, Ellwanger:2009dp} and the cosmological domain-wall problem~\cite{Abel:1996cr, Kolda:1998rm, Panagiotakopoulos:1998yw}.
Past studies~\cite{Abel:1996cr, Lee:2010gv,Lee:2011dya,Ross:2011xv,Ross:2012nr} demonstrated that the electroweak scale $\mu$ and $\nu$ may come from the discrete $\mathbb{Z}^R_4$ or $\mathbb{Z}^R_8$ symmetry breaking at high-energy scale. Furthermore, the scale-invariant $\mathbb{Z}_3$-NMSSM allows it to be embedded into canonical superconformal supergravity in the Jordan frame. The superconformal symmetry of matter multiplets can be broken via a non-minimal interaction $\chi \hat{H}_u \cdot \hat{H}_d R$, where $R$ is the supersymmetric version of the Ricci scalar. As shown in Ref.~\cite{Ferrara:2010in}, the dimensionless coupling $\chi$ can drive inflation in the early Universe, and can also provide a $\mu$ term correction to the $\mathbb{Z}_3$-NMSSM superpotential, where $\mu = \frac{3}{2} m_{3/2} \chi $, with $m_{3/2}$ denoting the gravitino mass.

\par This work treats $\mu$ and $\nu$ as free theoretical input parameters, irrespective of their physical origin. Particularly for the sake of brevity without loss of generality, a specific scenario of the GNMSSM characterized by $\nu \equiv 0$ is investigated~\cite{Cao:2021ljw}. This so-called $\mu$-term extended $\mathbb{Z}_3$-NMSSM, i.e., $\mu$NMSSM, is defined by its superpotential and the soft breaking Lagrangian as follows~\cite{Hollik:2018yek, Hollik:2020plc}
\begin{equation}\begin{split}
	W_{\rm \mu NMSSM} &= W_{\rm Yukawa} + (\lambda \hat{S}+\mu) \hat{H_u} \cdot \hat{H_d}+\frac{1}{3} \kappa \hat{S}^3,  \\
-\mathcal{L}_{\rm soft} &=\Bigg[A_{\lambda}\lambda S H_u \cdot H_d + \frac{1}{3} A_{\kappa} \kappa S^3+B_\mu \mu H_u\cdot H_d +h.c.\Bigg]  \\
& + m^2_{H_u}|H_u|^2 + m^2_{H_d}|H_d|^2 + m^2_{s}|S|^2 + \cdots,
\end{split}\end{equation}
where $H_u$, $H_d$, and $S$ are the scalar parts of superfields $\hat{H}_u$, $\hat{H}_d$, and $\hat{S}$, respectively. After the electroweak symmetry breaking, the neutral Higgs fields acquire non-zero vacuum expectation values (vevs),
\begin{equation}
	\left\langle H_u^0 \right\rangle 	= \frac{1}{\sqrt{2}} v_u, \quad
	\left\langle H_d^0 \right\rangle 	= \frac{1}{\sqrt{2}} v_d, \quad
	\left\langle S \right\rangle 		= \frac{1}{\sqrt{2}} v_s,
\end{equation}
where $v = \sqrt{v_u^2 + v_d^2} = 246~{\rm GeV}$. In practice, the free input parameters of the Higgs sector can be taken as follows\footnote{Since $B_{\mu}$ plays a minor role in the DM phenomenology~\cite{Hollik:2018yek,Cao:2021ljw} and muon $g-2$, $B_{\mu}$ is fixed as zero in this work.}
\begin{equation}
	\lambda, \quad
	\kappa,	\quad
	A_\lambda, \quad
	A_\kappa, \quad
	\mu, \quad
	\mu_{\rm eff} = \frac{1}{\sqrt{2}} \lambda v_s, \quad
	\tan{\beta} = \frac{v_u}{v_d}.
\end{equation}
\par In the field convention that $H_{\rm SM} \equiv \sin\beta {\rm Re}(H_u^0) + \cos\beta {\rm Re} (H_d^0)$, $H_{\rm NSM} \equiv \cos\beta {\rm Re}(H_u^0) - \sin\beta {\rm Re}(H_d^0)$, and $A_{\rm NSM} \equiv \cos\beta {\rm Im}(H_u^0) - \sin\beta  {\rm Im}(H_d^0)$~\cite{Miller:2003ay, Cao:2012fz}, the elements of the $CP$-even Higgs boson mass matrix $\mathcal{M}_S^2$ in the bases $\left(H_{\rm NSM}, H_{\rm SM}, {\rm Re}(S)\right)$ are read as
\begin{equation}\label{eq:mmeh2}
\begin{split}
\mathcal{M}^2_{S, 11}	&= \frac{2 \mu_{\rm eff} (\lambda A_\lambda + \kappa \mu_{\rm eff})}{\lambda \sin 2 \beta} + \frac{1}{2} (2 m_Z^2- \lambda^2v^2)\sin^22\beta,  \\
\mathcal{M}^2_{S, 12}	&= -\frac{1}{4}(2 m_Z^2-\lambda^2v^2)\sin4\beta,  \\
\mathcal{M}^2_{S, 13}	&= -\frac{1}{\sqrt{2}} ( \lambda A_\lambda + 2 \kappa \mu_{\rm eff}) v \cos 2 \beta,  \\
\mathcal{M}^2_{S, 22} 	&= m_Z^2\cos^2{2\beta}+ \frac{1}{2} \lambda^2v^2\sin^2{2\beta},  \\
\mathcal{M}^2_{S, 23}	&= \frac{v}{\sqrt{2}} \left( 2 \lambda \mu_{\rm eff} + 2 \lambda \mu - (\lambda A_\lambda + 2 \kappa \mu_{\rm eff}) \sin2\beta \right),  \\
\mathcal{M}^2_{S, 33}	&= \frac{\lambda A_\lambda \sin 2 \beta}{4 \mu_{\rm eff}} \lambda v^2   + \frac{\mu_{\rm eff}}{\lambda} \left(\kappa A_\kappa +  \frac{4 \kappa^2 \mu_{\rm eff}}{\lambda} \right) - \frac{\lambda \mu}{2 \mu_{\rm eff}} \lambda v^2.
\end{split}
\end{equation}
Dropping the Goldstone mode, those elements for $CP$-odd Higgs fields in the bases $\left( A_{\rm NSM}, {\rm Im}(S)\right)$ are given by
\begin{equation}\label{eq:mmeA2}
\begin{split}
\mathcal{M}^2_{P,11}&= \frac{2 \mu_{\rm eff} (\lambda A_\lambda + \kappa \mu_{\rm eff})}{\lambda \sin 2 \beta},   \quad\quad
\mathcal{M}^2_{P,12}= \frac{v}{\sqrt{2}} ( \lambda A_\lambda - 2 \kappa \mu_{\rm eff}) ,\\
\mathcal{M}^2_{P,22}&= \frac{(\lambda A_\lambda + 4 \kappa \mu_{\rm eff}) \sin 2 \beta }{4 \mu_{\rm eff}} \lambda v^2  - \frac{3 \mu_{\rm eff}}{\lambda} \kappa A_\kappa - \frac{\lambda \mu}{2 \mu_{\rm eff}} \lambda v^2.   \\
\end{split}
\end{equation}
Three $CP$-even mass eigenstates $h$, $H$, and $h_s$ are achieved by a unitary matrix $V$ to diagonalize $\mathcal{M}_S^2$. Similarly, two $CP$-odd mass eigenstates $A_H$ and $A_s$ are defined via rotation matrix $U$. Among them, $h$ corresponds to the scalar state discovered at the LHC, $H$ and $A_H$ represent the doublet dominated states which are preferred to be heavy by the LHC search for extra Higgs bosons, and $h_s$ and $A_s$ represent the singlet-dominated states. The mass of the charged Higgs state $H^\pm$ is expressed as
\begin{equation}
	m^2_{H^{\pm}} = \frac{2\mu_{\rm eff}}{\sin 2 \beta} \left(\frac{\kappa}{\lambda}\mu_{\rm eff} + A_{\lambda}\right) + m^2_W -\lambda^2 v^2.
\end{equation}
\par The fermion parts of Higgs superfields $\left(\tilde{H}_u, \tilde{H}_d, \tilde{S} \right)$ and gauginos $\left( \tilde{B}, \tilde{W} \right)$ form five neutralino states and two chargino states, and they are referred as electroweakinos (EWinos) in general. The symmetric neutralino mass matrix in the gauge eigenstate bases of $\psi^0 = \left( -i \tilde{B}, - i \tilde{W}^0, \tilde{H}_{d}^0, \tilde{H}_{u}^0, \tilde{S} \right)$ is
\begin{equation}\label{eq:mmn}
\mathcal{M}_{\tilde{N}} = \begin{pmatrix}
	M_1 			& 0 	& -c_\beta s_W m_Z  &  s_\beta s_W m_Z & 0 \\
 	0 				& M_2 	&  c_\beta c_W m_Z 	& -s_\beta c_W m_Z & 0 \\
	-c_\beta s_W m_Z& c_\beta c_W m_Z 	& 0 & -\mu - \mu_{\rm eff} & - \frac{1}{\sqrt{2}}s_\beta \lambda v  \\
	s_\beta s_W m_Z & -s_\beta c_W m_Z	& -\mu-\mu_{\rm eff} & 0 & -\frac{1}{\sqrt{2}} c_\beta \lambda v  \\
	0 & 0 & - \frac{1}{\sqrt{2}} s_\beta \lambda v & -\frac{1}{\sqrt{2}}c_\beta \lambda v & \frac{2\kappa}{\lambda}\mu_{\rm eff}
\end{pmatrix},
\end{equation}
where the abbreviations $s_W = \sin{\theta_W}$ and $c_W = \cos{\theta_W}$ are used, with $\theta_W$ being the weak mixing angle, and $s_\beta = \sin{\beta}$ and $c_\beta = \cos{\beta}$. Similarly, the chargino mass matrix in the bases $\psi^\pm = \left( \tilde{W}^+, \tilde{H}^+_u , \tilde{W}^-, \tilde{H}^-_d \right) $ is
\begin{equation}\label{eq:mmc}
	\mathcal{M}_{\tilde{C}} = \begin{pmatrix}
		0_{2\times 2}	& X^{T}_{2\times 2} \\
		X_{2\times 2}	& 0_{2\times 2}
	\end{pmatrix}, \quad {\rm with} \quad
	X_{2 \times 2} = \begin{pmatrix}
		M_2 & \sqrt{2} s_{\beta} m_W \\
		\sqrt{2} c_{\beta} m_W & \mu + \mu_{\rm eff}
	\end{pmatrix}.
\end{equation}
After diagonalization, one arrives at  the neutralino $\tilde{\chi}_i^0$ and chargino $\tilde{\chi}_i^\pm$ as mass eigenstates, with increasing mass for a higher label $i$.
\par The smuon mass matrix in the gauge eigenstate bases $\left( \tilde{\mu}_L, \tilde{\mu}_R \right)$ is given as
\begin{equation}\label{eq:mmsmu}
	\mathcal{M}_{\tilde{\mu}}^2 = \begin{pmatrix}
		m_\mu^2 + m_{\tilde{\mu}_L}^2 + (s_W^2 - \frac{1}{2}) m_Z^2 \cos{2\beta} & \quad m_\mu \left [ A_\mu - (\mu + \mu_{\rm eff}) \tan\beta \right ] \\
		m_\mu \left [ A_\mu - (\mu + \mu_{\rm eff}) \tan\beta \right ] & m_\mu^2 + m_{\tilde{\mu}_R}^2 - s_W^2 m_Z^2 \cos{2\beta}
	\end{pmatrix},
\end{equation}
where $A_\mu$, $m_{\tilde{\mu}_L}$, and $m_{\tilde{\mu}_R}$ are muon-type soft breaking parameters. Eq.~(\ref{eq:mmsmu}) indicates that the left-right mixing term is dominated by $(\mu + \mu_{\rm eff})\tan{\beta}$, so $A_\mu$ is fixed as zero in the following. The muon-type sneutrino mass is
\begin{equation}
	m_{\tilde{\nu}_\mu}^2 = m_{\tilde{\mu}_L}^2 + \frac{1}{2}m_Z^2 \cos{2\beta}.
\end{equation}

\subsection{Muon $g-2$ in $\mu$NMSSM}
The SUSY contribution $a_{\mu}^{\rm SUSY}$, in which the muon lepton number is carried by $\tilde{\mu}$ or $\tilde{\nu}_\mu$ in the loops\footnote{In the NMSSM, besides the contributions from the SM particles in the loops, heavy doublet-dominated Higgs bosons can also mediate the contribution to $a_\mu$. However, after considering the constraints from the LHC searches for extra Higgs bosons and the measurements of the branching ratios for $B \to X_s \gamma$ and $B_s \to \mu^+ \mu^-$, this contribution is negligibly small because the Higgs bosons should be very massive for a large $\tan \beta$. In addition, although the contribution from light singlet-dominated Higgs bosons might reach ${\cal{O}}(10^{-10})$ as pointed out in~\cite{Domingo:2008bb}, it is negligible in this study. The reason is that the constraints from the DM direct detection experiments strongly favor a small $\lambda$ for a singlino-dominated DM, and consequently, the singlet-doublet Higgs mixings and their related $\bar{\mu} \mu A_s$ and $\bar{\mu} \mu h_s$ couplings are suppressed significantly. It was testified numerically that the total Higgs-mediated contributions are less than $10^{-10}$ for the samples obtained in this study. Besides, $a_\mu^{\rm SUSY}$ has two-loop contribution~\cite{Heinemeyer:2003dq,Heinemeyer:2004yq,Fargnoli:2013zia,Fargnoli:2013zda,vonWeitershausen:2010zr,Arhrib:2001xx,Zhao:2013sou,Zhao:2012zze,Chang:2000ii,Fargnoli:2014rrd,Stockinger:2004vf,Feng:2008cn}. A recent analysis revealed that the correction is less than $4 \times 10^{-10}$~\cite{Zhao:2021eaa}. We anticipate that it can be further suppressed if the restrictions from the LHC search for SUSY and DM physics are considered.}, can be the source of $\Delta a_\mu$~\cite{Moroi:1995yh,Hollik:1997vb,Martin:2001st}. The expression of $a_\mu^{\rm SUSY}$ in the $\mu$NMSSM is similar to that in the MSSM~\cite{Domingo:2008bb}, which is given by~\cite{Martin:2001st}:
\begin{small}\begin{equation}\label{eq:ammususy}\begin{split}
	&a_{\mu}^{\rm SUSY} = a_{\mu}^{\tilde{\chi}^0 \tilde{\mu}} + a_{\mu}^{\tilde{\chi}^{\pm} \tilde{\nu}},\\
    a_{\mu}^{\tilde{\chi}^0 \tilde{\mu}} &= \frac{m_{\mu}}{16 \pi^2}\sum_{i,l}\left\{
    -\frac{m_{\mu}}{12 m_{\tilde{\mu}_l}^2} \left( |n_{il}^{\rm L}|^2 + |n_{il}^{\rm R}|^2 \right) F_1^{\rm N}(x_{il}) + \frac{m_{\tilde{\chi}_i^0}}{3 m_{\tilde{\mu}_l}^2} {\rm Re}(n_{il}^{\rm L} n_{il}^{\rm R}) F_2^{\rm N}(x_{il})
    \right\},\\
    a_{\mu}^{\tilde{\chi}^\pm \tilde{\nu}} &= \frac{m_{\mu}}{16 \pi^2}\sum_{k}\left\{
    \frac{m_{\mu}}{12 m_{\tilde{\nu}_{\mu}}^2} \left( |c_{k}^{\rm L}|^2 + |c_{k}^{\rm R}|^2 \right) F_1^{\rm C}(x_{k}) + \frac{2 m_{\tilde{\chi}_k^\pm}}{3 m_{\tilde{\nu}_{\mu}}^2} {\rm Re}(c_{k}^{\rm L}c_{k}^{ \rm R}) F_2^{\rm C}(x_{k})
    \right\},
\end{split}
\end{equation}\end{small}
where $i=1,\cdots,5$, $k=1,2$, and $l=1,2$ denote the neutralino, chargino, and smuon index, respectively, and
\begin{equation}
    \begin{split}
        n_{il}^{\rm L} 	= \frac{1}{\sqrt{2}}\left( g_2 N_{i2} + g_1 N_{i1} \right)X^*_{l1} -y_{\mu} N_{i3}X^*_{l2}, \quad
        &n_{il}^{\rm R} = \sqrt{2} g_1 N_{i1} X_{l2} + y_{\mu} N_{i3} X_{l1},\\
        c_{k}^{\rm L}  	= -g_2 V^{\rm c}_{k1}, \quad
        &c_{k}^{\rm R} 	= y_{\mu} U^{\rm c}_{k2}.\\
    \end{split}
\end{equation}
Here, $N$ is the neutralino mass rotation matrix, $X$ the smuon mass rotation matrix, and $U^{\rm c}$ and $V^{\rm c}$ the chargino mass rotation matrices defined by ${U^{\rm c}}^* X_{2\times2} {V^{\rm c}}^\dag = m_{\tilde{\chi}^\pm}^{\rm diag}$. The kinematic loop functions $F(x)$s depend on the variables $x_{il} \equiv m_{\tilde{\chi}_i^0}^2 / m_{\tilde{\mu}_l}^2$ and $x_{k} \equiv m_{\tilde{\chi}_k^\pm}^2 / m_{\tilde{\nu}_{\mu}}^2$, and are given by
\begin{eqnarray}
F^N_1(x) & = &\frac{2}{(1-x)^4}\left[ 1-6x+3x^2+2x^3-6x^2\ln x\right], \\
F^N_2(x) & = &\frac{3}{(1-x)^3}\left[ 1-x^2+2x\ln x\right], \\
F^C_1(x) & = &\frac{2}{(1-x)^4}\left[ 2+ 3x - 6 x^2 + x^3 +6x\ln x\right], \\
F^C_2(x) & = & -\frac{3}{2(1-x)^3}\left[ 3-4x+x^2 +2\ln x \right].
\end{eqnarray}
They satisfy $F^N_1(1) = F^N_2(1) = F^C_1(1) = F^C_2(1) = 1$ for mass-degenerate sparticle case.

It is instructive to point out that, although the $\mu$NMSSM predicts five neutralinos, the singlino-induced contribution never makes sense, and the $\mu$NMSSM prediction of $a_\mu^{\rm SUSY}$ is roughly the same as that of the MSSM except that the $\mu$ parameter of the MSSM should be replaced by $\mu + \mu_{eff}$. This feature can be understood by noting the fact that the field operator for $a_\mu$ involves chiral flipped Muon leptons and adopting the mass insertion approximation in the calculation of $a_\mu$~\cite{Moroi:1995yh}. Specifically, the contributions to $a_\mu^{\rm SUSY}$ in the MSSM can be classified into four types: "WHL", "BHL", "BHR", and "BLR", where $W$, $B$, $H$, $L$, and $R$ stands for wino, bino, higgsino, left-handed and right-handed smuon field, respectively. They arise from the Feynman diagrams involving $\tilde{W}-\tilde{H}_d$, $\tilde{B}-\tilde{H}_d^0$, $\tilde{B}-\tilde{H}_d^0$, and $\tilde{\mu}_L-\tilde{\mu}_R$ transitions, respectively. Their concrete expressions are~\cite{Athron:2015rva, Moroi:1995yh,Endo:2021zal}
\begin{eqnarray}
a_{\mu, \rm WHL}^{\rm SUSY}
    &=&\frac{\alpha_2}{8 \pi} \frac{m_\mu^2 \mu M_2 \tan \beta}{m_{\tilde{\nu}_\mu}^4} \left \{ 2 f_C\left(\frac{M_2^2}{m_{\tilde{\nu}_{\mu}}^4}, \frac{\mu^2}{m_{\tilde{\nu}_{\mu}}^2} \right) - \frac{m_{\tilde{\nu}_\mu}^4}{m_{\tilde{\mu}_L}^4} f_N\left(\frac{M_2^2}{m_{\tilde{\mu}_L}^2}, \frac{\mu^2}{m_{\tilde{\mu}_L}^2} \right) \right \}\,, \quad \quad
    \label{eq:WHL} \\
a_{\mu, \rm BHL}^{\rm SUSY}
  &=& \frac{\alpha_Y}{8 \pi} \frac{m_\mu^2 \mu M_1 \tan \beta}{m_{\tilde{\mu}_L}^4} f_N\left(\frac{M_1^2}{m_{\tilde{\mu}_L}^2}, \frac{\mu^2}{m_{\tilde{\mu}_L}^2} \right)\,,
    \label{eq:BHL} \\
a_{\mu, \rm BHR}^{\rm SUSY}
  &=& - \frac{\alpha_Y}{4\pi} \frac{m_{\mu}^2 \mu M_1 \tan \beta}{m_{\tilde{\mu}_R}^4} f_N\left(\frac{M_1^2}{m_{\tilde{\mu}_R}^2}, \frac{\mu^2}{m_{\tilde{\mu}_R}^2} \right)\,,
    \label{eq:BHR} \\
a_{\mu \rm BLR}^{\rm SUSY}
  &=& \frac{\alpha_Y}{4\pi} \frac{m_{\mu}^2 \mu M_1 \tan \beta}{M_1^4}
    f_N\left(\frac{m_{\tilde{\mu}_L}^2}{M_1^2}, \frac{m_{\tilde{\mu}_R}^2}{M_1^2} \right)\,,
    \label{eq:BLR}
\end{eqnarray}
where the loop functions are given by
\begin{eqnarray}
    \label{eq:loop-aprox}
    f_C(x,y)
    &=&  \frac{5-3(x+y)+xy}{(x-1)^2(y-1)^2} - \frac{2\ln x}{(x-y)(x-1)^3}+\frac{2\ln y}{(x-y)(y-1)^3} \,,
      \\
    f_N(x,y)
    &=&
      \frac{-3+x+y+xy}{(x-1)^2(y-1)^2} + \frac{2x\ln x}{(x-y)(x-1)^3}-\frac{2y\ln y}{(x-y)(y-1)^3} \,,
\end{eqnarray}
and they satisfy $f_C(1,1) = 1/2$ and $f_N(1,1) = 1/6$. In the $\mu$NMSSM, the singlino field $\tilde{S}$ can also enter the insertions, but because both the $\tilde{W}-\tilde{S}$ and $\tilde{B}^0-\tilde{S}$ transitions and the $\bar{\mu} \tilde{S} \tilde{\mu}_{L,R}$ couplings vanish, it only appears in the "WHL", "BHL" and "BHR" loops by two more insertions at the lowest order, which corresponds to the $\tilde{H}_d^0-\tilde{S}$ and $\tilde{S}-\tilde{H}_d^0$ transitions in the neutralino mass matrix in Eq.~(\ref{eq:mmn}), respectively. Since a massive singlino and  a small $\lambda$ are preferred by DM physics (see discussion below), the singlino-induced contribution can not be significantly large\footnote{In fact, we performed a comprehensive study about the characteristics of the samples that survive all the experimental constraints. We found that, for all the sample obtained in this work, resetting $\lambda = 0.001$ (note that the finiteness of $m_{\tilde{S}}$ requires $\lambda \neq 0$) and keeping the other parameters unchanged increase $a_\mu^{\rm SUSY}$ by less than $10\%$.}.

It should be emphasized that, although $a_\mu^{\rm SUSY}$ has roughly the same properties in the $\mu$NMSSM and MSSM, it is subject to significantly relaxed experimental and theoretical limitations in the $\mu$NMSSM (see following discussions). Thus, the $\mu$NMSSM is more readily to explain the $a_\mu$ discrepancy, which is the main motivation of this work. In addition, the WHL contribution is usually much larger than the other contributions if $\tilde{\mu}_L$ is not significantly heavier than $\tilde{\mu}_R$.

\subsection{Singlino-dominated DM}
This work aims to reveal DM physics in the natural parameter space of interpreting $\Delta a_\mu$. In the $\mathbb{Z}_3$-NMSSM, the properties of the singlino-dominated DM are mainly determined by three parameters: $\lambda$, $\mu$, and $m_{\tilde{\chi}_1^0}$~\cite{Zhou:2021pit}. The Yukawa coupling $\kappa$ is related with $m_{\tilde{\chi}_1^0}$ and satisfies $2|\kappa| < \lambda$ to ensure that $\tilde{\chi}_1^0$ is singlino-dominated. The DM obtained the correct relic abundance mainly through co-annihilation with higgsinos in the early Universe, and  $\lambda$ should be less than 0.1 to suppress the DM direct detection rate~\cite{Cao:2018rix, Zhou:2021pit}. As a result, the parameters in the $\mathbb{Z}_3$-NMSSM are highly constrained. In contrast, due to the introduction of the additional $\mu$ term in the $\mu$NMSSM, the DM properties are described by four Higgs parameters: $\lambda$, $\kappa$, $\mu$, and $m_{\tilde{\chi}_1^0} \simeq \frac{2\kappa}{\lambda} \mu_{\rm eff}$, and a singlino-dominated DM does not require $2|\kappa| < \lambda$~\cite{Cao:2021ljw}. This causes the singlino-dominated DM properties in the $\mu$NMSSM to be significantly different from those in the $\mathbb{Z}_3$-NMSSM.

Assuming a standard thermal history of the Universe with a singlino-dominated DM candidate in thermal equilibrium until it freezed out, DM annihilation rate must be sufficiently large to be consistent with the Planck observation $\Omega_{\rm DM}h^2 = 0.120 \pm 0.001$. Various processes, as follows, may provide a sufficiently large annihilation cross-section~\cite{Cao:2021ljw}.
\begin{itemize}
	\item $\tilde{\chi}_1^0 \tilde{\chi}_1^0 \to h_s A_s$. This process is mainly carried out through $s$-channel exchange of $Z$ and $CP$-odd Higgs $A_s$ and $t$-channel exchange of neutralinos~\cite{Baum:2017enm, Griest:1990kh}. As shown in Eq.~(2.19) in~\cite{Cao:2021ljw}, the $t$-channel annihilation cross-section is roughly proportional to $\kappa^4$. In the $t$-channel dominated case, the parameters should satisfy the relationship~\cite{Baum:2017enm}
	\begin{equation}
		|\kappa| \sim 0.15 \times \left ( \frac{m_{\tilde{\chi}_1^0}}{300~{\rm GeV}} \right )^{1/2}
	\end{equation}
to obtain the measured density. For the case of $m_{A_s} \simeq 2\times m_{\tilde{\chi}_1^0}$, the cross-section is enhanced by the $s$-channel pole. However, the $A_s$ pole enhancement may lead to a very light singlet Higgs boson $h_s$. This $h_s$ must satisfy the constraints from LEP Higgs searches and predict a small ${\rm BR}(h\to h_s h_s)$ to fit SM-like Higgs data. Besides, the SI direct detection rate mediated by $h_s$ must satisfy the current bound from Xenon-1T~\cite{Aprile:2018dbl} and PandaX-II~\cite{Wang:2020coa}. A comparative study of light $h_s$ scenario ($m_{h_s} < m_h$) and heavy $h_s$ scenario ($m_{h_s} > m_h$) indicated that the latter scenario is preferred by a Bayes factor 2.42~\cite{Cao:2021ljw}\footnote{A Bayes factor is the ratio of the likelihood of one particular hypothesis to the likelihood of another. It can be interpreted as a measure of the strength of evidence in favor of one theory among two competing theories~\cite{Tanabashi:2018oca, Kass:1995loi}. A factor of 2.42 is generally considered as a decisive result to indicate
the preference of one theory over another.}. For the above reasons, only the $m_{h_s} > m_{h}$ scenario will be considered (i.e., $h_1=h$ and $h_2=h_s$) in the following numerical study.
	\item $\tilde{\chi}_1^0 \tilde{\chi}_1^0 \to t\bar{t}$. This process is mediated by the $s$-channel exchange of $Z$ and Higgs bosons. It is significant only when $\lambda$ is sizable, but this usually leads to a sizable DM-nucleon scattering rate~\cite{Zhou:2021pit}.
	\item Co-annihilation with EWinos. In principle, this channel affects the abundance when the mass splitting between $\tilde{\chi}_1^0$ and a co-annihilation particle is less than approximately $10\%$~\cite{Zhou:2021pit}.
	\item Co-annihilation with smuons $\tilde{\mu}_L$/$\tilde{\mu}_R$ or $\mu$-type sneutrino $\tilde{\nu}_\mu$. Within the interpretation of $\Delta a_\mu$, smuons should not be too heavy, so this co-annihilation channel can be opened when $m_{\tilde{\mu}} \gtrapprox m_{\tilde{\chi}_1^0}$.
\end{itemize}

In the heavy squark limits, SI DM-nucleon scattering is mainly from $t$-channel exchange of $CP$-even Higgs bosons, and the cross-section is given as~\cite{Badziak:2016qwg, Pierce:2013rda}
\begin{equation}
	\sigma^{\rm SI}_{N} = \frac{4 \mu_r^2}{\pi} |f^{(N)}|^2, \quad
	f^{(N)} =  \sum_{i}^3 f^{(N)}_{h_i} = \sum_{i}^3 \frac{C_{ \tilde {\chi}^0_1 \tilde {\chi}^0_1 h_i} C_{N N h_i }}{2m^2_{h_i} },
\end{equation}
where $\mu_r = m_N m_{\tilde{\chi}_1^0}/(m_N + m_{\tilde{\chi}_1^0} )$ is the reduced mass of the DM-nucleon system, and $C_{N N h_i }$ is the coupling of a Higgs boson with a nucleon,
\begin{equation}
C_{NN h_i} = -\frac{m_N}{v}\left( F^{(N)}_d \left( V_{i2}- \tan\beta V_{i1} \right)+F^{(N)}_u \left(V_{i2} +\frac{1}{\tan\beta} V_{i1} \right)
\right).
\end{equation}
Here $F^{(N)}_d = f^{(N)}_d+f^{(N)}_s+\frac{2}{27}f^{(N)}_G$ and $F^{(N)}_u = f^{(N)}_u+\frac{4}{27}f^{(N)}_G$ with form factor $f^{(N)}_q =m_N^{-1}\langle N | m_qq\bar{q} |N\rangle$ and $f^{(N)}_G=1-\sum_{q=u,d,s} f^{(N)}_q$.
In the heavy $H$ case, the SI cross-section is dominated by two light Higgs contributions, and the $h$-mediated contribution is usually significantly larger than the $h_s$-mediated contribution.
The couplings $C_{\tilde {\chi}^0_1 \tilde {\chi}^0_1 h}$ and $C_{\tilde {\chi}^0_1 \tilde {\chi}^0_1 h_s}$ are given by~\cite{Cao:2021ljw}
\begin{small}\begin{equation}
\begin{split}
	C_{\tilde {\chi}^0_1 \tilde {\chi}^0_1 h} &\simeq
	\frac{\mu + \mu_{\rm eff}}{ v}\,\left( \frac{\lambda v}{\mu + \mu_{\rm eff}} \right)^2\, \frac { N_{15}^2 V_{12}(m_{\tilde{\chi}_1^0}/(\mu + \mu_{\rm eff}) -\sin 2 \beta)}{1-(m_{\tilde{\chi}_1^0}/(\mu + \mu_{\rm eff}))^2}\\
	&+ \frac{\lambda}{2 \sqrt{2}} \left( \frac{\lambda v}{\mu + \mu_{\rm eff}} \right)^2 \frac{N_{15}^2 V_{13} \sin2\beta}{ 1-(m_{\tilde{\chi}_1^0}/(\mu + \mu_{\rm eff}))^2 } \\
	& -\sqrt{2}\kappa N_{15}^2 V_{13} \left[1+ \left( \frac{\lambda v}{\sqrt{2}(\mu + \mu_{\rm eff})} \right)^2\frac{1}{1-(m_{\tilde{\chi}_1^0}/(\mu + \mu_{\rm eff}))^2} \frac{\mu_{\rm eff}}{\mu + \mu_{\rm eff}} \right],\\
	C_{\tilde {\chi}^0_1 \tilde {\chi}^0_1 h_{s}} &\simeq
	\frac{\mu + \mu_{\rm eff}}{ v}\,\left( \frac{\lambda v}{\mu + \mu_{\rm eff}} \right)^2\, \frac { N_{15}^2 V_{22}(m_{\tilde{\chi}_1^0}/(\mu + \mu_{\rm eff}) -\sin 2 \beta)}{1-(m_{\tilde{\chi}_1^0}/(\mu + \mu_{\rm eff}))^2}\\
	&+ \frac{\lambda}{2 \sqrt{2}} \left( \frac{\lambda v}{\mu + \mu_{\rm eff}} \right)^2 \frac{N_{15}^2 V_{23} \sin2\beta}{ 1-(m_{\tilde{\chi}_1^0}/(\mu + \mu_{\rm eff}))^2 } \\
	& -\sqrt{2}\kappa N_{15}^2 V_{23} \left[1+ \left( \frac{\lambda v}{\sqrt{2} (\mu + \mu_{\rm eff})} \right)^2\frac{1}{1-(m_{\tilde{\chi}_1^0}/(\mu + \mu_{\rm eff}))^2} \frac{\mu_{\rm eff}}{\mu + \mu_{\rm eff}}  \right].
	 \end{split}
\end{equation}
\end{small}
In contrast, the SD scattering cross-section takes the following simple from~\cite{Badziak:2015exr, Badziak:2017uto}:
\begin{equation}\begin{split}
	\sigma_{N}^{\rm SD} &\simeq C_N \times 10^{-4}~{\rm pb} \times \left(\frac{N_{13}^2-N_{14}^2}{0.1}\right)^2, \\
	&\simeq C_N \times 10^{-2}~{\rm pb} \times \left( \frac{\lambda v}{\sqrt{2}(\mu + \mu_{\rm eff})} \right)^4 \left( \frac{N_{15}^2 \cos{2\beta}} {1-(m_{\tilde{\chi}_1^0}/(\mu + \mu_{\rm eff}))^2} \right)^2,
\end{split}\end{equation}
with $C_p \simeq 4.0$ for the proton and $C_n \simeq 3.1 $ for the neutron.
The above formulae indicate that the DM direct detection rate is positively related to the Higgs coupling $\lambda$, roughly in terms of $\lambda v/(\sqrt{2} (\mu + \mu_{eff}))$.

\section{\label{sec:ana}Explaining $\Delta a_\mu$ in $\mu$NMSSM  }

\subsection{Research strategy}
The following relevant parameter space was scanned using the \texttt{MultiNest} technique with the setting $n_{\rm live}=30000$~\cite{Feroz:2008xx,Feroz:2013hea}\footnote{The
parameter $n_{\rm live}$ in the MultiNest algorithm denotes the number of active or live points used to determine the iso-likelihood contours in each iteration~\cite{Feroz:2008xx,Feroz:2013hea}. The larger it is, the more accurate the results, and correspondingly, more samples are obtained in the scan.} to explore the features of the $\mu$NMSSM interpretation of $\Delta a_\mu$:
\begin{equation}\begin{split}
	|M_1| &\leq 1.5~{\rm TeV}, \quad
	100~{\rm GeV} \leq M_2 \leq 1.5~{\rm TeV}, \\
	0 \leq \lambda \leq 0.5, \quad
	&|\kappa| \leq 0.5, \quad
	1\leq \tan{\beta} \leq 60, \quad
	2~{\rm TeV} \leq A_{t} \leq 5~{\rm TeV}, \\
	10~{\rm GeV} \leq \mu \leq 1~{\rm TeV}&, \quad
	100~{\rm GeV} \leq \mu + \mu_{\rm eff} \leq 1~{\rm TeV},\quad
	|A_\kappa| \leq 700~{\rm GeV}, \\
	100~{\rm GeV} &\leq m_{\tilde{\mu}_L} \leq 1~{\rm TeV}, \quad
	100~{\rm GeV} \leq m_{\tilde{\mu}_R} \leq 1~{\rm TeV}.  \label{Scan-range}
\end{split}\end{equation}
All of the input parameters are flatly distributed beforehand. Other SUSY parameters, such as $A_\lambda$, the parameters for the first and third generation sleptons, three generation squarks, and gluino, are fixed at 2 TeV. In the numerical calculation, the model file of the $\mu$NMSSM is constructed through the package \texttt{SARAH-4.14.3}~\cite{Staub:2008uz, Staub:2012pb, Staub:2013tta, Staub:2015kfa}. The particle mass spectra and low-energy observables, such as $a_\mu^{\rm SUSY}$, are generated by the codes \texttt{SPheno-4.0.3}~\cite{Porod:2003um, Porod:2011nf} and \texttt{FlavorKit}~\cite{Porod:2014xia}.  The DM relic density and direct detection cross-sections are calculated using package~\texttt{MicrOMEGAs-5.0.4}~\cite{Belanger:2001fz, Belanger:2005kh, Belanger:2006is, Belanger:2010pz, Belanger:2013oya, Barducci:2016pcb}. The following likelihood function
\begin{eqnarray}
\cal{L} = \left \{ \begin{aligned} &\exp\left[-\frac{1}{2} \left( \frac{a_{\mu}^{\rm SUSY}- 2.51\times 10^{-9}}{5.9\times 10^{-10} }\right)^2\right],\ & &{\rm if\ restrictions\ satisfied}; \\ &\exp\left[-100\right], & &{\rm if\ restrictions\ unsatisfied}. \end{aligned} \right . \label{eq:likeliamm}
 \end{eqnarray}
was constructed to guide the scan, where the restrictions on each sample include:
\begin{itemize}
	\item Higgs data fit. As mentioned above, the lightest $CP$-even Higgs boson $h_1$ corresponds to the SM-like state, and this state must satisfy the constraints from the LHC data using the code \texttt{HiggsSignal-2.2.3}~\cite{Bechtle:2014ewa}. The $p$ value in the fit is required to be larger than 0.05, which implies that the Higgs property coincides with the data at the 95\% confidence level. The extra Higgs states must pass the constraints from the direct searches at the LEP, Tevatron, and LHC, which is implemented by the code \texttt{HiggsBounds-5.3.2}~\cite{Bechtle:2015pma}.
	\item Constraints from $B$-physics observation ${\rm BR}(B_s \to \mu^+ \mu^-)$ and ${\rm BR}(B_s \to X_s \gamma)$ are taken into consideration~\cite{Tanabashi:2018oca}. These two $B$-physics observables must fall into the $2\sigma$ bounds.
	\item DM relic density constraints. The samples are required to have a neutralino LSP, and  the predicted relic density of samples must agree with the Planck measurement~\cite{Aghanim:2018eyx}\footnote{Note that the uncertainty of $\Omega h^2$ is dominated by the $10\%$ theoretical uncertainties in calculating the density, which are much larger than the uncertainty of the Planck measurement.}, i.e., $0.096 \leq \Omega h^2 \leq 0.144$.
	\item Direct detection limits on DM. The SI DM scattering cross-section $\sigma_{p}^{\rm SI}$ is required to be below the constraint from the Xenon-1T experiment~\cite{Aprile:2018dbl}. The SD cross-section $\sigma_{n}^{\rm SD}$ is needed to pass the limits of the Xenon-1T report \cite{Aprile:2019dbj}.
	\item Constraints from LHC sparticles direct searches. In the $\mu$NMSSM explanation of $\Delta a_\mu$, the EWinos and sleptons can be produced at the LHC, and thus restricted by the searches for multi-lepton signals. The constraints implemented in this step are produced using the code \texttt{SModelS-1.2.3}~\cite{Khosa:2020zar}, which contained the experimental analyses in
 simplified models that are summarized in Appendix~\ref{app:smodels}.
\end{itemize}

For each sample obtained in the scan, the stability of its vacuum for the scalar potential consisting of Higgs and the last two generation slepton fields was finally checked by the code Vevacious~\cite{Camargo-Molina:2013qva,Camargo-Molina:2014pwa}. Compared with the MSSM, the vacuum in the $\mu$NMSSM is more stable due to the addition of the singlet Higgs field as dynamical degree of freedom, especially in the case of a small $\lambda$ and a large $\mu_{eff}$~\cite{Hollik:2018yek}, but tremendously large soft-breaking trilinear coefficients may still cause its destabilization~\cite{Beuria:2016cdk}. Generally speaking, the vacuum destabilization occurs in the following situations:
\begin{itemize}
\item One or more tachyonic Higgs masses are predicted. Tachyonic masses are related to the fact that the electroweak point, around which the potential is
expanded, is not a local minimum in the scalar potential, but rather resembles a saddle point or even a local maximum~\cite{Hollik:2018yek}. In this case, the true vacuum lies
at a deeper point along this tachyonic direction. Consequently, the true vacuum has vevs different from the input values, and the electroweak breaking condition does not select a minimum.

In this study, it was found that more than half of the samples encountered in the scan correspond to the tachyonic mass case. They are abandoned directly in the calculation since they can not predict physical mass spectra.

\item The formation of non-standard minima which break the electric and/or color charges, known as charge- and color-breaking (CCB) minima. In the MSSM, the following condition should be satisfied to avoid the CCB vacuum~\cite{Kitahara:2013lfa,Endo:2013lva}:
\begin{eqnarray}
& & \left|y_{\ell, \rm eff} \mu v \tan \beta \right|
\leq  \eta_\ell \bigg[
1. 01  \times 10^2 {\rm GeV} \sqrt{m_{\tilde{\ell}_L} m_{\tilde{\ell}_R}}
+ 1.01 \times 10^2 {\rm GeV}  (m_{\tilde{\ell}_L} + 1.03 m_{\tilde{\ell}_R})
\nonumber \\
& & -2.27 \times 10^4 {\rm GeV}^2
+ \frac{2.97 \times 10^6 {\rm GeV}^3}{m_{\tilde{\ell}_L} + m_{\tilde{\ell}_R}}
- 1.14 \times 10^8 {\rm GeV}^4
  \left( \frac{1}{m^2_{\tilde{\ell}_L}} +  \frac{0.983}{m^2_{\tilde{\ell}_R}} \right)
\bigg], 
\end{eqnarray}
where $y_{\ell, \rm eff}$ with $\ell =\tau, \mu$ are the lepton Yukawa couplings including radiative corrections~\cite{Carena:1999py}, $\eta_\tau \simeq 1$ and $\eta_\mu \simeq 0.88$.
If the singlet-field direction were neglected, the formula could be directly applied to the $\mu$NMSSM, keeping $v_s \neq 0~{\rm GeV}$ and replacing $\mu \to \mu + \mu_{eff}$.
However, with the singlet as dynamical degree of freedom, the stability
of the electroweak vacuum is improved as the only singlet$-$slepton contribution is actually a quadrilinear
term $\lambda Y_\ell S H_u^0 \tilde{\ell}_L^\ast \tilde{\ell}_R$, and the occurrence of a vacuum with $\langle \tilde{\ell}_L \rangle \neq 0$ and  $\langle \tilde{\ell}_R \rangle \neq 0$
can enhance the scalar potential~\cite{Hollik:2018yek}.

It should be pointed out that, for the parameter space in Eq.~(\ref{Scan-range}), samples obtained in the scan satisfy the inequality automatically. The reason is the ATLAS measurement of the properties of the discovered
Higgs particle has required $0.77 \leq Y_{\tau, \rm eff}/Y_{\tau, \rm SM} \leq 1.37$~\cite{ATLAS:2019nkf} and $Y_{\mu, \rm eff}/Y_{\mu, \rm SM} \leq 2.4$~\cite{ATLAS:2020fzp} at $2 \sigma$ confidence level.
This conclusion translates the inequality into simple forms: $(\mu + \mu_{eff}) \tan \beta \lesssim 2.5 \times 10^5~{\rm GeV}$ for $\tilde{\tau}$-Higgs potential, and  $(\mu + \mu_{eff}) \tan \beta \lesssim 1.9 \times 10^5~{\rm GeV}$ for $\tilde{\mu}$-Higgs potential with $m_{\tilde{\mu}_L} = m_{\tilde{\mu}_R} = 200~{\rm GeV}$. For $\tan \beta =60$, they read $(\mu + \mu_{eff}) \lesssim 4.2 \times 10^3~{\rm GeV}$ and $(\mu + \mu_{eff}) \lesssim 3.1 \times 10^3~{\rm GeV}$, respectively. In fact, our calculation with the code Vevacious found no CCB global minima in the scan.

\item The electroweak vacuum of the sample, which was called the desired symmetry breaking (DSB) vacuum in literature~\cite{Beuria:2016cdk,Hollik:2018yek}, corresponds to a local minimum of the scalar potential instead of a global minimum. In this case, the DSB vacuum could undergo quantum tunneling to the true vacuum. If the tunneling time is short enough in reference to the age of Universe, the DSB vacuum would decay completely~\cite{Camargo-Molina:2013qva,Camargo-Molina:2014pwa}. Such vacuum was called metastable but short-lived. Evidently, the occurrence of the metastable vacuum depends on the contour of the scalar potential, which is mainly decided by the parameters $\lambda$, $\kappa$, $\mu_{eff}$, and $A_\kappa$~\cite{Hollik:2018yek}. The calculation of the code Vevacious indicated that only about $1\%$ of the scanned samples predict short-lived vacuum.

\end{itemize}

\begin{figure}[t]
	\centering
	\includegraphics[width=0.505\textwidth]{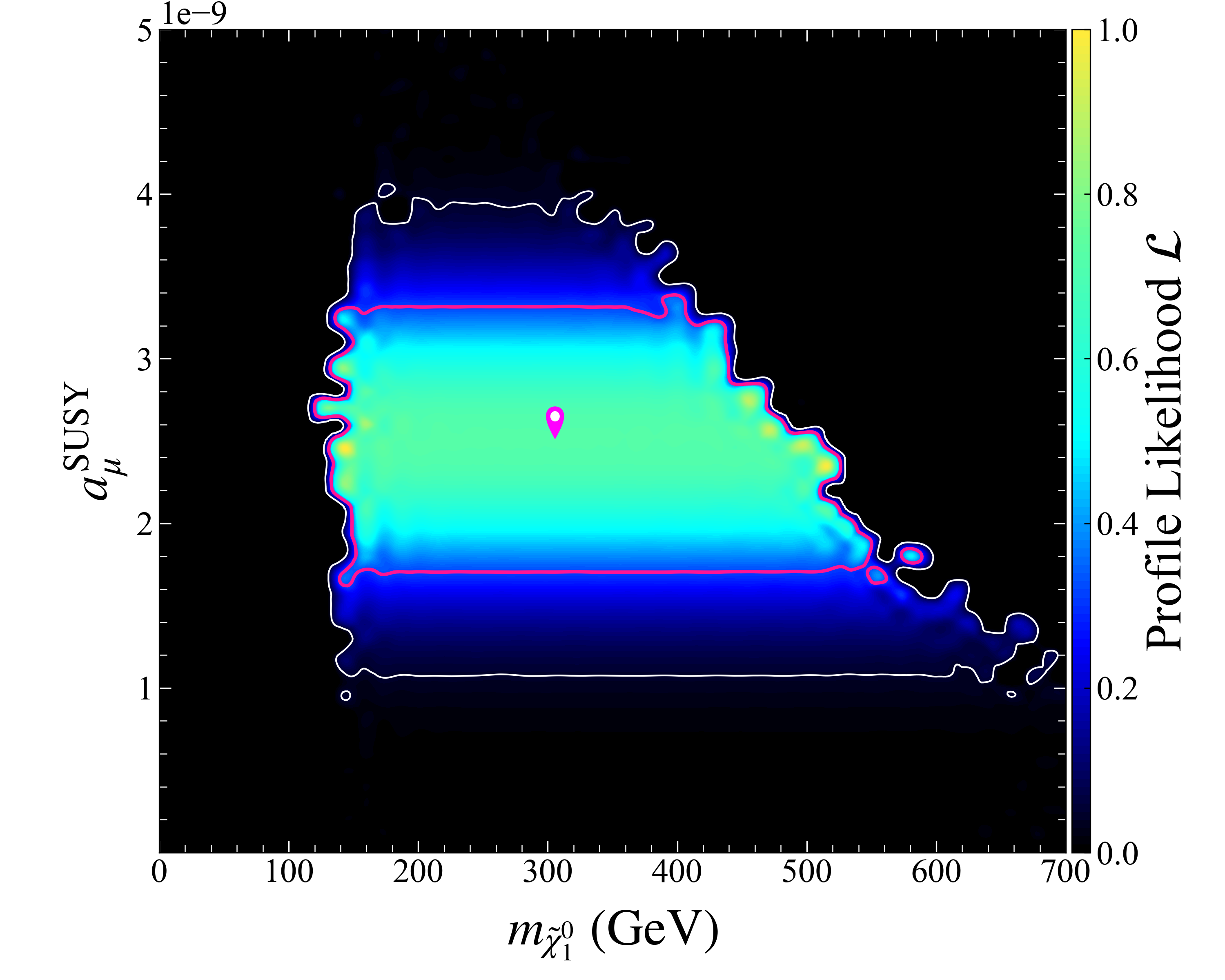}\hspace{-0.3cm}
	\includegraphics[width=0.505\textwidth]{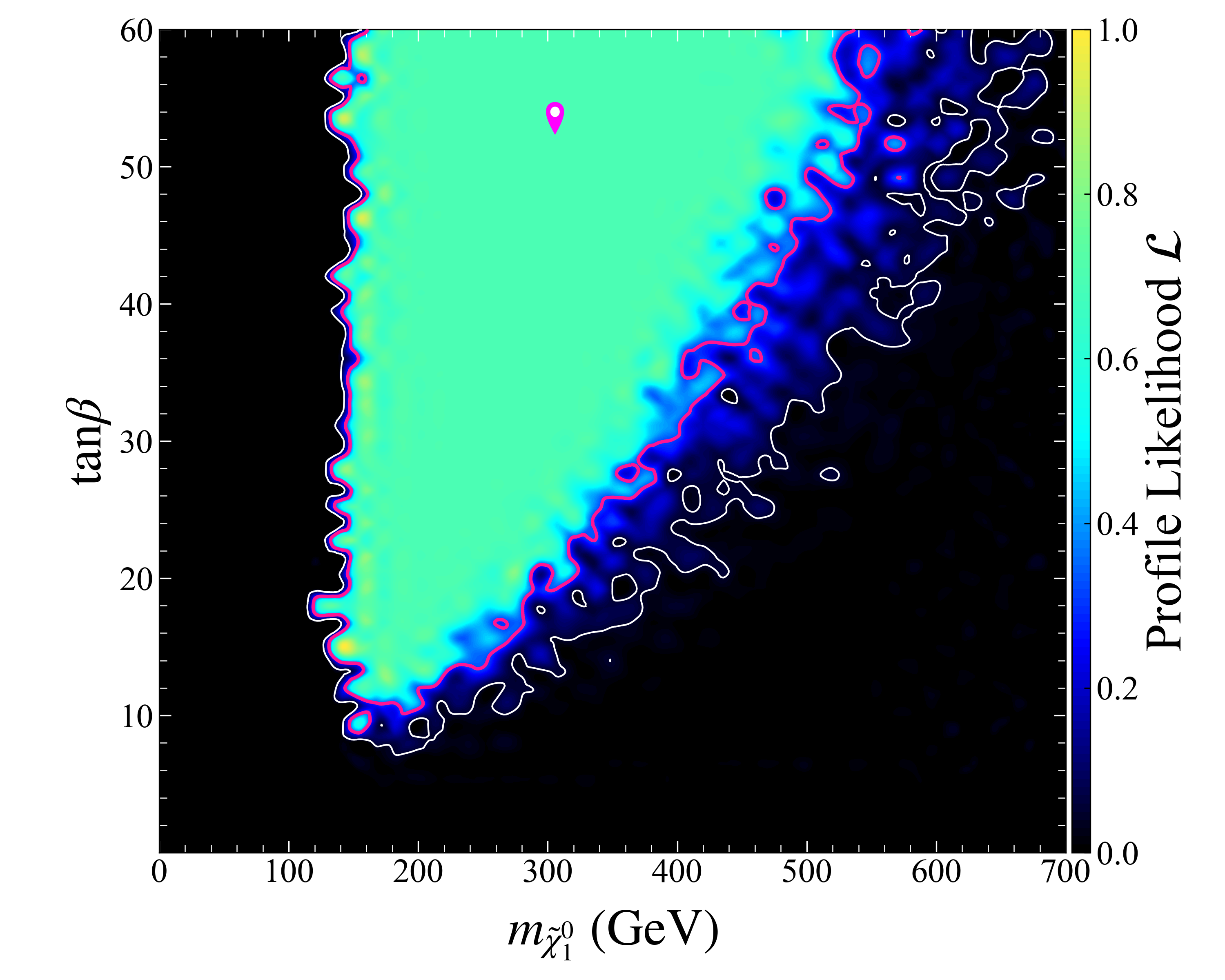}\\
	\includegraphics[width=0.505\textwidth]{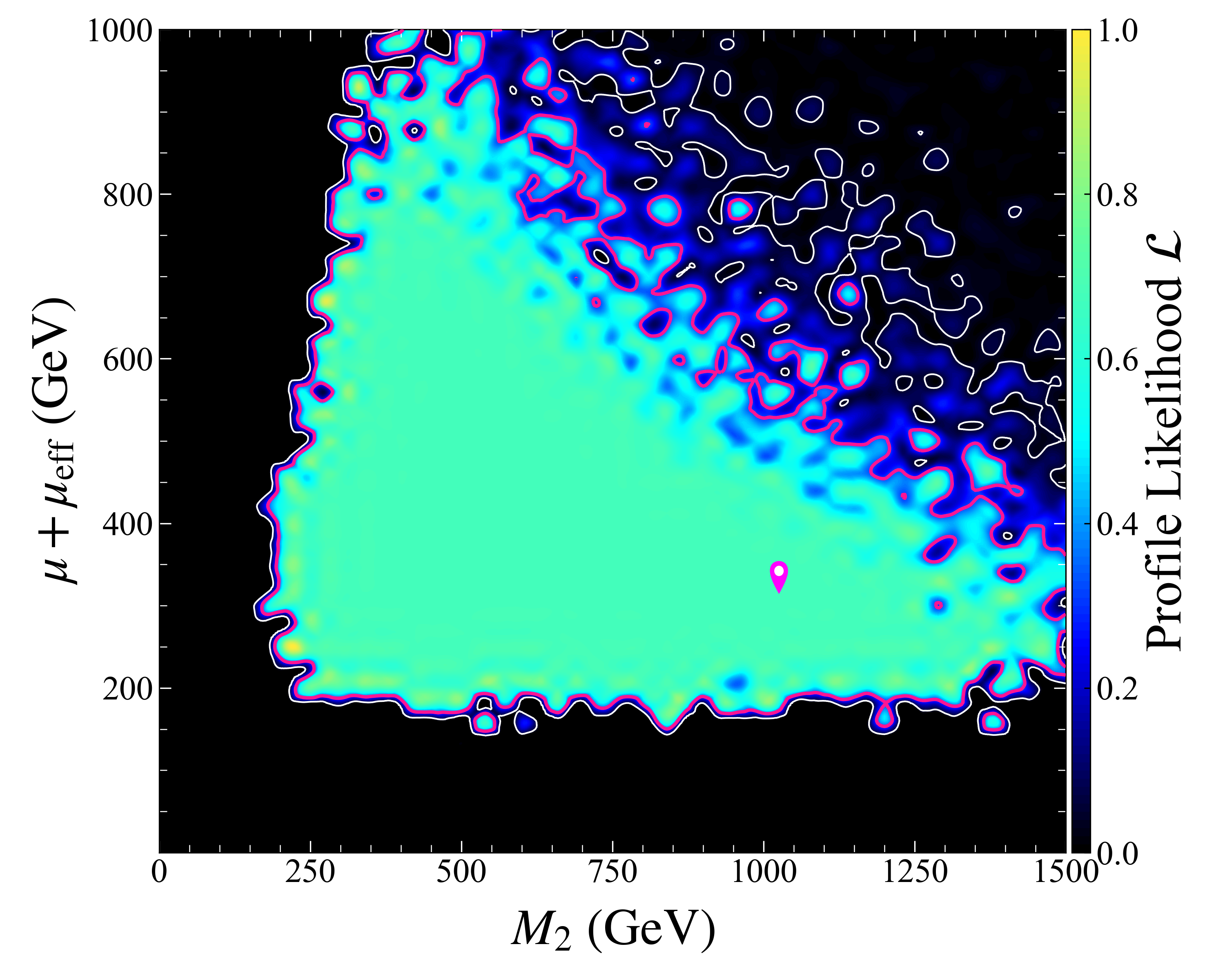}\hspace{-0.3cm}
	\includegraphics[width=0.505\textwidth]{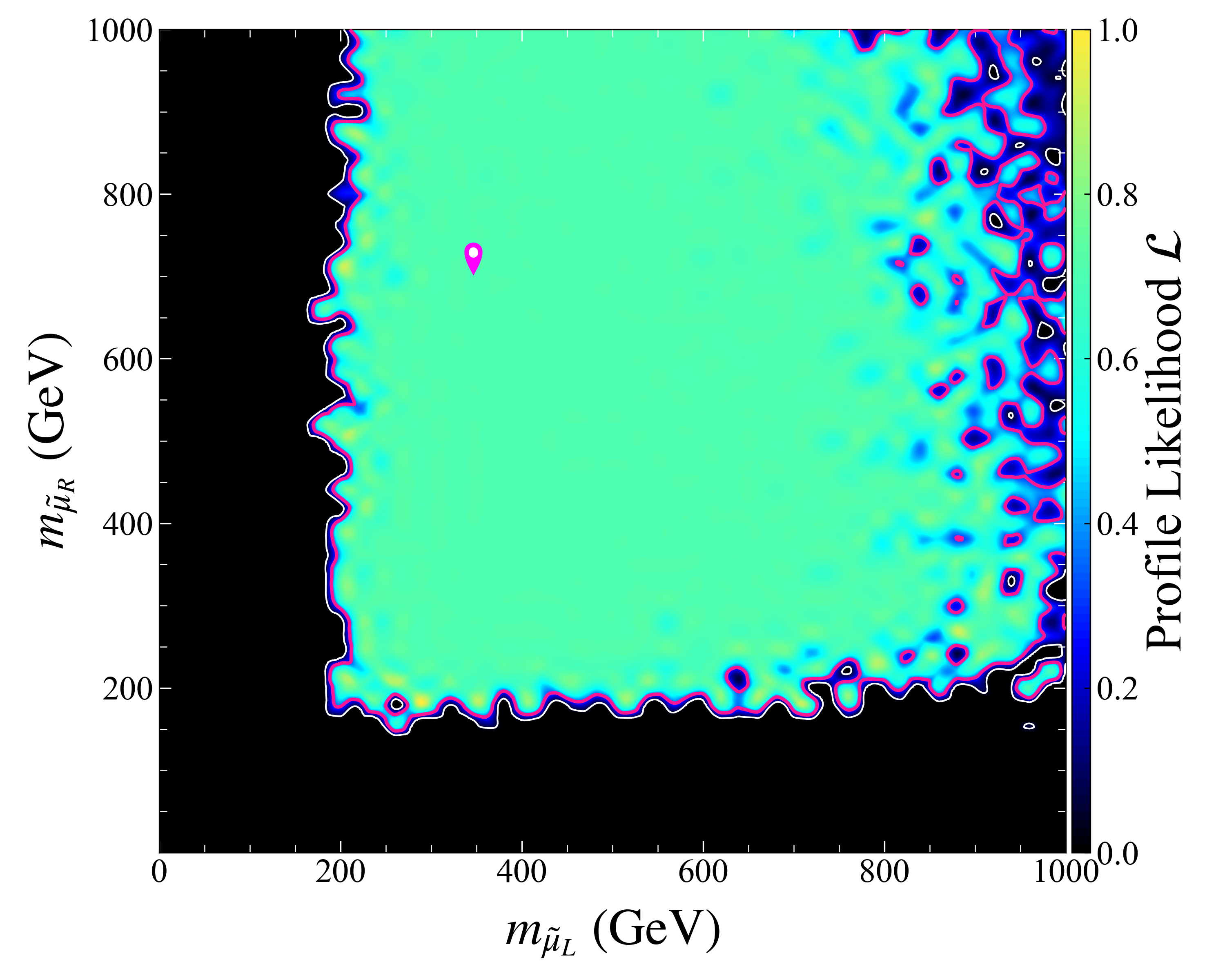}
	\caption{\label{fig:plamm}Two-dimensional profile likelihood map of the function in Eq.~(\ref{eq:likeliamm}) projected onto $m_{\tilde{\chi}_1^0}-a_{\mu}^{\rm SUSY}$, $m_{\tilde{\chi}_1^0}-\tan{\beta}$, $M_2-(\mu+\mu_{\rm eff})$, and $m_{\tilde{\mu}_L} - m_{\tilde{\mu}_R}$ planes. $1\sigma$ ($2\sigma$) confidence regions are shown by pink (white) contour lines. The best point locates at $m_{\tilde{\chi}_1^0} = 305~{\rm GeV}$, $\mu + \mu_{\rm eff} \simeq 340~{\rm GeV}$, $M_2 \simeq 1050~{\rm GeV}$, $m_{\tilde{\mu}_L} \simeq 350~{\rm GeV}$, and $m_{\tilde{\mu}_R} \simeq 740~{\rm GeV}$, and is marked by the pin symbol. Its $R$ value defined in Sec. 4 is about 0.3.}
\end{figure}
\begin{figure}[t]
	\centering
	\includegraphics[width=0.505\textwidth]{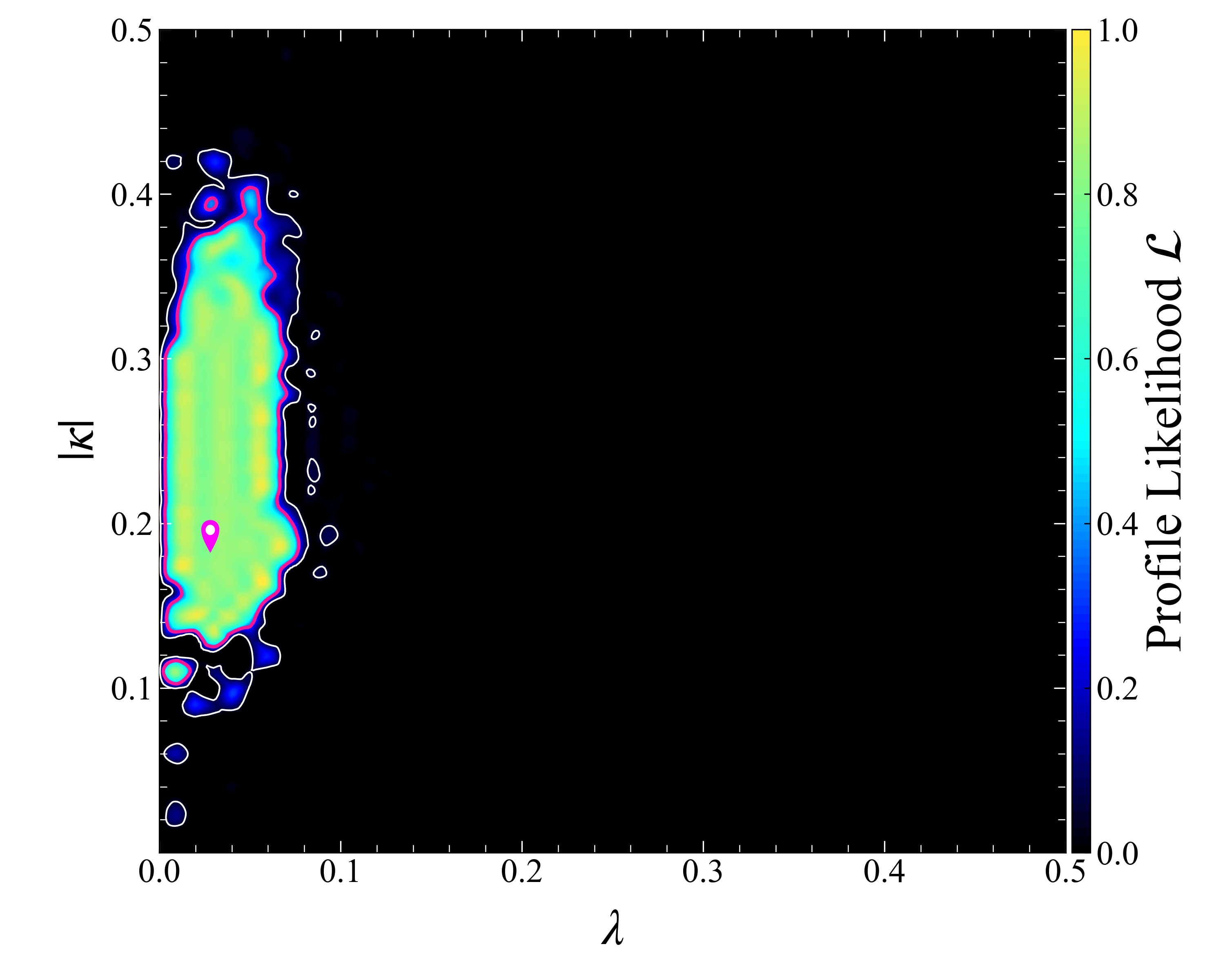}\hspace{-0.3cm}
	\includegraphics[width=0.505\textwidth]{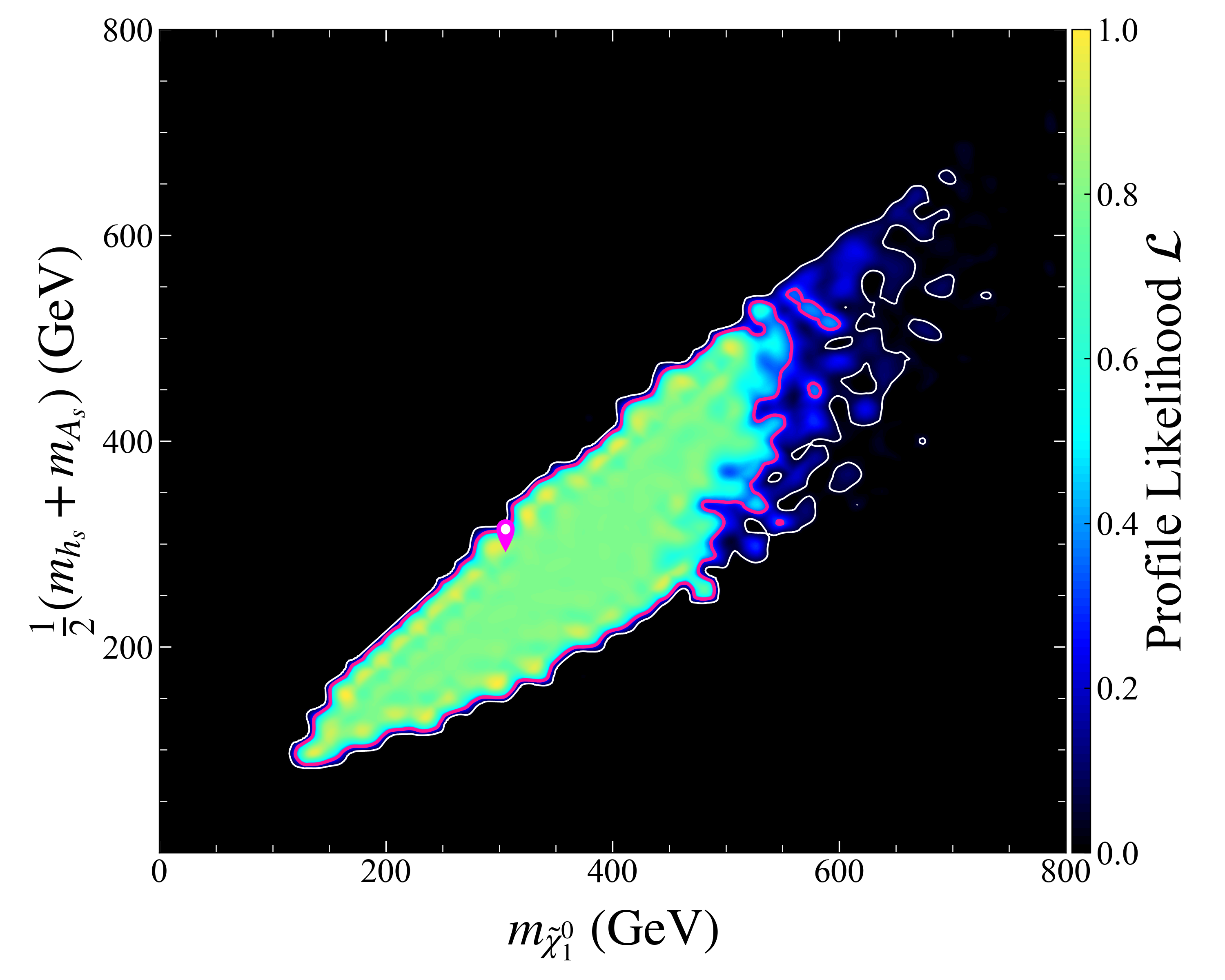}\\
	\includegraphics[width=0.505\textwidth]{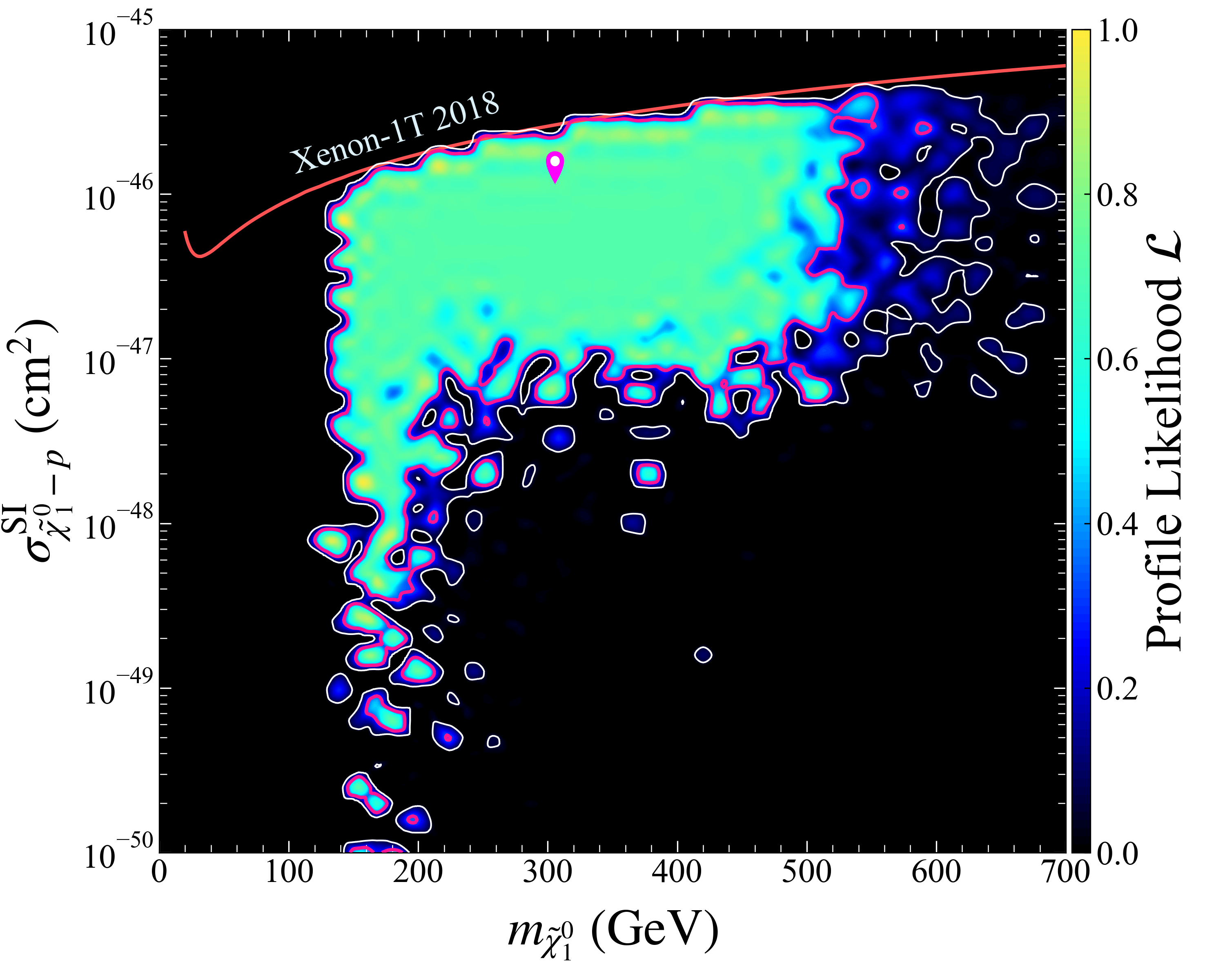}\hspace{-0.3cm}
	\includegraphics[width=0.505\textwidth]{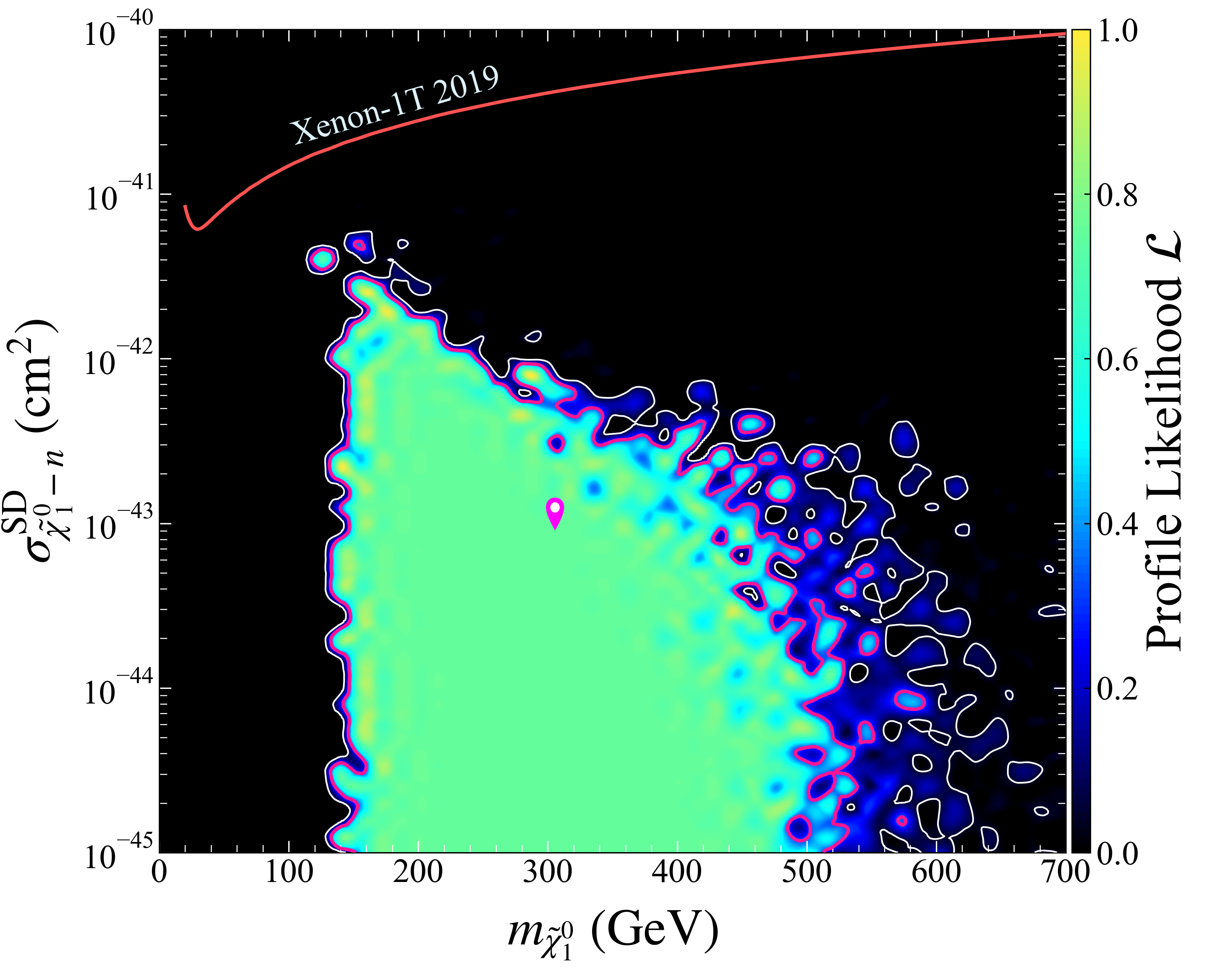}
	\caption{\label{fig:dm}Similar to Fig.~\ref{fig:plamm}, but onto $\lambda-|\kappa|$,  $m_{\tilde{\chi}_1^0} -\frac{1}{2}(m_{h_s} + m_{A_s})$, $m_{\tilde{\chi}_1^0} - \sigma_{n}^{\rm SD}$, and $m_{\tilde{\chi}_1^0} - \sigma_{p}^{\rm SI}$ planes, respectively. }
\end{figure}

\begin{figure}
	\centering
	\includegraphics[width=0.9\textwidth]{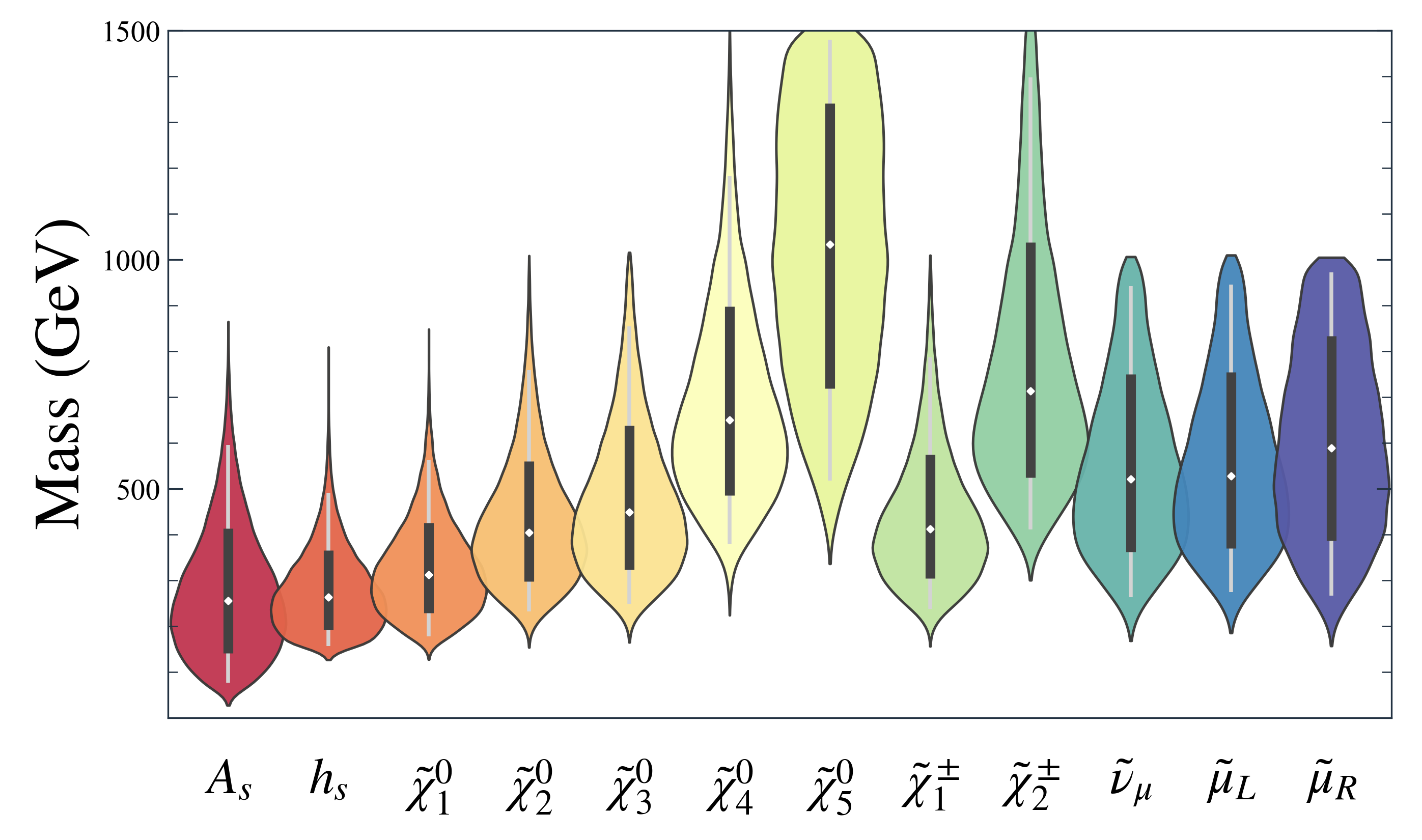}
	\caption{\label{fig:smassviolin} Violin plots showing mass distributions of singlet Higgs, EWinos, and $\mu$-type sparticles. Smuons are labeled by their dominated components. The violins are scaled by count. Thick vertical bar in center indicates interquartile range with white dot representing the median; long vertical line represents $95\%$ confidence interval.}
\end{figure}

\subsection{Numerical results}
The discussion begins with the normalized two-dimensional profile likelihood (PL), $\mathcal{L}(\theta_1^\prime, \theta_2^\prime)$, for the likelihood function in Eq.~(\ref{eq:likeliamm}), where $\mathcal{L}(\theta_1^\prime, \theta_2^\prime)$ is defined as the maximum value of $\mathcal{L}$ in the position $\theta_1 = \theta_1^\prime, \theta_2 = \theta_2^\prime$. In Fig.~{\ref{fig:plamm}}, all of the samples are projected onto $m_{\tilde{\chi}_1^0}-a_{\mu}^{\rm SUSY}$, $m_{\tilde{\chi}_1^0}-\tan{\beta}$, $(\mu+\mu_{\rm eff}) - M_2$, and $m_{\tilde{\mu}_L} - m_{\tilde{\mu}_R}$ planes to obtain the PLs. From Fig.~\ref{fig:plamm}, one can find that the $\mu$NMSSM can interpret $\Delta a_\mu$ in broad parameter space satisfying all of the experimental constraints. The lower mass bounds of sparticles come from the assumption of $m_{h_s} \geq 125~{\rm GeV}$, so the DM masses are often larger than $150~{\rm GeV}$ (See the following discussion. A similar result can be found in Fig.~1 of Ref.~\cite{Cao:2021ljw}). The upper bound of the LSP mass $m_{\tilde{\chi}_1^0}$ is mainly determined by the value of $\tan{\beta}$ and $\Delta a_\mu$, e.g., $a_\mu^{\rm SUSY} = 2\times 10^{-9}$ and $\tan{\beta}=60$ set up an upper limit of approximately $550~{\rm GeV}$ for $m_{\tilde{\chi}_1^0}$. For most samples, the wino-higgsino loop provides the dominant contribution in $a_{\mu}^{\rm SUSY}$. Within the $1\sigma$ level in $\Delta a_\mu$, the contours in the $(\mu+\mu_{\rm eff}) - M_2$ plane imply that the mass of chargino $\tilde{\chi}_1^\pm$ cannot be larger than approximately $700~{\rm GeV}$ (in particular, the higgsino mass may be less than $500~{\rm GeV}$ to predict $m_Z$ naturally), while both $m_{\tilde{\mu}_L}$ and $m_{\tilde{\mu}_R}$ can be larger than  $800~{\rm GeV}$.

\par As shown in Fig.~\ref{fig:dm}, $\lambda$ is less than 0.1, the absolute values of $\kappa$ vary from 0.1 to 0.4, and all of the samples meet the condition $m_{\tilde{\chi}_1^0} \gtrsim (m_{h_s} + m_{A_s})/2$.  This feature indicates that the strengths of DM coupled to other non-singlet fields are relatively weak, and the main DM annihilation channel is $\tilde{\chi}_1^0 \tilde{\chi}_1^0 \to h_s A_s$. Thus, the singlet-dominated particles form a secluded DM sector where the singlet Higgs states $h_s$ and $A_s$ act as the mediators between DM and SM particles~\cite{Pospelov:2007mp}, which was pointed out in our previous work~\cite{Cao:2021ljw}. Owing to the smallness of $\lambda$, the DM direct detection rates $\sigma_p^{\rm SI}$ and $\sigma_n^{\rm SD}$ in the $\Delta a_\mu$ favored parameter space can be far below the current detection limits. {\bf{The above discussion shows that the DM phenomenology and natural interpretations of $\Delta a_\mu$ are very weakly connected in the $\mu$NMSSM}}. This situation is significantly different from the MSSM case, where the bino-dominated $\tilde{\chi}_1^0$ mainly co-annihilated with the other sparticles to obtain the measured relic density, and consequently the parameter space to explain $\Delta a_\mu$ is limited~\cite{Chakraborti:2020vjp, Chakraborti:2021dli}.

\par In Fig.~\ref{fig:smassviolin}, the mass distributions of the singlet Higgs states and SUSY particles are shown with violin plots\footnote{A violin plot combines the advantages of the box plot and probability density distribution plot~\cite{hintze1998violin}.}. From Fig.~\ref{fig:smassviolin}, one can find that the masses of sleptons and EWinos can be lower than $500~{\rm GeV}$. They can be produced at the LHC, and they are thus restricted by searching for multi-lepton signals. Compared with the MSSM prediction, the sparticles in the $\mu$NMSSM have the following distinct features.
\begin{itemize}
	\item As shown in Figs.~\ref{fig:plamm} and~\ref{fig:dm}, $\tilde{\chi}_1^0$ is moderately massive $m_{\tilde{\chi}_1^0} > 150~{\rm GeV}$ from the current results because it must be heavier than $(m_{h_s} + m_{A_S})/2$ to proceed with $\tilde{\chi}_1^0 \tilde{\chi}_1^0 \to h_s A_s$.
	\item Since $\tilde{\chi}_1^0$ is singlino-dominated, and thus coupled very weakly to the other sparticles, heavy sparticles prefer to decay into next-to-LSP (NLSP) or next-next-to-LSP (NNLSP) first. Consequently, their decay chain is lengthened and their signals become complicated.
	\item Since singlet-dominated Higgs bosons are preferred light, they may serve as sparticle decay products and enrich the decay channels of sparticles.
\end{itemize}
These characteristics make sparticle detection at the LHC rather tricky, and the corresponding constraints should be weaker than those in the simplified models and in the MSSM. This issue will be intensively studied in the following.

\section{\label{sec:LHC} LHC constraints}

Given that the capability of the \texttt{SModelS} package in implementing the LHC constraints is limited by its database and the strict prerequisites to use it, the LHC detection of sparticles is further studied by Monte Carlo event simulations\footnote{More than 88000 samples were obtained in the previous scan. In order to save time and at the same time make our conclusions as general as possible, a smaller $n_{\rm live}$ ($n_{\rm live} =3000$) was chosen to repeat the scan and obtained about 7500 samples for studying the LHC constraints. These samples satisfy the vacuum stability requirement and can interpret $\Delta a_\mu$ within $2 \sigma$ level.
Besides, the analyses of $\sqrt{s} = 8~{\rm TeV}$ pp collisions were not considered in this work because the LSP is relatively heavy so that their constraints on the theory are weak.}. The following processes are considered in the simulation:
\begin{equation}\begin{split}
    pp \to \tilde{\chi}_i^0\tilde{\chi}_j^{\pm} &, \quad i = 2, 3, 4, 5; \quad j = 1, 2 \\
    pp \to \tilde{\chi}_i^{\pm}\tilde{\chi}_j^{\mp} &, \quad i = 1, 2; \quad j = 1, 2 \\
    pp \to \tilde{\chi}_i^{0}\tilde{\chi}_j^{0} &, \quad i = 2, 3, 4, 5; \quad j = 2, 3, 4, 5\\
    pp \to \tilde{\mu}_i\tilde{\mu}_j &,\quad i = L, R;\quad j = L, R
\end{split}\end{equation}
The cross-sections at $\sqrt{s}$ = 13 TeV were obtained at the NLO using the package \texttt{Prospino2}~\cite{Beenakker:1996ed}. The signal events were generated by \texttt{MadGraph\_aMC@NLO}~\cite{Alwall:2011uj, Conte:2012fm} with the package \texttt{PYTHIA8}~\cite{Sjostrand:2014zea} for parton showers, hadronization, and sparticle decay. Finally, the event files were put into \texttt{CheckMATE-2.0.29}~\cite{Drees:2013wra,Dercks:2016npn, Kim:2015wza} with the embedded package \texttt{Delphes}\cite{deFavereau:2013fsa} for detector simulation.

In the simulations, $10^5$ signal events were generated for each $\mu$NMSSM parameter point, and the following latest LHC searches among the analyses mentioned before were found crucial in testing the samples:
\begin{itemize}
	  \item ATLAS search for chargino and slepton pair production in two lepton final states (see Report No. CERN-EP-2019-106)~\cite{Aad:2019vnb}.
	  \item Search for chargino-neutralino production with involved mass splittings near the electroweak scale in three-lepton final states in $\sqrt{s} = 13~{\rm TeV}$ pp collisions with the ATLAS detector (see Report No. CERN-EP-2019-263)~\cite{Aad:2019vvi}.
	  \item Search for direct production of electroweakinos in final states with one lepton, missing transverse momentum and a Higgs boson decaying into two bb jets in pp collisions at $\sqrt{s} = 13~{\rm TeV}$ with the ATLAS detector (see Report No. CERN-EP-2019-188)~\cite{Aad:2019vvf}.
	  \item CMS combined search for charginos and neutralinos (see Report No. CMS-SUS-17-004)~\cite{Sirunyan:2018ubx}.
\item CMS search for final states with two opposite sign same-flavor leptons, jets, and missing transverse momentum (see Report No. CMS-SUS-20-001)~\cite{CMS:2020bfa}.
\end{itemize}

\par The $R$ values obtained from the code \texttt{CheckMATE} were applied to implement the LHC constraints. Here, $R \equiv \max\{S_i/S_{{\rm obs},i}^{95}\}$ for all the involved analyses, where $S_i$ represents the simulated event number of the $i$th signal region (SR), and $S_{{\rm obs},i}^{95}$ is the corresponding 95$\%$ confidence level upper limit. Therefore, $R > 1$ indicates that the sample is excluded by the LHC searches if the involved uncertainties are not considered (see the discussion in footnote 9 of this work).

\begin{figure}[t]
	\centering
	\includegraphics[width=0.495\textwidth]{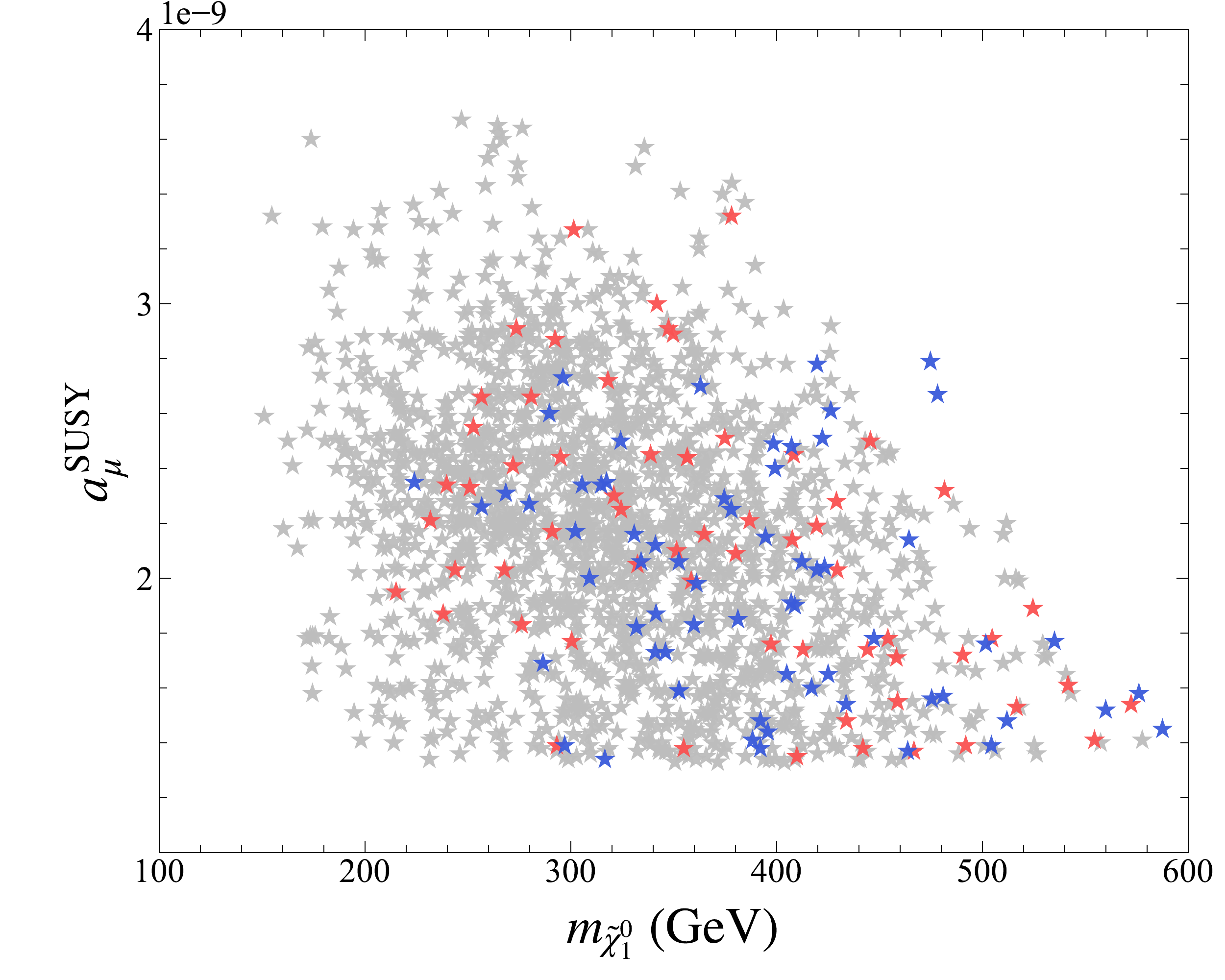}
	\includegraphics[width=0.495\textwidth]{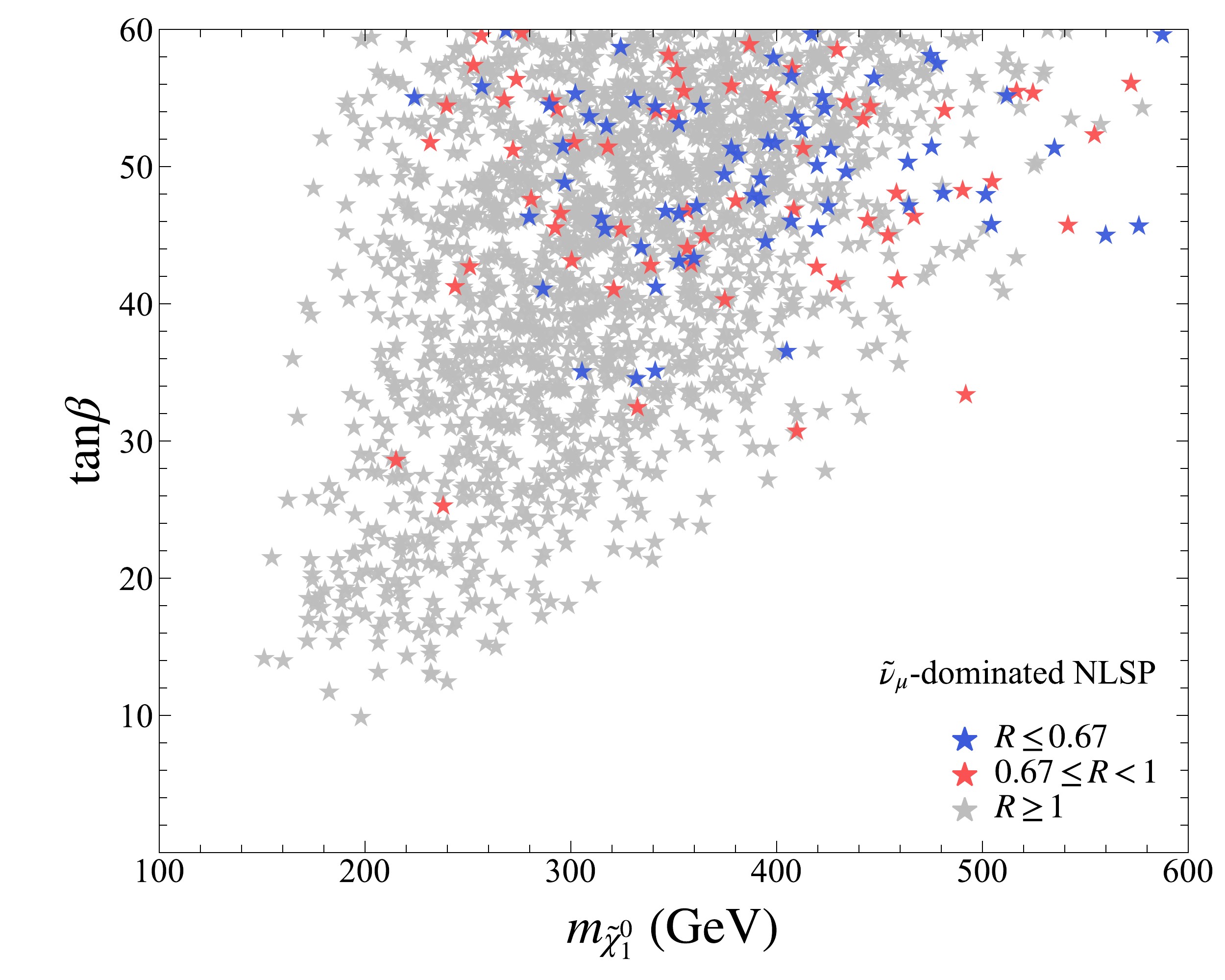}\\
	\includegraphics[width=0.495\textwidth]{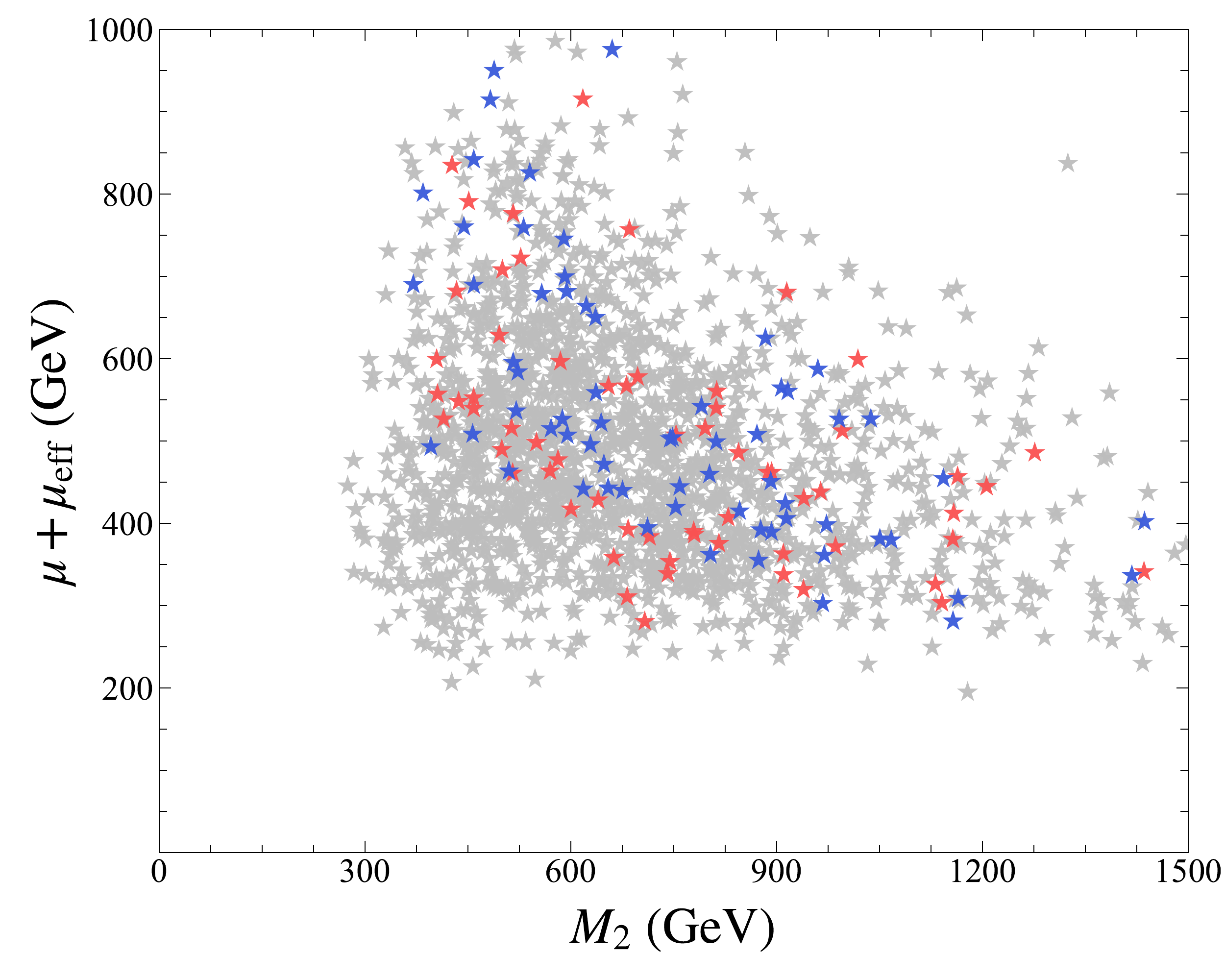}
	\includegraphics[width=0.495\textwidth]{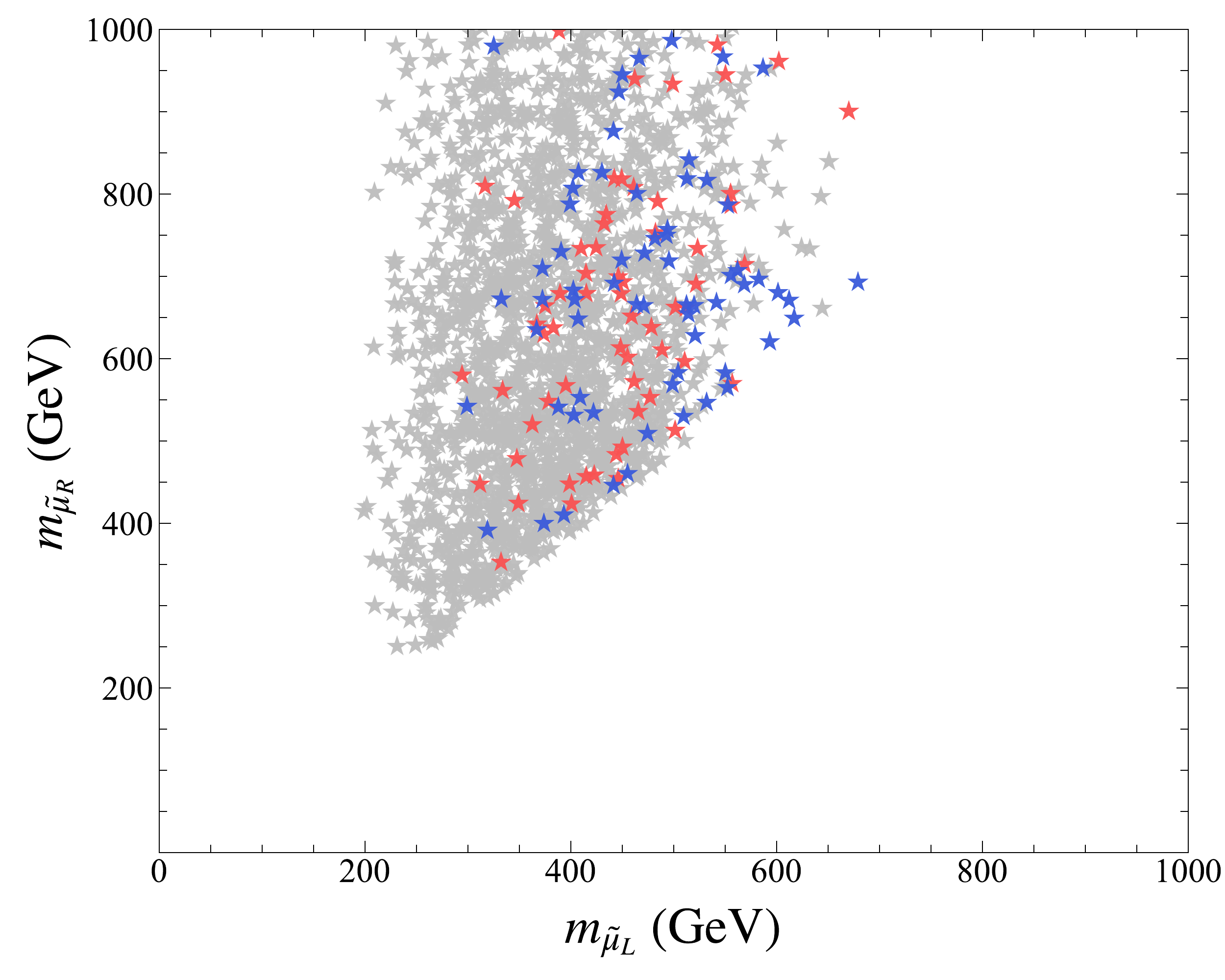}
	\caption{\label{fig:sct}The LHC constraints on the samples with $\tilde{\nu}_\mu$-dominated NLSP.  The samples are projected on $m_{\tilde{\chi}_1^0}-a_\mu^{\rm SUSY}$, $m_{\tilde{\chi}_1^0} - \tan{\beta}$, $M_2 - (\mu+\mu_{\rm eff})$ and $m_{\tilde{\mu}_L} - m_{\tilde{\mu}_R}$ planes, and those satisfying $R \geq 1$, $0.67 \leq  R < 1$ and $R < 0.67$ are marked by grey, red, and blue colors, respectively. }
\end{figure}

\begin{figure}[t]
	\centering
	\includegraphics[width=0.495\textwidth]{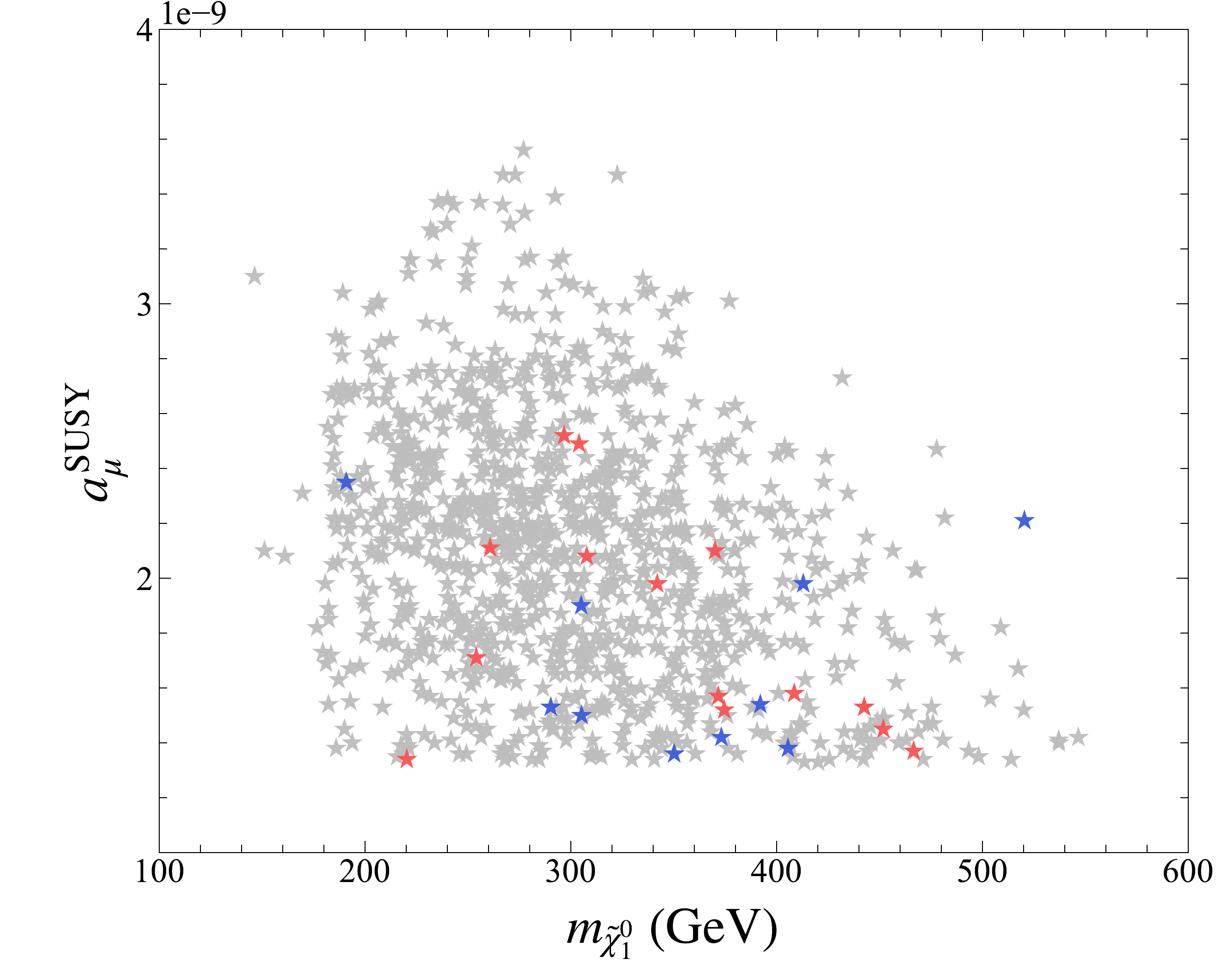}
	\includegraphics[width=0.495\textwidth]{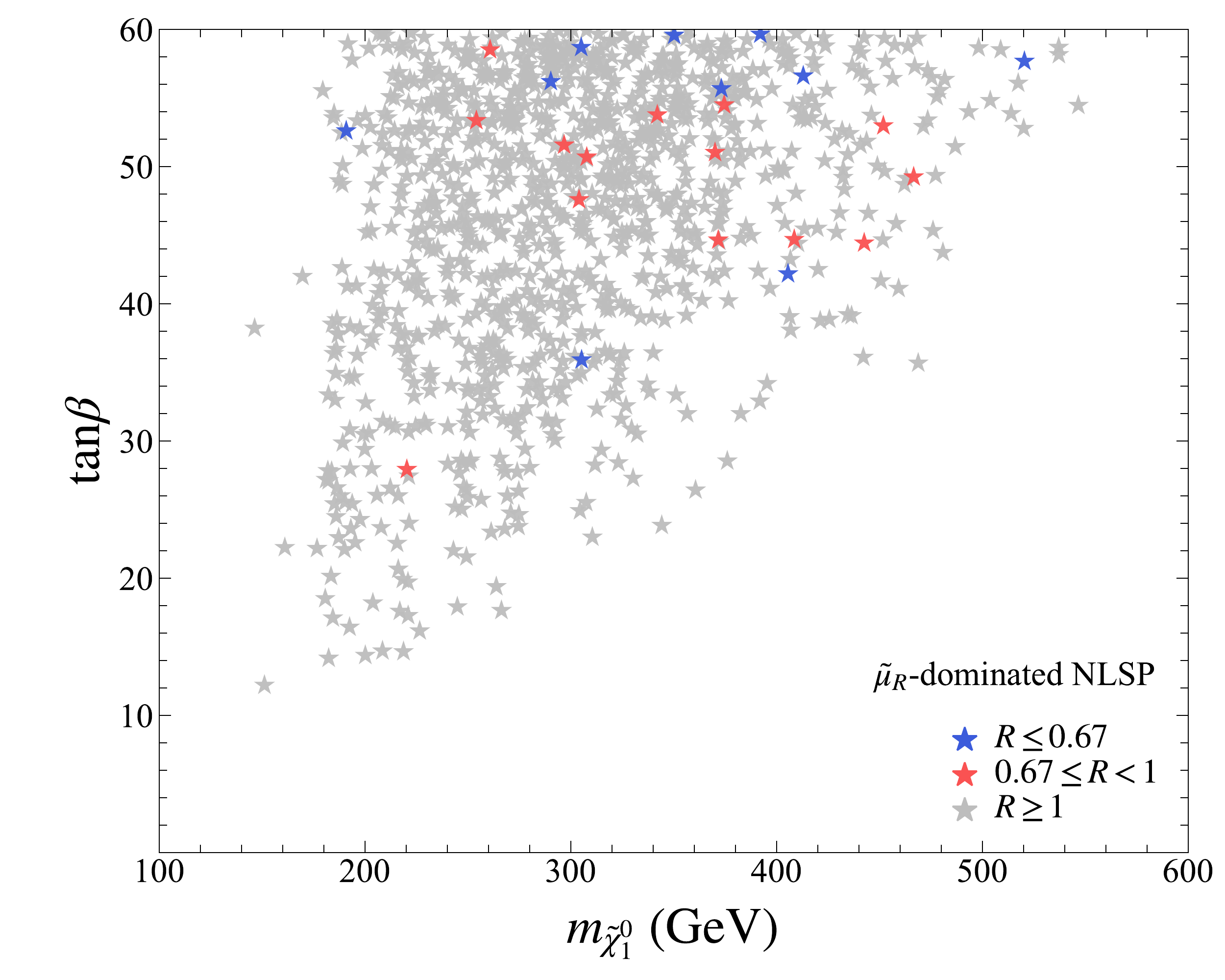}\\
	\includegraphics[width=0.495\textwidth]{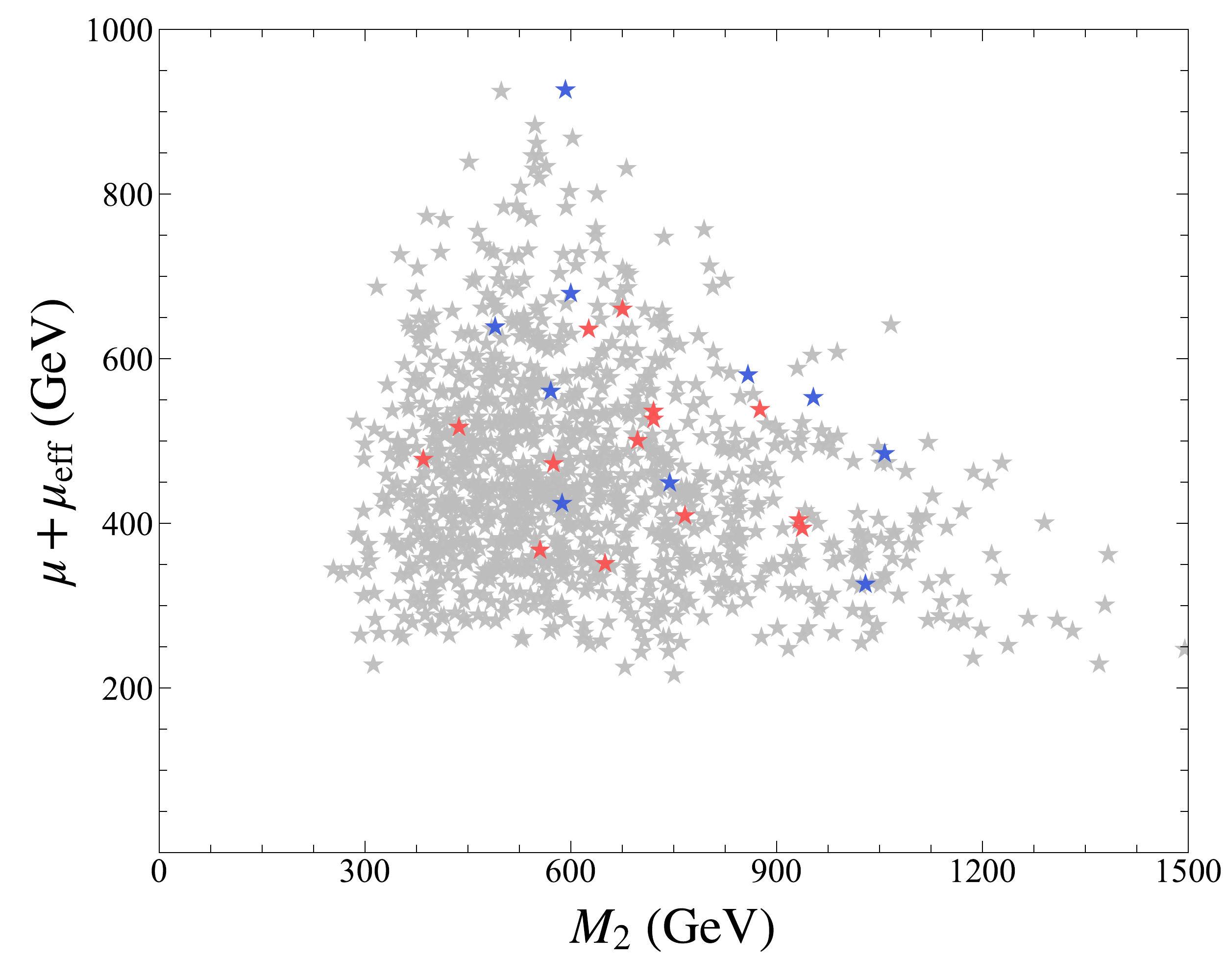}
	\includegraphics[width=0.495\textwidth]{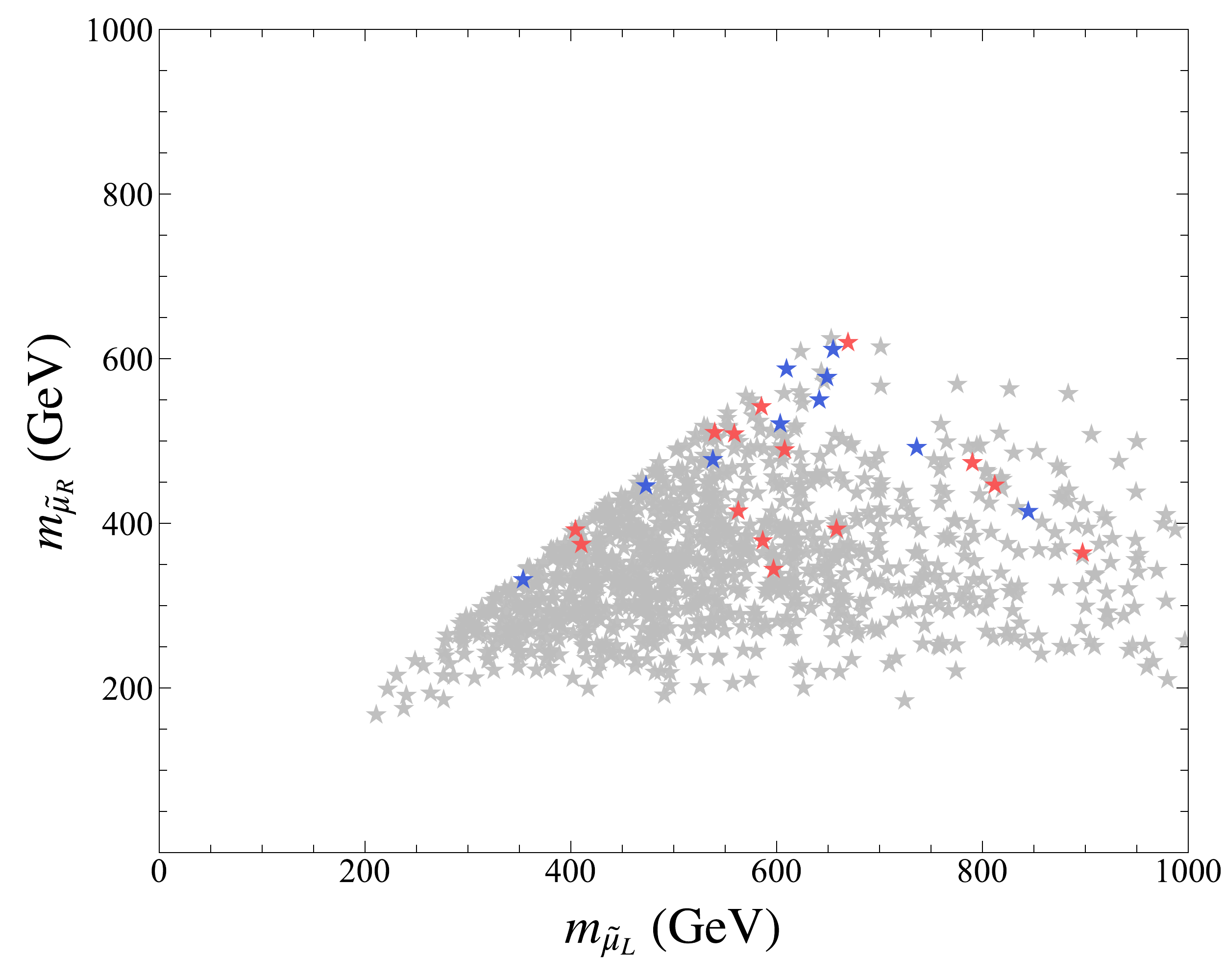}
	\caption{\label{fig:sct1} Similar to Fig.~\ref{fig:sct}, but for the samples with $\tilde{\mu}_R$-dominated NLSP. }
\end{figure}

\begin{figure}[t]
	\centering
	\includegraphics[width=0.495\textwidth]{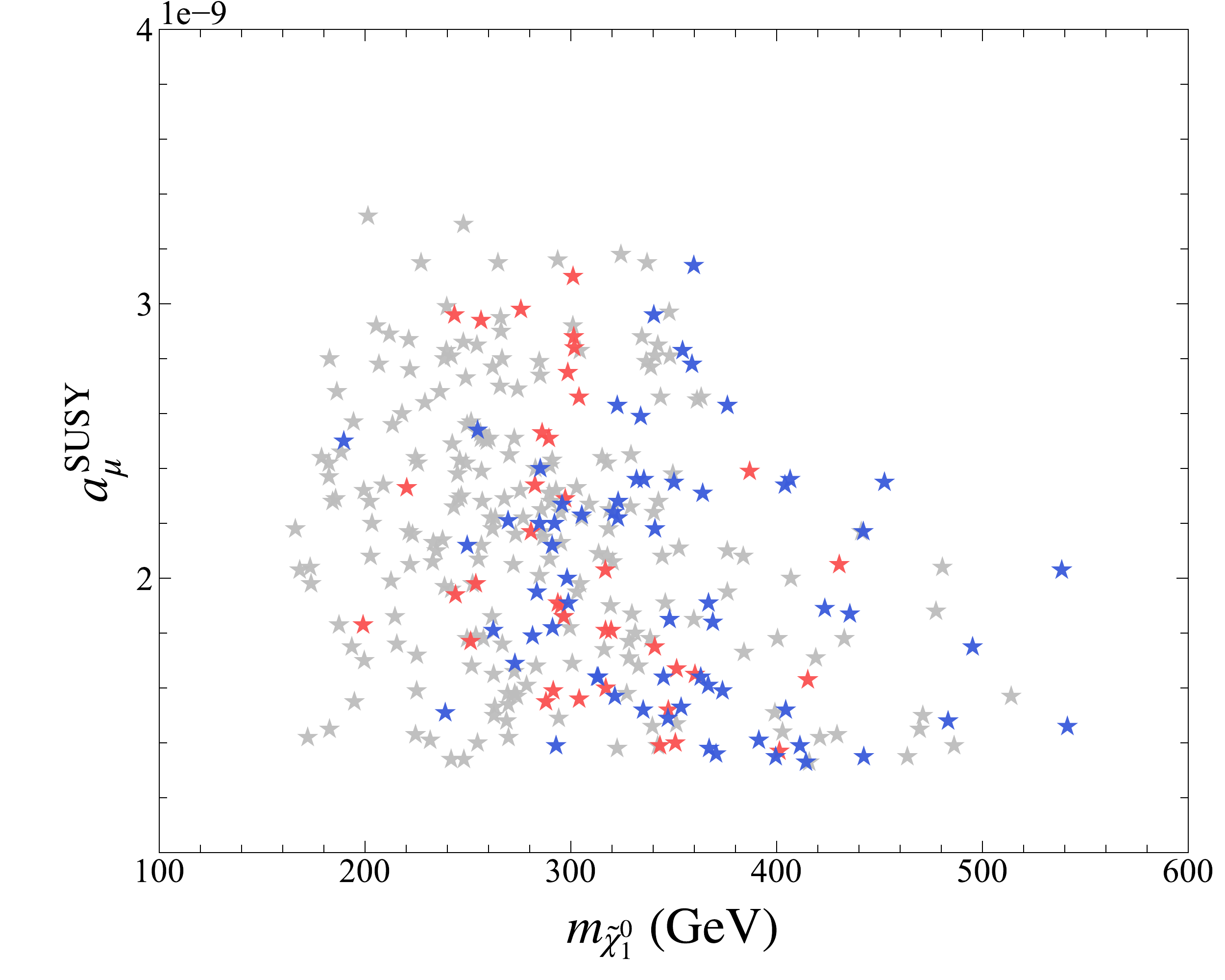}
	\includegraphics[width=0.495\textwidth]{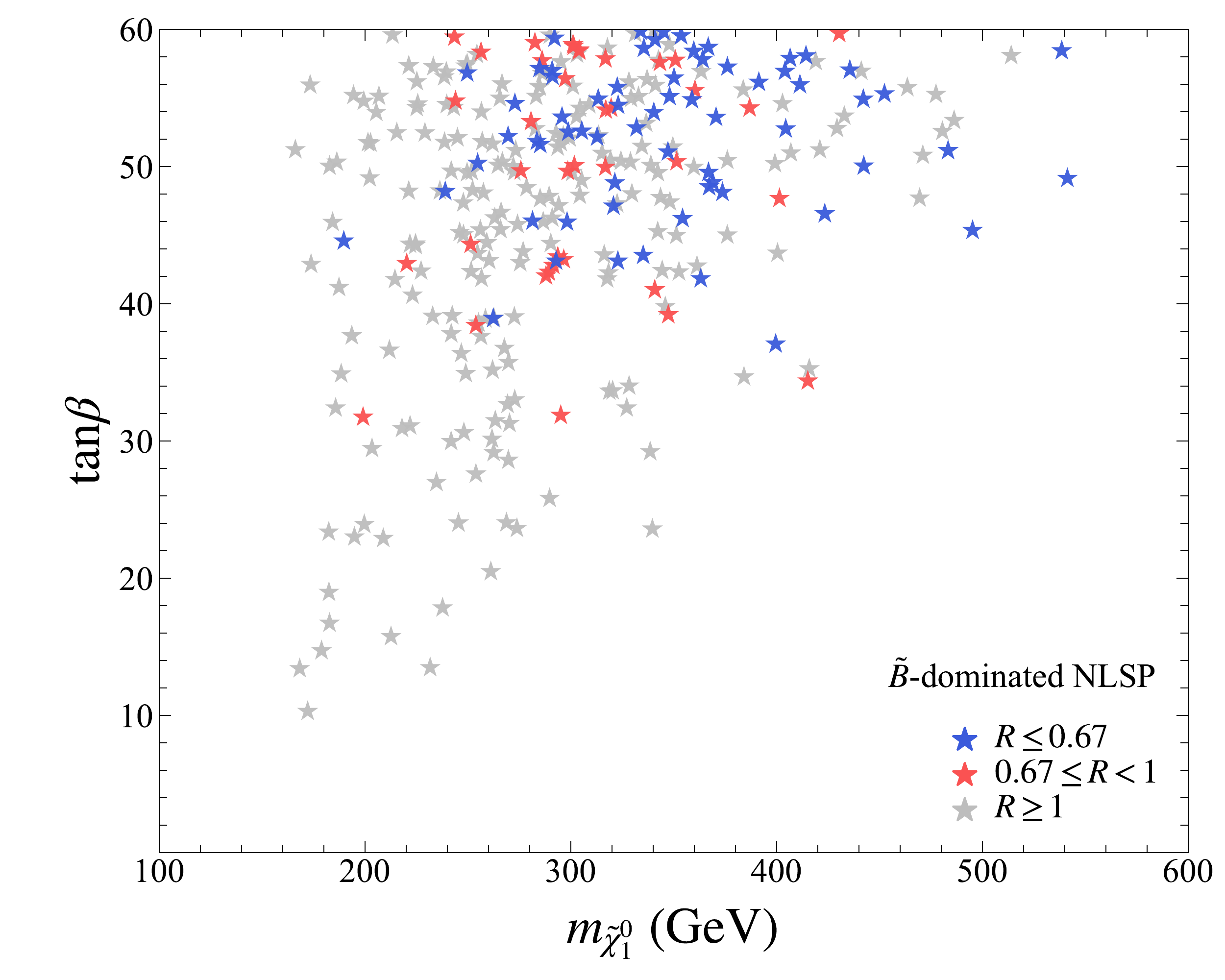}\\
	\includegraphics[width=0.495\textwidth]{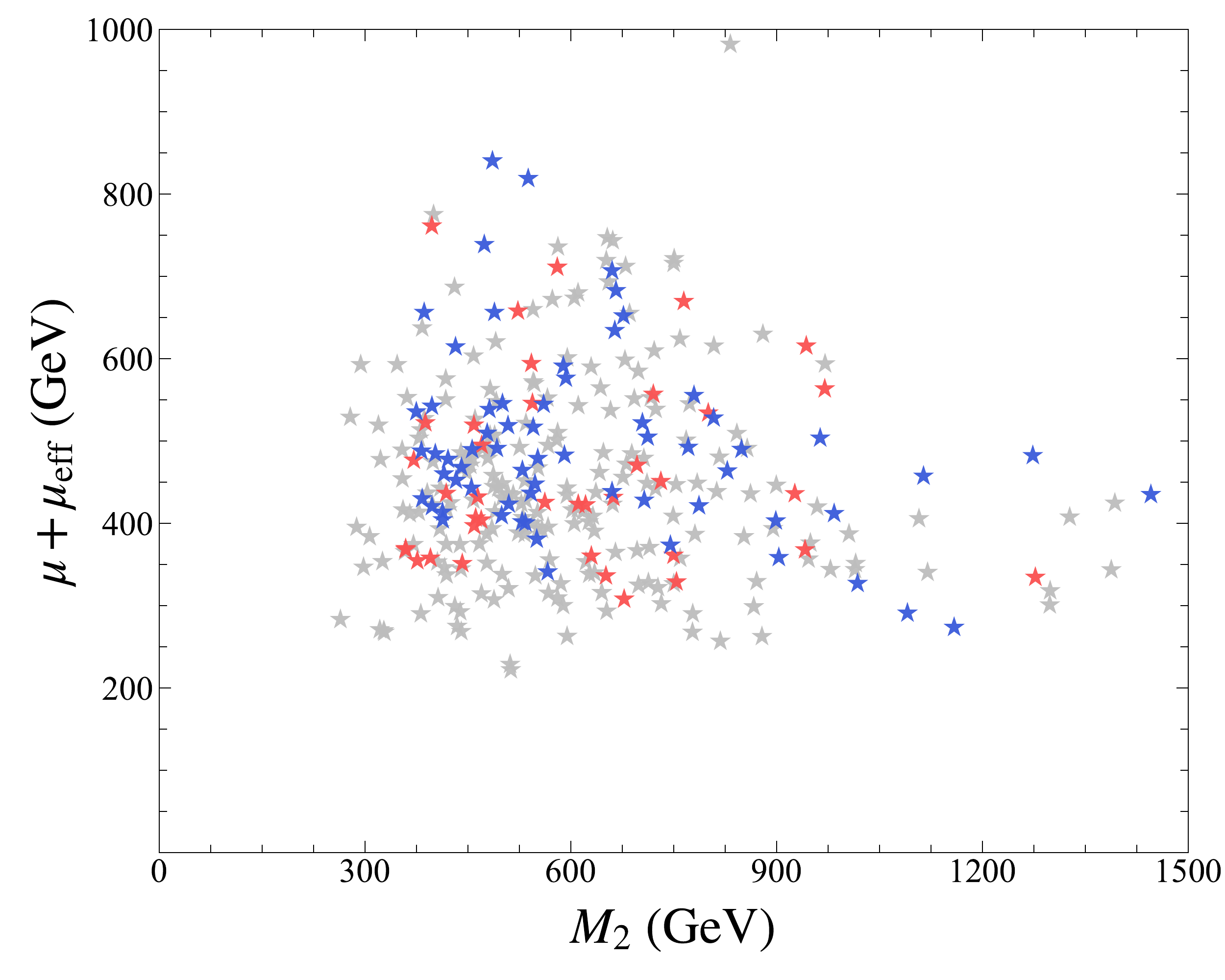}
	\includegraphics[width=0.495\textwidth]{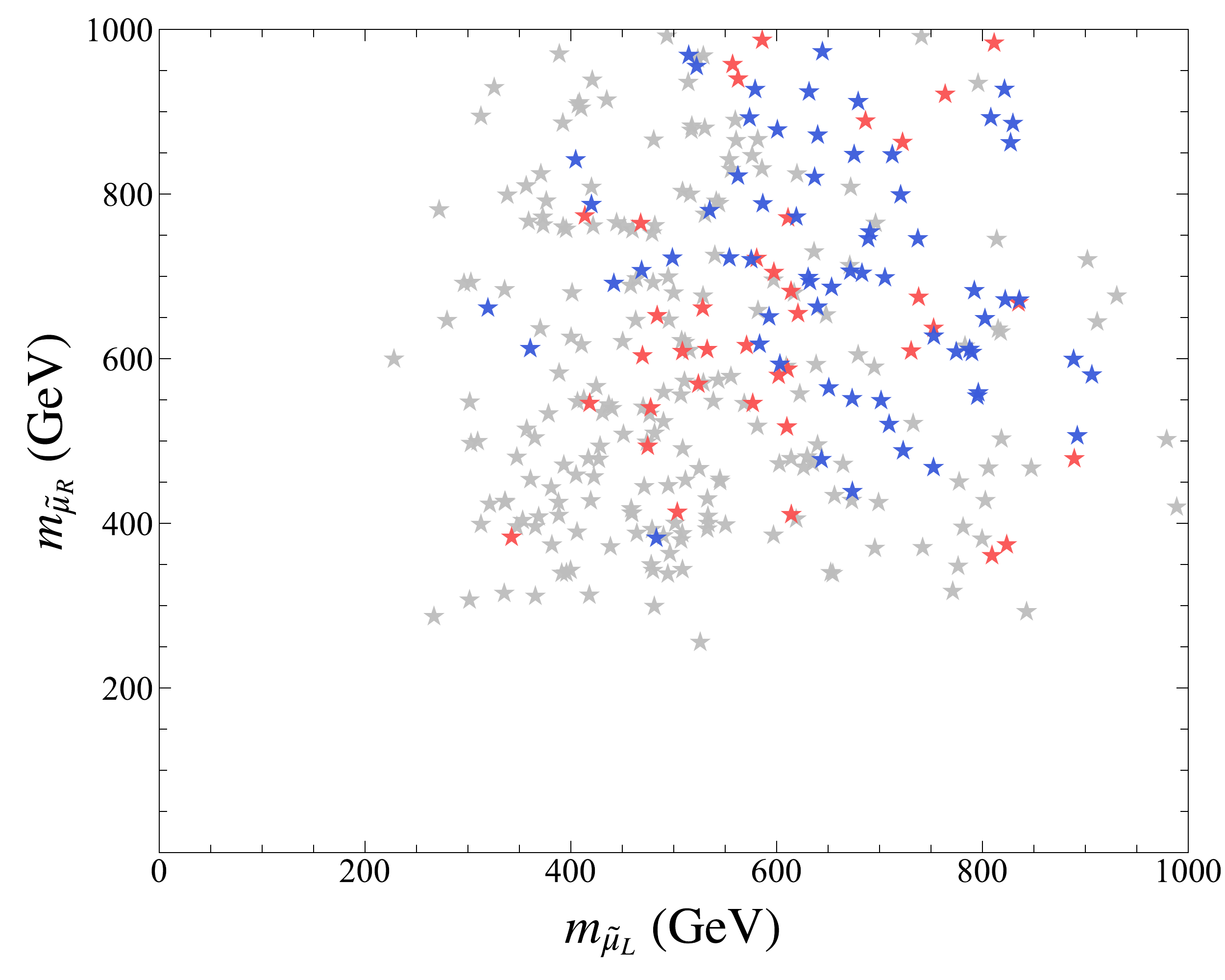}
	\caption{\label{fig:sct2} Similar to Fig.~\ref{fig:sct}, but for the samples with $\tilde{B}$-dominated NLSP. }
\end{figure}

\begin{figure}[t]
	\centering
	\includegraphics[width=0.495\textwidth]{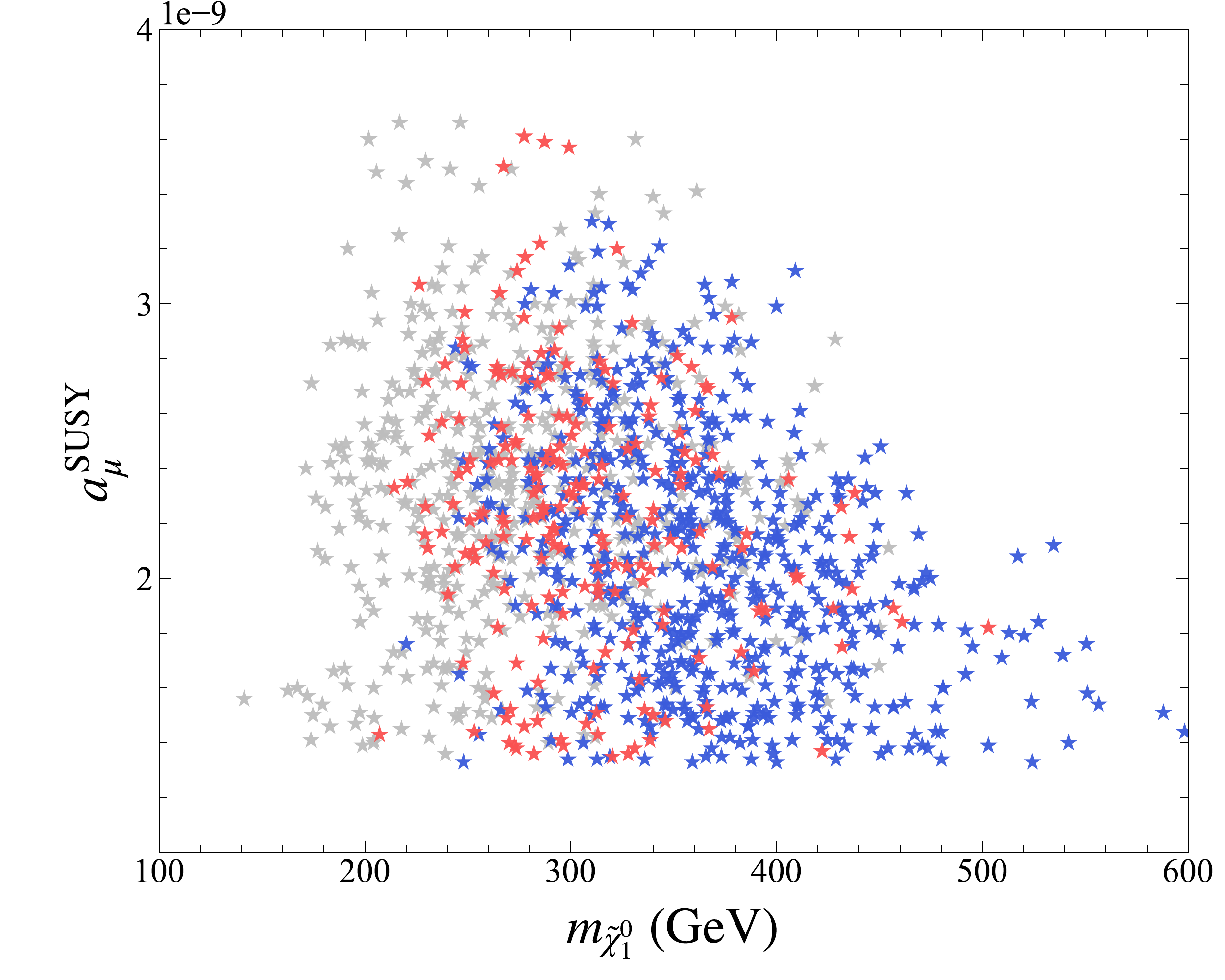}
	\includegraphics[width=0.495\textwidth]{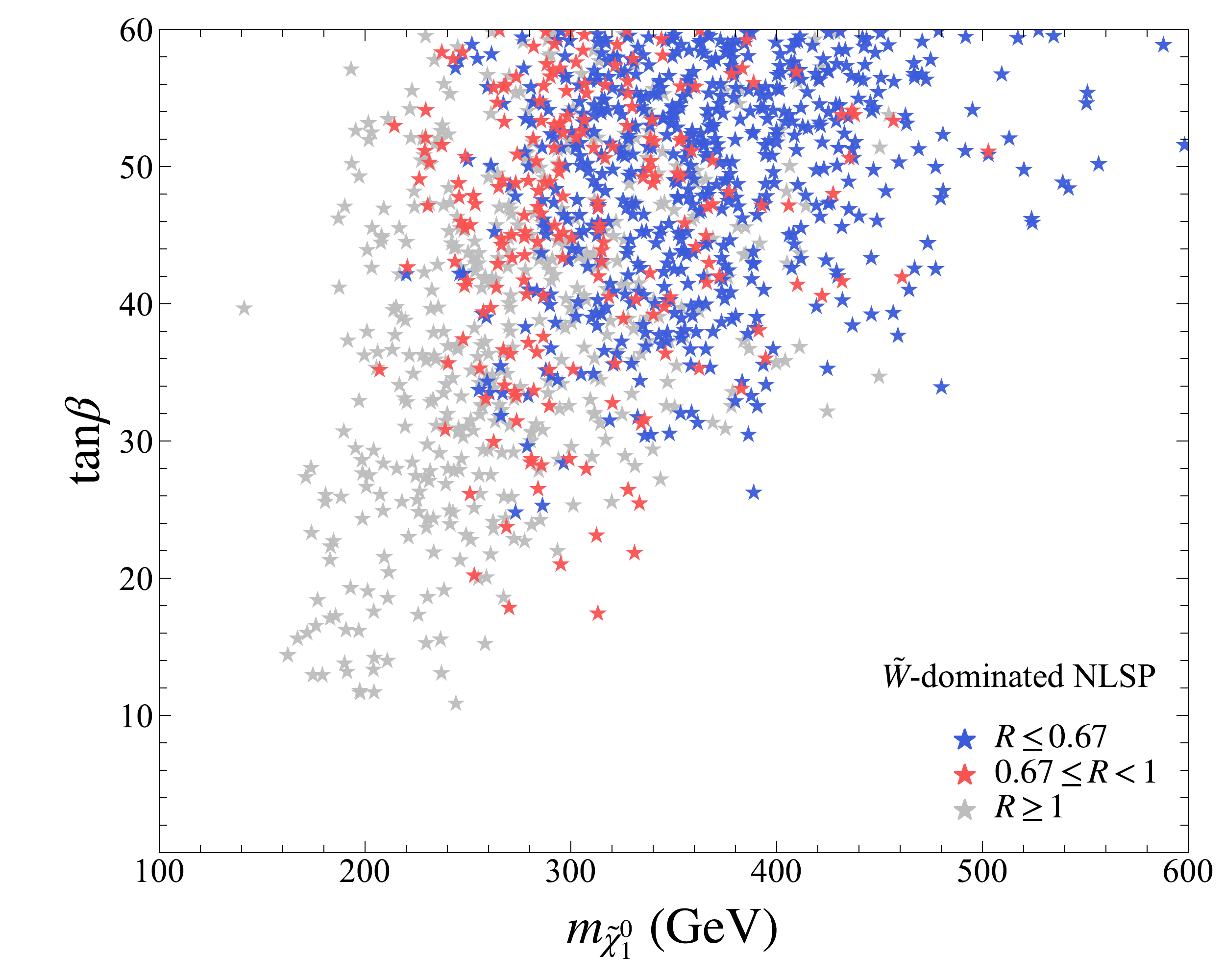}\\
	\includegraphics[width=0.495\textwidth]{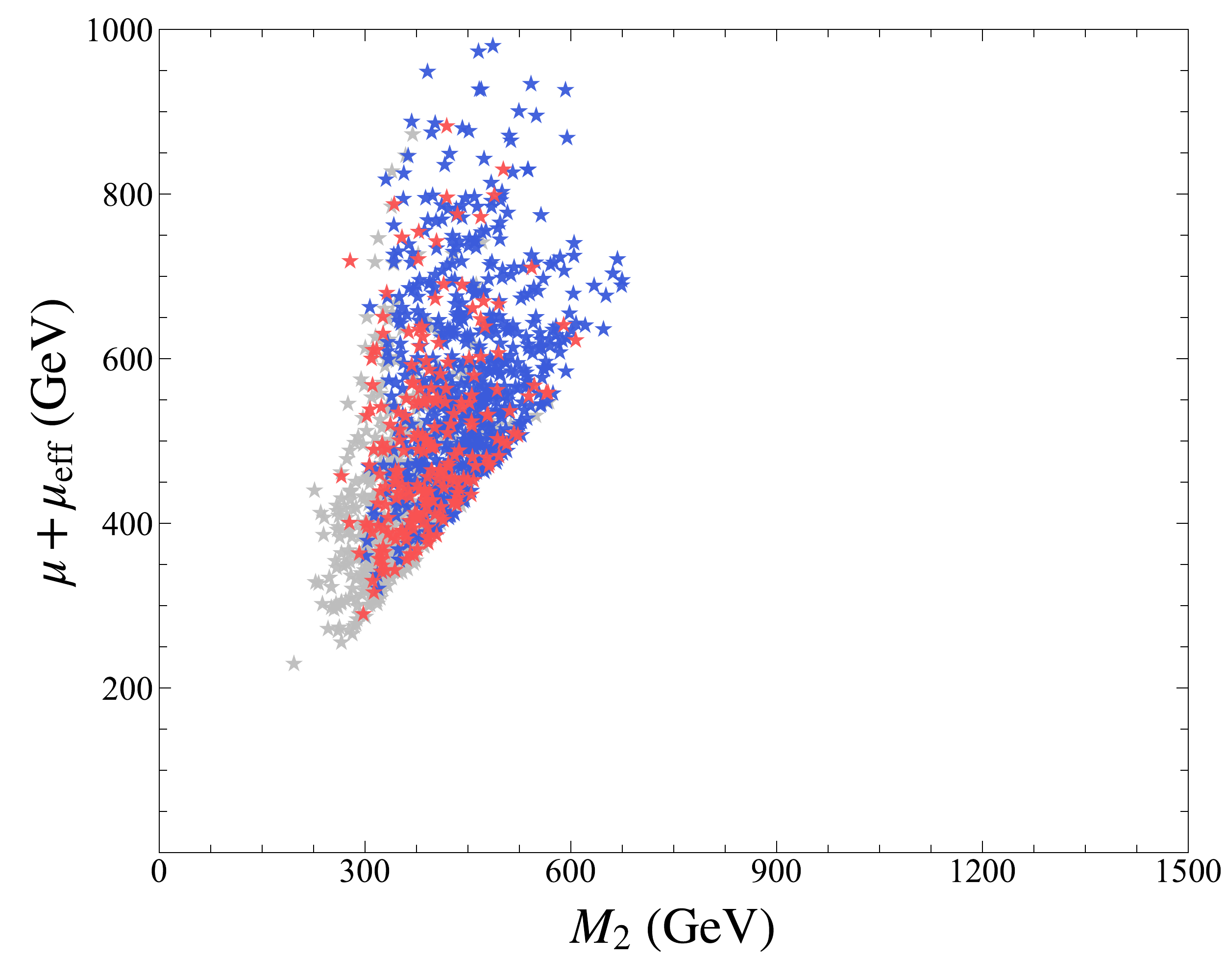}
	\includegraphics[width=0.495\textwidth]{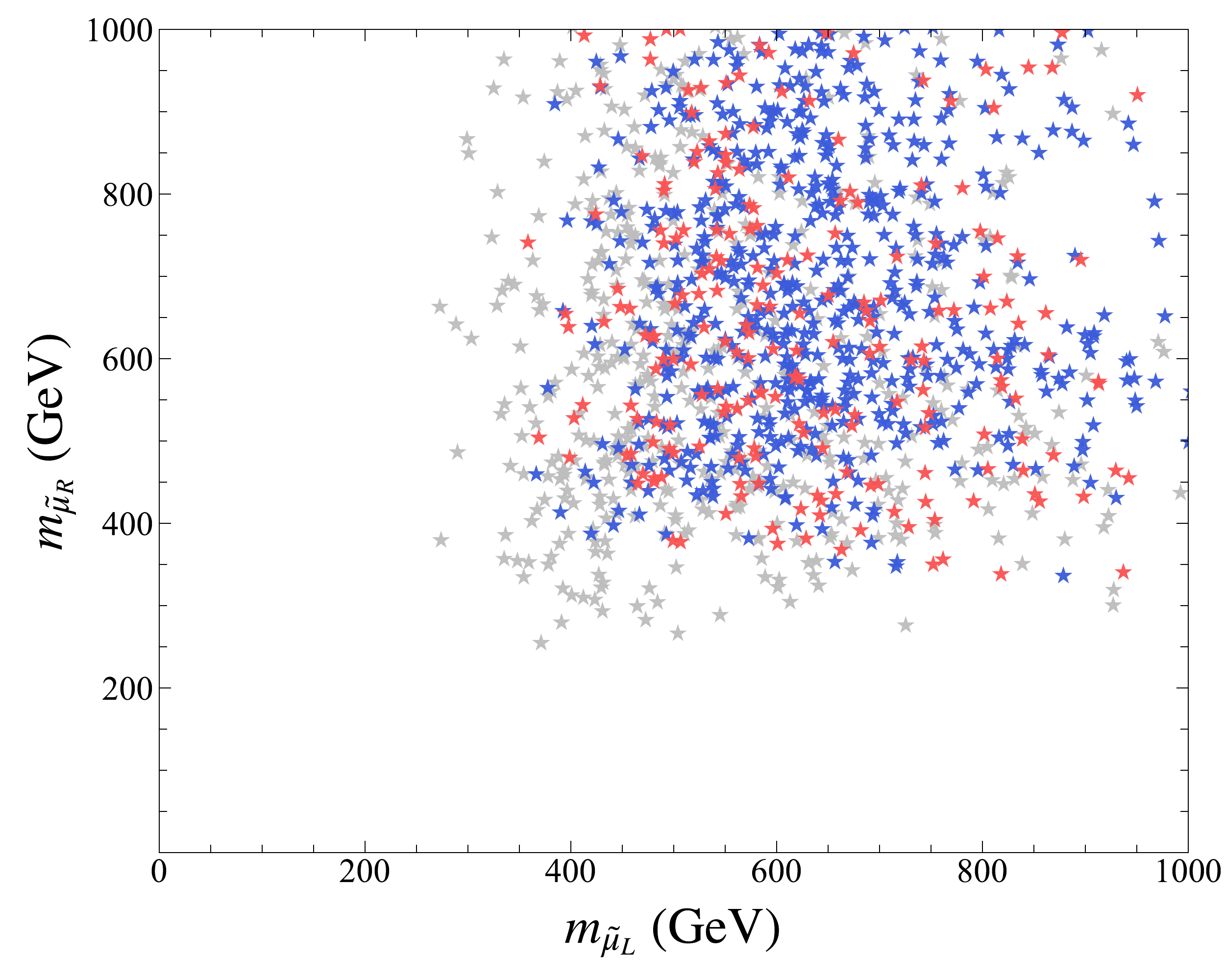}
	\caption{\label{fig:sct3} Similar to Fig.~\ref{fig:sct}, but for the samples with $\tilde{W}$-dominated NLSP. }
\end{figure}

\begin{figure}[t]
	\centering
	\includegraphics[width=0.495\textwidth]{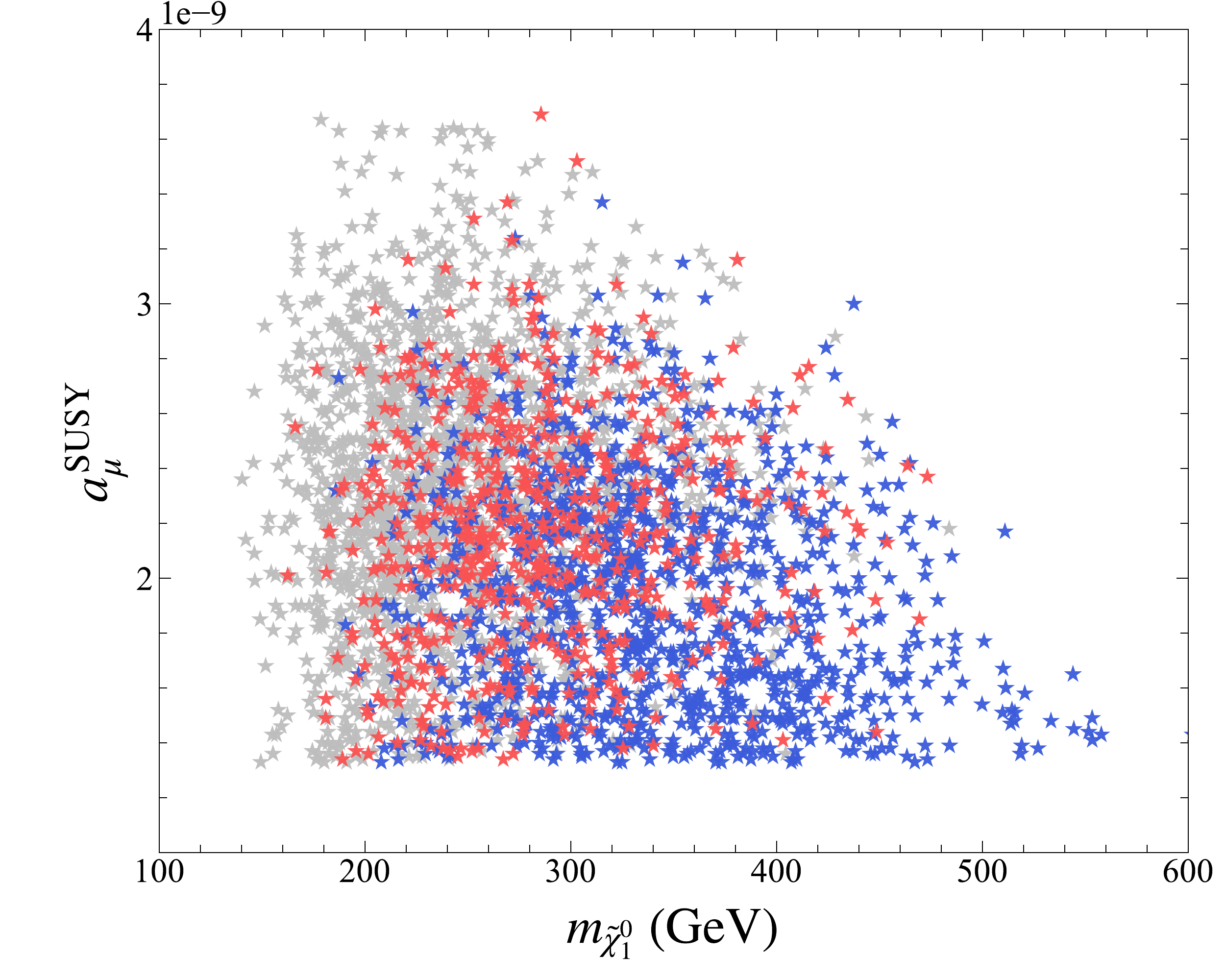}
	\includegraphics[width=0.495\textwidth]{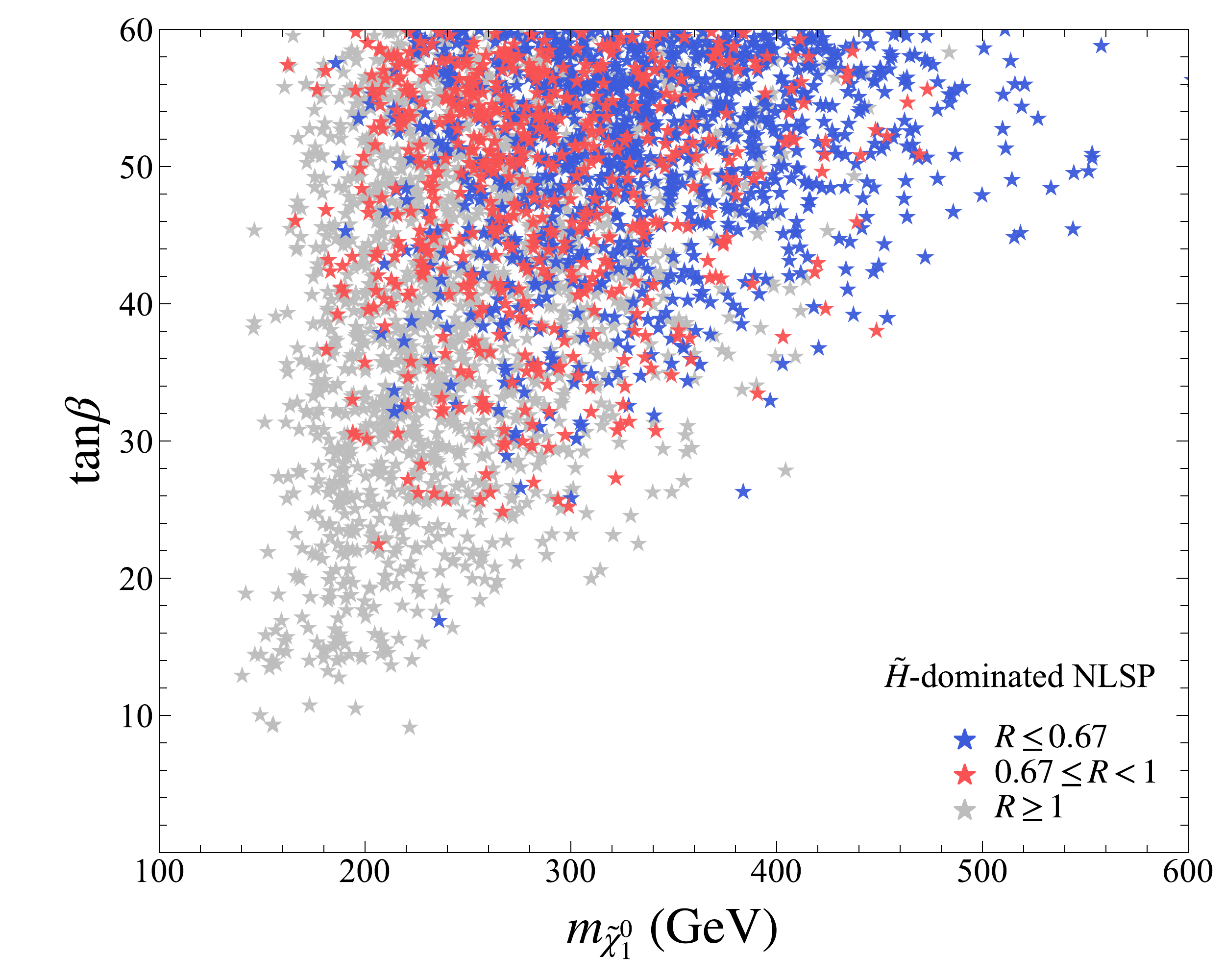}\\
	\includegraphics[width=0.495\textwidth]{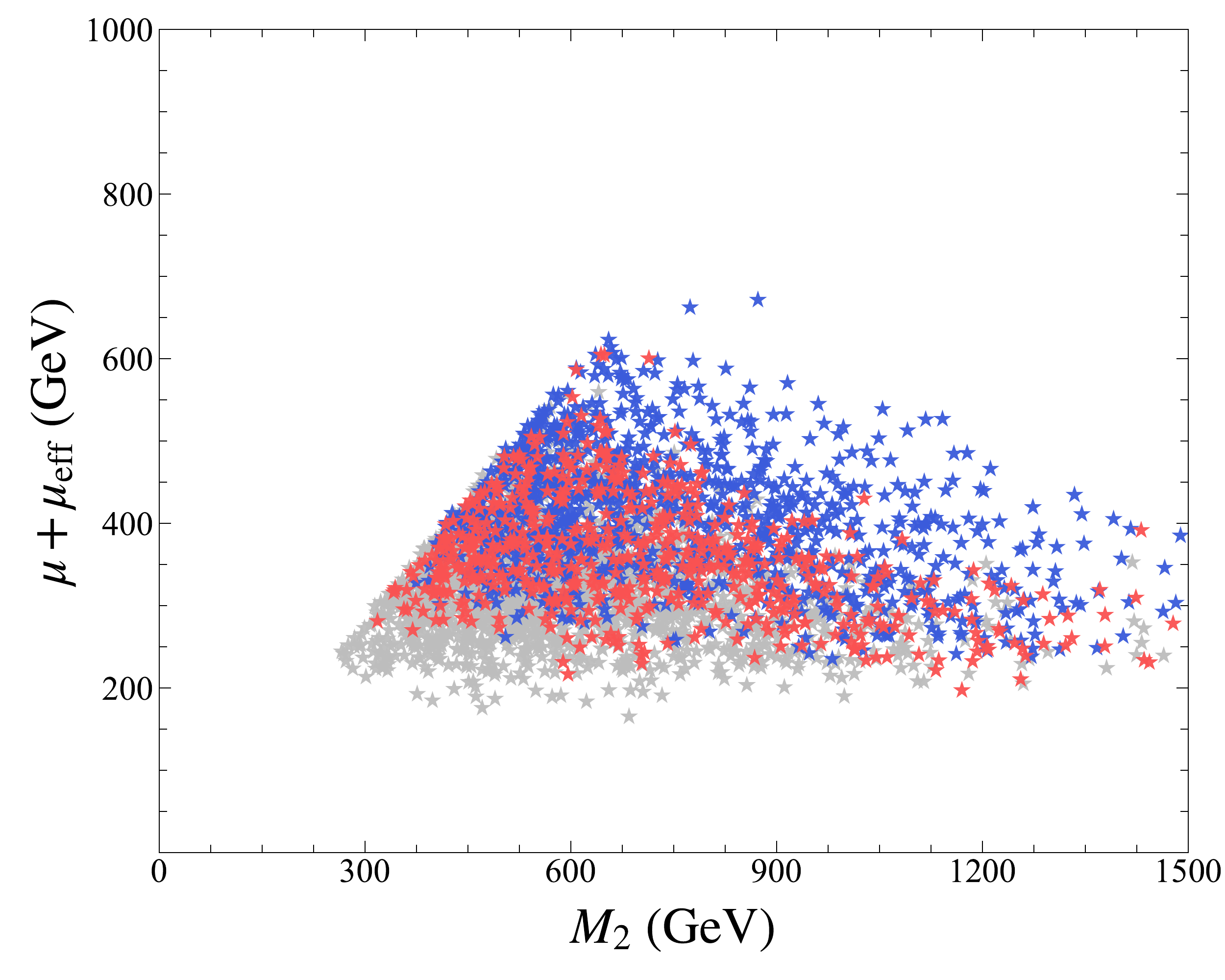}
	\includegraphics[width=0.495\textwidth]{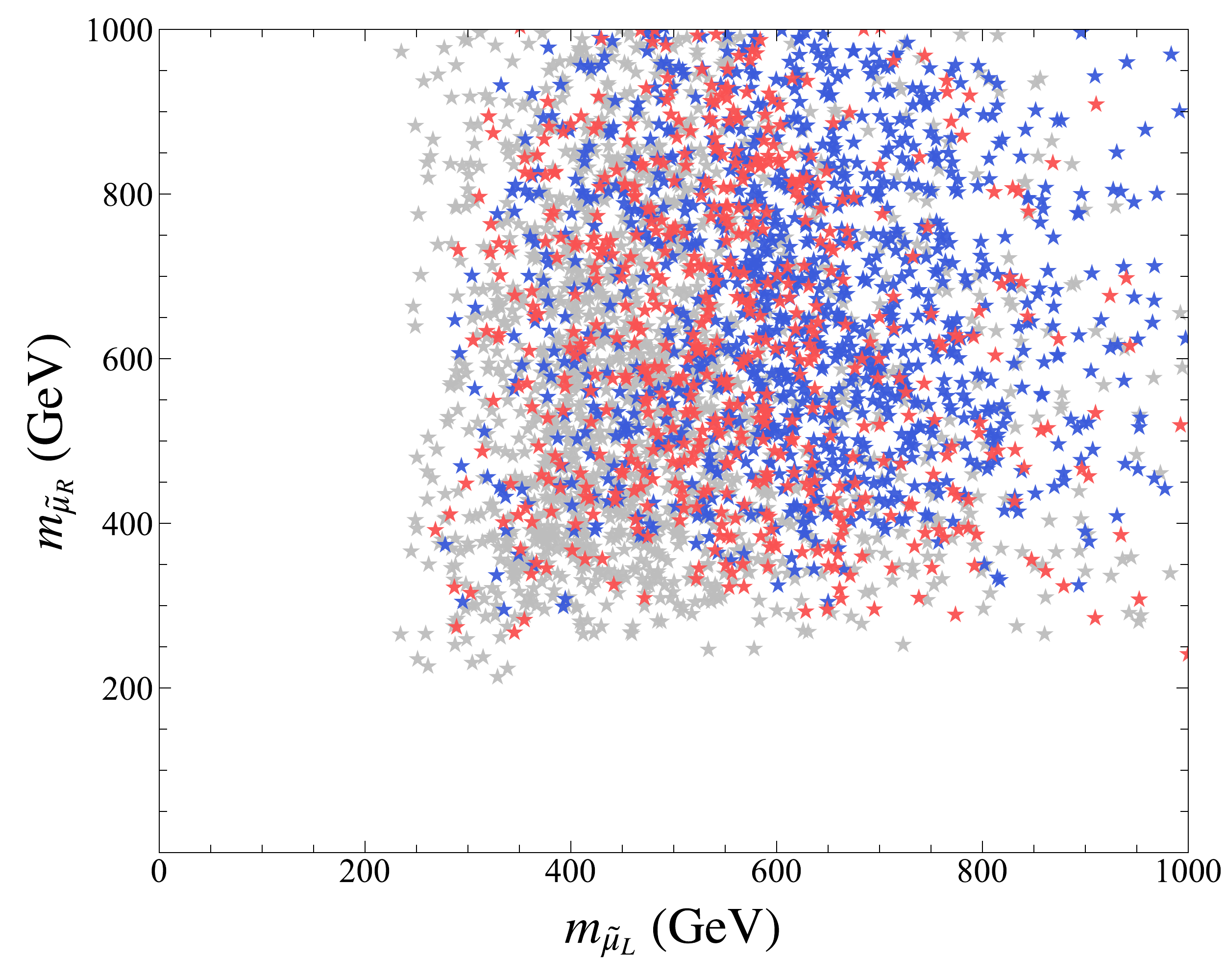}
	\caption{\label{fig:sct4} Similar to Fig.~\ref{fig:sct}, but for the samples with $\tilde{H}$-dominated NLSP. }
\end{figure}

\begin{table}[t]
\centering
\begin{tabular}{crcrrr|rr}
\hline\hline
$\rm NLSP$ 			& $m_{\tilde{\chi}^0_1}$ & $\mu + \mu_{\rm eff}$ & $M_2$ & $m_{\tilde{\mu}_L}$ & $m_{\tilde{\mu}_R}$ & $N_{\rm tot}$ & $N_{\rm pass}$ \\ \hline
$\tilde{\nu}_\mu$ 	& 200 & 250 & 370 & 250 & 300 & 1748 & 124  \\
$\tilde{\mu}_R$ 	& 200 & 300 & 350 & 350 & 300 & 1071 & 24 	\\
$\tilde{B}$ 		& 200 & 300 & 300 & 350 & 350 & 310  & 103 	\\
$\tilde{W}$ 		& 200 & 300 & 250 & 350 & 350 & 1238 & 784 	\\
$\tilde{H}$ 		& 160 & 200 & 300 & 250 & 250 & 3162 & 1606 \\
\hline \hline
\end{tabular}
\caption{\label{tab:mcsum} Summarization of the samples classified by their NLSP's dominant component. $N_{\rm tot}$ represents the total number of each type of samples surveyed by specific Monte Carlo simulations. $N_{\rm pass}$ represents the corresponding number satisfying $R < 1$.  The lower limits of parameters $(\mu + \mu_{\rm eff})$, $M_2$, $m_{\tilde{\chi}_1^0}$, $m_{\tilde{\mu}_L}$, and $m_{\tilde{\mu}_R}$ for the samples surviving the constraints are given in units of $\rm GeV$ in each row. }
\end{table}

\begin{figure}[t]
	\centering
	\includegraphics[width=0.9\textwidth]{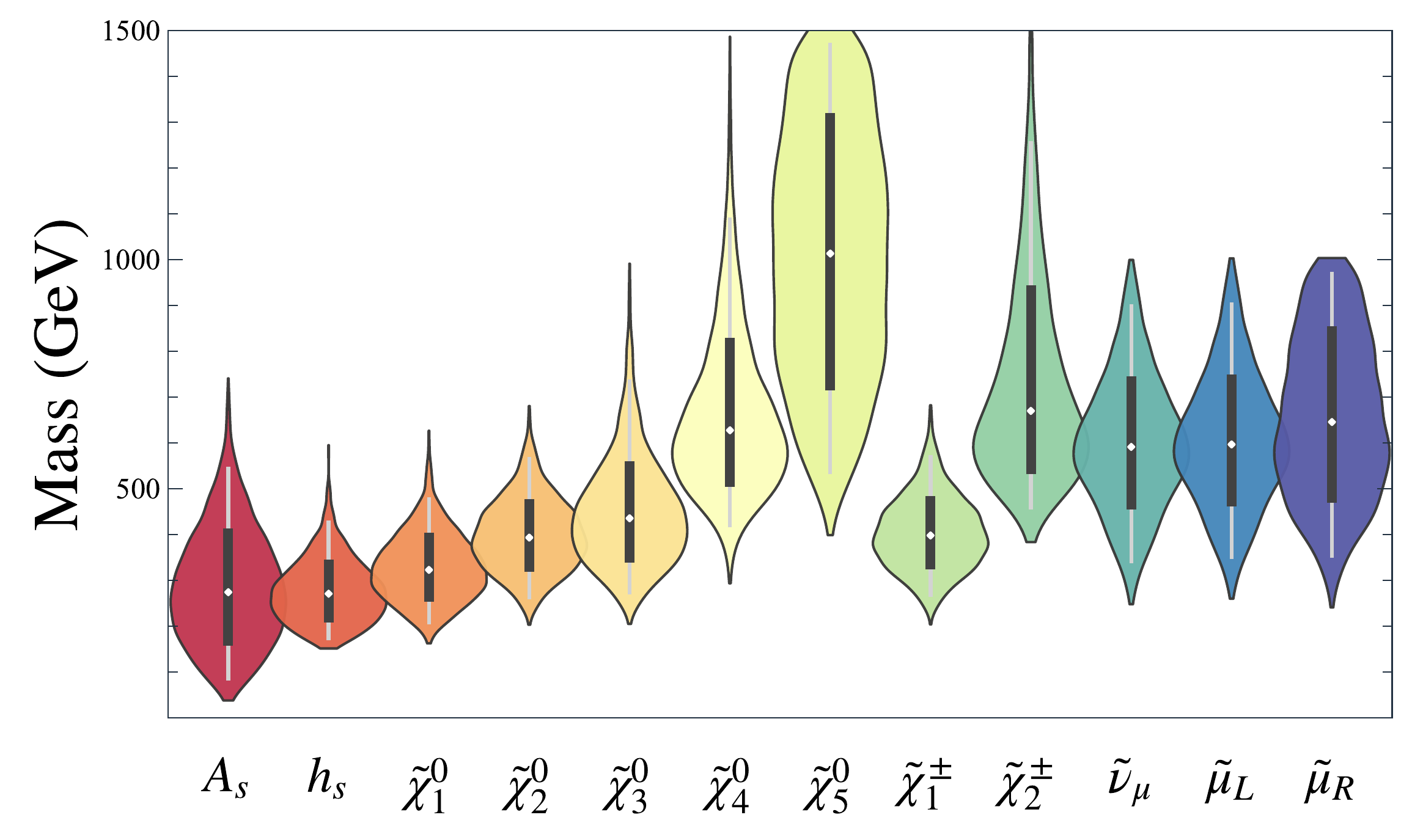}
	\caption{\label{fig:smassviolin1} Same as Fig.~\ref{fig:smassviolin}, but for the samples surviving the LHC constraints.}
\end{figure}

\par The collider simulation results via the \texttt{CheckMATE} show that the LHC searches for supersymmetry set clearly limits on the mass spectrum. In order to show details of the results, the samples are classified by their NLSP's dominant component, which may be $\tilde{\nu}_\mu$, $\tilde{\mu}_R$, $\tilde{B}$, $\tilde{W}$, or $\tilde{H}$.  In Figs.~\ref{fig:sct},~\ref{fig:sct1}, and~\ref{fig:sct2}, samples featured by $\tilde{\nu}_\mu$-, $\tilde{\mu}_R$-, and $\tilde{B}$-dominated NLSP, respectively, are projected on $m_{\tilde{\chi}_1^0}-a_\mu^{\rm SUSY}$ (upper left panel), $m_{\tilde{\chi}_1^0} - \tan{\beta}$ (upper right panel), $M_2 - (\mu+\mu_{\rm eff})$ (lower left panel) and $m_{\tilde{\mu}_L} - m_{\tilde{\mu}_R}$ (lower right panel) planes. Most of the samples, which are marked by grey color, are excluded by the LHC experiments. The rest points marked by red color and blue color stand for those which survive the LHC experiments with $0.67 \leq R < 1$ and $R < 0.67$, respectively\footnote{$0.67 \leq R < 1$ means that the sample's signal is close to exclusion, but a full accounting of uncertainties (originating from e.g., parton distribution function sets, the choice of renormalization and factorisation scale, the details of parton showering or the finite Monte Carlo statistics) would certainly place it within error
bars~\cite{Domingo:2018ykx}. On the other hand, if $R < 0.67$, the sample appears to be essentially compatible with
the experimental results. It corresponds to the number of signal events which is below the $95\%$ C.L. upper bound divided by 1.5.}. In Figs.~\ref{fig:sct3} and~\ref{fig:sct4}, similar diagrams
are plotted for the samples with $\tilde{W}$- and $\tilde{H}$-dominated NLSP, respectively. The total number of each type of samples surveyed by specific Monte Carlo simulations and the corresponding number satisfying $R < 1$  are summarized in Table~\ref{tab:mcsum}, which are denoted by $N_{\rm tot}$ and $N_{\rm pass}$, respectively.
The lower limits of the representative parameters $(\mu + \mu_{\rm eff})$, $M_2$, $m_{\tilde{\chi}_1^0}$, $m_{\tilde{\mu}_L}$ and $m_{\tilde{\mu}_R}$ after considering the LHC constraints are also presented in the table.

From Figs.~\ref{fig:sct}-\ref{fig:sct4} and Table~\ref{tab:mcsum}, the following conclusions are inferred:
\begin{itemize}
	\item Due to the singlet nature of the LSP, sparticle prefers to decay first into lighter sparticles other than the LSP. When $\tilde{\nu}_{\mu}$ or $\tilde{\mu}_R$ acts as NLSP, wino-dominated and higgsino-dominated EWinos decay mostly via slepton and/or sneutrino into leptonic final states, which can enhance the multi-lepton signals. The sparticle's signal is similar to the prediction of the simplified models adopted by ATLAS and CMS collaborations in analyzing experimental data, and it is highly restricted by the current LHC searches.
	\item When the bino-dominated $\tilde{\chi}_2^0$ is NLSP, EWinos and sleptons prefer to decay into it with significant branching ratios because $\tilde{\chi}_2^0$ couples to these particles with unsuppressed gauge interactions. For samples featured by $\Delta \equiv m_{\tilde{\chi}_2^0} - m_{\tilde{\chi}_1^0}$ less than dozens of GeV,  $\tilde{\chi}_2^0$ appears as a missing track at the collider detector. The signal is similar to the MSSM prediction with a bino-dominated LSP, and it leads to strong constraints on the samples from the LHC searches for supersymmetry. However, $\tilde{\chi}_2^0$ will decay by $\tilde{\chi}_2^0 \to \tilde{\chi}_1^0 h$, $\tilde{\chi}_1^0 Z$ if $\Delta > m_h$, and the $\tilde{\chi}_1^0 h$ channel is usually dominant. In this case, it is hard to detect the sparticles partially due to the complexity of the decay chain.
\item For samples with wino- or higgsino-dominated $\tilde{\chi}_2^0$ acting as NLSP, the collider constraints are relatively weak. This observation comes from at least three facts. First, since the wino- or higgsino-dominated $\tilde{\chi}_2^0$ can not decay into sleptons, the leptonic signal rate is usually much smaller than the case where $\tilde{\nu}_{\mu}$ or $\tilde{\mu}_R$ acts as NLSP. Second, the collider sensitive signal events are often diluted by the complicated decay chains of sparticles given that heavy sparticles prefer to decay into the NLSP first. Third, the interpretation of $\Delta a_\mu$ requires that EWinos and smuons are in several hundred GeVs. So for most of the samples surviving the LHC constraints, the mass splitting between sparticles is not large enough to produce high-$p_{\rm T}$ signal objects which can be significantly distinguished from the background in the collider. For example, it was found that the samples featured by wino-dominated NLSP receive the weakest restriction, and most of the surviving samples roughly satisfy $m_{\tilde{\chi}_1^0} \gtrsim 200~{\rm GeV}$, $M_2 - m_{\tilde{\chi}_1^0} \lesssim 150~{\rm GeV}$ and $(\mu + \mu_{\rm eff}) - M_2 \lesssim 200~{\rm GeV}$. This parameter configuration is not sensitive to the current LHC EWino direct searches (see, e.g., point P2 in Table~\ref{tab:benchmark-points}).

   In addition, it is noticeable that most of the surviving samples are characterized by $\mu + \mu_{tot} < 600~{\rm GeV}$, which can predict $m_Z$ naturally. It is a distinct difference between the $\mu$NMSSM and MSSM.

\item The $a_\mu^{\rm SUSY}$ enhancement factor $\tan{\beta}$ is apparently restricted by the current LHC constraints. As indicated by the upper right panel of Figs.~\ref{fig:sct}-\ref{fig:sct4}, the LHC observations require  $\tan{\beta}$ larger than about 20 to interpret $\Delta a_\mu$ within $2\sigma$ level, and about 30 within $1\sigma$ level.
	\item It should be noted that some of the latest LHC analyses, e.g., the analysis in~\cite{ATLAS:2021moa}, are not included in this work. However, one can make some rough estimations by thinking that the red points in Figs.~\ref{fig:sct}-\ref{fig:sct4} will be excluded by the latest or the near future LHC analyses. There is still a relatively large parameter space of the $\mu$NMSSM that can interpret $\Delta a_\mu$ without conflicting with the newest LHC analyses. So the LHC constraints considered in this article are of a reference value.
\end{itemize}

\begin{table}[t]
\centering
\resizebox{1\textwidth}{!}
{
\begin{tabular}{crcr|crcr}
\hline \hline
\multicolumn{4}{c|}{\bf Benchmark Point P1}                                                                                                & \multicolumn{4}{c}{\bf Benchmark Point P2}                                                                                                \\ \hline
$\lambda$             & 0.019     & $m_{h_s}$                & 364.4~GeV & $\lambda$             & 0.017 & $m_{h_s}$                & 233.1~GeV \\
$\kappa$              & -0.267     & $m_{A_s}$                & 262.1~GeV & $\kappa$              & 0.234 & $m_{A_s}$               & 315.8~GeV \\
$\tan{\beta}$         & 54.99     & $m_{h}$                  & 125.5~GeV & $\tan{\beta}$         & 59.87 & $m_{h}$                  & 124.7~GeV \\
$\mu$                 & 423.3~GeV & $m_{H}$                  & 1970~GeV & $\mu$                 & 640.6~GeV & $m_{H}$                  & 2388~GeV \\
$\mu + \mu_{\rm eff}$ & 437.6~GeV & $m_{A_H}$                & 1970~GeV & $\mu + \mu_{\rm eff}$ & 651.4~GeV & $m_{A_H}$                & 2388~GeV \\
$A_t$                 & 2545~GeV & $m_{\tilde{\chi}_1^0}$   & 394.5~GeV & $A_t$                  & 2570~GeV & $m_{\tilde{\chi}_1^0}$   & 295.2~GeV \\
$A_\kappa$            & 116.6~GeV  & $m_{\tilde{\chi}_2^0}$   & 425.9~GeV & $A_\kappa$                  & -225.9~GeV   & $m_{\tilde{\chi}_2^0}$   & 445.2~GeV \\
$M_1$                 & 1086~GeV          & $m_{\tilde{\chi}_3^0}$   & 452.6~GeV & $M_1$                & -1239~GeV & $m_{\tilde{\chi}_3^0}$   & 665.7~GeV \\
$M_2$                 & 541.7~GeV          & $m_{\tilde{\chi}_4^0}$   & 591.0~GeV & $M_2$            & 443.0~GeV & $m_{\tilde{\chi}_4^0}$   & 679.9~GeV \\
$m_L$                 & 609.6~GeV          & $m_{\tilde{\chi}_5^0}$   & 1091~GeV & $m_L$                 & 539.5~GeV & $m_{\tilde{\chi}_5^0}$   & 1244~GeV \\
$m_E$                 & 992.4~GeV          & $m_{\tilde{\chi}_1^\pm}$ & 429.9~GeV & $m_E$                 & 697.9~GeV & $m_{\tilde{\chi}_1^\pm}$ & 445.5~GeV \\
$a_\mu^{\rm SUSY}$    & $2.514\times 10^{-9}$          & $m_{\tilde{\chi}_2^\pm}$ & 592.0~GeV & $a_\mu^{\rm SUSY}$    & $2.510\times 10^{-9}$ & $m_{\tilde{\chi}_2^\pm}$ & 682.2~GeV \\
$\Omega h^2$          &  0.142    & $m_{\tilde{\mu}_L}$      & 617.5~GeV & $\Omega h^2$          & 0.108 & $m_{\tilde{\mu}_L}$      & 546.3~GeV \\
$\sigma_p^{\rm SI}$   & $2.02\times 10^{-47}~{\rm cm}^2$          & $m_{\tilde{\mu}_R}$      & 927.3~GeV & $\sigma_p^{\rm SI}$   & $1.47\times 10^{-46}~{\rm cm}^2$ & $m_{\tilde{\mu}_R}$      & 703.8~GeV \\
$\sigma_n^{\rm SD}$   & $1.84\times 10^{-45}~{\rm cm}^2$          & $m_{\tilde{\nu}_\mu}$    & 612.0~GeV & $\sigma_n^{\rm SD}$   & $1.83\times 10^{-47}~{\rm cm}^2$ & $m_{\tilde{\nu}_\mu}$    & 540.3~GeV \\ \hline
\multicolumn{2}{c}{$N_{11},~N_{12},~N_{13},~N_{14},~N_{15}$} & \multicolumn{2}{c|}{~0.000,~~0.002,~~0.027,~~0.031,~-0.999} & \multicolumn{2}{c}{$N_{11},~N_{12},~N_{13},~N_{14},~N_{15}$} & \multicolumn{2}{c}{~0.000,~~0.003,~-0.003,~~0.006,~~0.999}\\
\multicolumn{2}{c}{$N_{21},~N_{22},~N_{23},~N_{24},~N_{25}$} & \multicolumn{2}{c|}{~0.043,~-0.365,~~0.674,~-0.640,~-0.003} & \multicolumn{2}{c}{$N_{21},~N_{22},~N_{23},~N_{24},~N_{25}$} & \multicolumn{2}{c}{-0.004,~-0.969,~~0.206,~-0.138,~-0.004}\\
\multicolumn{2}{c}{$N_{31},~N_{32},~N_{33},~N_{34},~N_{35}$} & \multicolumn{2}{c|}{-0.020,~~0.054,~~0.703,~~0.708,~~0.041} & \multicolumn{2}{c}{$N_{31},~N_{32},~N_{33},~N_{34},~N_{35}$} & \multicolumn{2}{c}{-0.053,~~0.049,~-0.705,~-0.705,~-0.002}\\
\multicolumn{2}{c}{$N_{41},~N_{42},~N_{43},~N_{44},~N_{45}$} & \multicolumn{2}{c|}{~0.026,~~0.929,~~0.224,~-0.293,~-0.001} & \multicolumn{2}{c}{$N_{41},~N_{42},~N_{43},~N_{44},~N_{45}$} & \multicolumn{2}{c}{~0.016,~-0.242,~-0.678,~~0.693,~~0.005}\\
\multicolumn{2}{c}{$N_{51},~N_{52},~N_{53},~N_{54},~N_{55}$} & \multicolumn{2}{c|}{~0.999,~-0.007,~-0.021,~~0.049,~~0.000} & \multicolumn{2}{c}{$N_{51},~N_{52},~N_{53},~N_{54},~N_{55}$} & \multicolumn{2}{c}{~0.998,~-0.002,~-0.025,~-0.049,~~0.000}\\ \hline
\multicolumn{2}{c}{ Annihilations }                                                                                  & \multicolumn{2}{c|}{Fractions [\%]} & \multicolumn{2}{c}{Annihilations}                                                                                  & \multicolumn{2}{c}{Fractions [\%]}                                                                                  \\
\multicolumn{2}{c}{$\tilde{\chi}_1^0\tilde{\chi}_1^0 \to h_s A_s / h_s h_s / \cdots  $} & \multicolumn{2}{l|}{78.7~/~4.1~/~$\cdots$}        & \multicolumn{2}{c}{$\tilde{\chi}_1^0\tilde{\chi}_1^0 \to h_s A_s / h_s h_s / h A_s$} & \multicolumn{2}{l}{94.5~/~4.4~/~1.1}        \\ \hline
\multicolumn{2}{c}{ Decays }                                                                                  & \multicolumn{2}{c|}{Branching ratios [\%]} & \multicolumn{2}{c}{Decays}                                                                                  & \multicolumn{2}{c}{Branching ratios [\%]}                                                                                  \\
\multicolumn{2}{l}{$\tilde{\chi}_2^0 \to \tilde{\chi}_1^0 Z^*( \to \tilde{\chi}_1^0 f \bar{f})$} & \multicolumn{2}{l|}{100}        & \multicolumn{2}{l}{$\tilde{\chi}_2^0 \to \tilde{\chi}_1^0 h / \tilde{\chi}_1^0 Z $} & \multicolumn{2}{l}{99.8~/~0.2}        \\
\multicolumn{2}{l}{$\tilde{\chi}_3^0 \to \tilde{\chi}_1^0 Z^*( \to \tilde{\chi}_1^0 f \bar{f}) / \tilde{\chi}_1^\pm W^*(\to \tilde{\chi}_1^\pm f f^\prime)  $} & \multicolumn{2}{l|}{46.3~/~53.6}        & \multicolumn{2}{l}{$\tilde{\chi}_3^0 \to \tilde{\chi}_1^\pm W^\mp / \tilde{\chi}_2^0 Z $} & \multicolumn{2}{l}{68.3~/~30.3}        \\
\multicolumn{2}{l}{$\tilde{\chi}_4^0 \to \tilde{\chi}_1^\pm W^\mp / \tilde{\chi}_{3}^0 Z / \tilde{\chi}_{3}^0 h $} & \multicolumn{2}{l|}{70.0~/~15.1~/~14.0}        & \multicolumn{2}{l}{$\tilde{\chi}_4^0 \to \tilde{\chi}_1^\pm W^\mp / \tilde{\chi}_2^0 h / \tilde{\chi}_2^0 Z$} & \multicolumn{2}{l}{69.2~/~27.5~/~~1.6}        \\
\multicolumn{2}{l}{$\tilde{\chi}_5^0 \to \tilde{\chi}_1^\pm W^\mp / \tilde{\chi}_{2,3}^0 Z / \tilde{\chi}_{2,3}^0 h / \tilde{\mu}_{L,R} \mu / \tilde{\nu}_\mu \nu_\mu$} & \multicolumn{2}{l|}{21.2~/~14.7~/~12.1~/~29.8~/~~9.9}        & \multicolumn{2}{l}{$\tilde{\chi}_5^0 \to \tilde{\mu}_L \mu / \tilde{\mu}_R \mu / \tilde{\nu}_\mu \nu_\mu / \tilde{\chi}_2^\pm W^\mp / \tilde{\chi}_3^0 h $} & \multicolumn{2}{l}{13.3~/~38.1~/~13.7~/~15.3~/~~8.1}        \\
\multicolumn{2}{l}{$\tilde{\chi}_1^\pm \to \tilde{\chi}_1^0 W^\ast  $} & \multicolumn{2}{l|}{100}        & \multicolumn{2}{l}{$\tilde{\chi}_1^\pm \to \tilde{\chi}_1^0 W^\pm $} & \multicolumn{2}{l}{100}        \\
\multicolumn{2}{l}{$\tilde{\chi}_2^\pm \to \tilde{\chi}_{2,3}^0 W^\pm /\tilde{\chi}_1^\pm Z$} & \multicolumn{2}{l|}{55~/~23}        & \multicolumn{2}{l}{$\tilde{\chi}_2^\pm \to \tilde{\chi}_{2}^0 W^\pm / \tilde{\chi}_1^\pm h / \tilde{\chi}_1^\pm Z / \tilde{\mu}_L \nu_\mu$} & \multicolumn{2}{l}{36.1~/~33.5~/~28.5~/~~1.1}        \\
\multicolumn{2}{l}{$\tilde{\mu}_L \to \tilde{\chi}_1^- \nu_\mu / \tilde{\chi}_2^0 \mu / \tilde{\chi}_2^- \nu_\mu / \tilde{\chi}_4^0 \mu$} & \multicolumn{2}{l|}{52.0~/~30.4~/~10.8~/~5.8}        & \multicolumn{2}{l}{$\tilde{\mu}_L \to \tilde{\chi}_1^- \nu_\mu /\tilde{\chi}_2^0 \mu  $} & \multicolumn{2}{l}{65.9~/~34.1}        \\
\multicolumn{2}{l}{$\tilde{\mu}_R \to \tilde{\chi}_{2,3}^0 \mu / \tilde{\chi}_1^- \nu_{\mu} / \tilde{\mu}_L h / \tilde{\mu}_L Z / \tilde{\nu}_\mu W^- $} & \multicolumn{2}{l|}{48.3~/~29.9~/~~3.3~/~~3.2~/~~7.0}        & \multicolumn{2}{l}{$\tilde{\mu}_R \to \tilde{\nu}_\mu W^- / \tilde{\mu}_L h / \tilde{\mu}_L Z / \tilde{\chi}_3^0 \mu $} & \multicolumn{2}{l}{54.0~/~21.8~/~19.8~/~~2.2}        \\
\multicolumn{2}{l}{$\tilde{\nu}_\mu \to \tilde{\chi}_1^+ \mu / \tilde{\chi}_2^0 \nu_\mu /\tilde{\chi}_2^+ \mu / \tilde{\chi}_4^0 \nu_\mu /  $} & \multicolumn{2}{l|}{64.5~/~27.7~/~~4.6~/~~2.6}        & \multicolumn{2}{l}{$\tilde{\nu}_\mu \to \tilde{\chi}_2^0 \nu_\mu / \tilde{\chi}_1^+ \mu$} & \multicolumn{2}{l}{32.8~/~67.2}        \\ \hline
\multicolumn{2}{c}{$R$ value} & \multicolumn{2}{c|}{0.207}        & \multicolumn{2}{c}{$R$ value} & \multicolumn{2}{c}{0.393}        \\ \hline \hline
\end{tabular}}
\caption{\label{tab:benchmark-points}Detailed information of two benchmark points that agree well with all of the DM and Higgs experiments and predict $a_\mu^{\rm SUSY} \simeq 2.51 \times 10^{-9}$. Numbers after annihilation processes represent their fractions in contributing to total DM annihilation cross-section at freeze-out temperature. Numbers after sparticle decay channels denote their branching ratios. }
\end{table}

In order to emphasize the properties of the samples with higgsino-dominated $\tilde{\chi}_2^0$ and wino-dominated $\tilde{\chi}_2^0$, two benchmark points, P1 and P2, are chosen to present their detailed information in Table~\ref{tab:benchmark-points}. Both points predict $a_\mu^{\rm SUSY}$ values approximately equal to $2.51\times 10^{-9}$ and pass all the experimental constraints. These two benchmark points verify our previous discussions.

Finally, the mass spectra of the sparticles surviving the LHC constraints are presented in Fig.~\ref{fig:smassviolin1}. Comparing it with Fig.~\ref{fig:smassviolin}, it was found that the LHC constraints are very effective in excluding relatively light sparticles, especially light sleptons. In fact, this conclusion is also reflected in the last three panels of Figs.~\ref{fig:sct}-\ref{fig:sct4}. Another important conclusion is that the $\tilde{W}$ or $\tilde{H}$-dominated $\tilde{\chi}_1^\pm$ is always lighter than about $700~{\rm GeV}$. Taking into account its production cross-section at linear $e^+ e^-$ colliders~\cite{Heinemeyer:2017izw}, one can infer that $\tilde{\chi}_1^\pm$ is very likely to be discovered at the CLIC, whose collision energy can reach $3~{\rm TeV}$~\cite{linssen2012physics,CLICDetector:2013tfe,Moortgat-Pick:2015lbx,CLICdp:2018cto}. This point was recently emphasized by the authors of~\cite{Chakraborti:2020vjp,Chakraborti:2021dli}. Moreover, as mentioned in~\cite{Endo:2021zal,Endo:2020mqz}, future high luminosity LHC can significantly extend the LHC Run-II's capability in sparticle detection. The preliminary analyses carried out in, e.g.,~\cite{ATL-PHYS-PUB-2018-048,CidVidal:2018eel,Beresford:2018pbt}, have proven this point. These machines provide an opportunity to test the $\mu$NMSSM interpretation of $\Delta a_\mu$ once the deviation of $a_\mu^{exp}$ from the SM prediction is confirmed. This issue will be studied in detail in our future work.

\section{\label{sec:summary}Summary}
In this work, the phenomenology of the new Fermilab result of $\Delta a_\mu$ interpreted in the $\mu$NMSSM was investigated. The obtained results show the following features:
\begin{itemize}
	\item Compared with the MSSM or the $\mathbb{Z}_3$-NMSSM, the strong exclusivity from DM physics and natural interpretations of $\Delta a_\mu$ is weak in the $\mu$NMSSM.
	\item A singlino-dominated DM candidate is preferred in the interpretation. Owing to the smallness of the singlet-doublet Higgs coupling $\lambda$, the singlino-dominated neutralino and singlet-dominated Higgs bosons may form a secluded DM sector in which the annihilation channel $\tilde{\chi}_1^0 \tilde{\chi}_1^0 \to h_s A_s$ is responsible for the measured abundance by adopting an appropriate singlet Yukawa coupling $\kappa$.
	\item The secluded sector communicates with the SM sector by weak singlet-doublet Higgs mixing, so the scatterings of singlino-dominated DM with nucleons are suppressed below the current experimental limits.
	\item The mass of DM $m_{\tilde{\chi}_1^0}$ must be heavier than about $150~{\rm GeV}$ to proceed with the annihilation, and must be lighter than about $550~{\rm GeV}$ to explain $\Delta a_\mu$ at $1\sigma$ level.
	\item Owing to the singlet nature of DM and the complex mass hierarchy of sparticles, the decay chains of EWinos and sleptons are lengthened in comparison with the MSSM prediction. Moreover, the singlet Higgs bosons $h_s$ and $A_s$ in the final states of the searching channels weaken the LHC detection capability. These characteristics make sparticle detection at the LHC rather tricky.
\end{itemize}
\par This study shows that the proposed theory can readily explain the discrepancy of the muon anomalous magnetic moment between its SM prediction and experimentally measured value, without conflicting with DM and Higgs experimental results and the LHC searches for supersymmetry. Among the interpretations, it is remarkable that the higgsino mass is less than $500~{\rm GeV}$ in most cases so that the $Z$ boson mass can be naturally predicted.

\appendix
\section{\label{app:smodels}Fast simulation via SModelS}
\begin{table}[t]
	\centering
	\resizebox{1\textwidth}{!}{
	\begin{tabular}{llcr}
	\hline\hline  \addlinespace
	\bf Analysis & \bf Simplified Scenario  & \bf Signal of Final State & \bf Luminosity \\ \addlinespace \hline \addlinespace
	\tabincell{c}{\bf CMS-SUS-17-010~\cite{Sirunyan:2018lul}\\ (arXiv:1807.07799)}  &\tabincell{l}{
			$ \tilde{\chi}_1^{\pm} \tilde{\chi}_1^{\mp} \to W^{\pm} \tilde{\chi}_1^0 W^{\mp}\tilde{\chi}_1^0 $\\
			$ \tilde{\chi}_1^{\pm} \tilde{\chi}_1^{\mp} \to \nu\tilde{\ell} / \ell\tilde{\nu} \to \ell \ell \nu \nu \tilde{\chi}_1^0 \tilde{\chi}_1^0 $
		}
		& $2\ell + E_{\rm T}^{\rm miss}$
		& $35.9~{\rm fb}^{-1}$  \\ \addlinespace
	\tabincell{c}{\bf CMS-SUS-17-009~\cite{Sirunyan:2018nwe}\\ (arXiv:1806.05264)}
		& ${\tilde{\ell}\tilde{\ell}} \to \ell \ell \tilde{\chi}_1^0 \tilde{\chi}_1^0$
		& $2\ell + E_{\rm T}^{\rm miss}$
		& $35.9~{\rm fb}^{-1}$               \\ \addlinespace
	\tabincell{c}{\bf CMS-SUS-17-004~\cite{Sirunyan:2018ubx}\\ (arXiv:1801.03957)}
		& $\tilde{\chi}_{2}^0\tilde{\chi}_1^{\pm}\to Wh(Z)\tilde{\chi}_1^0\tilde{\chi}_1^0$
		& ${\rm n}\ell (\geq 0) + {\rm n}j ( \geq 0) +  E_{\rm T}^{\rm miss}$
		& $35.9~{\rm fb}^{-1}$               \\ \addlinespace
	\tabincell{c}{\bf CMS-SUS-16-045~\cite{Sirunyan:2017eie}\\ (arXiv:1709.00384)}
		& $\tilde{\chi}_2^0\tilde{\chi}_1^{\pm}\to W^{\pm}\tilde{\chi}_1^0h\tilde{\chi}_1^0$
		& $1 \ell 2b + E_{\rm T}^{\rm miss}$
		& $35.9~{\rm fb}^{-1}$               \\ \addlinespace
	\tabincell{c}{\bf CMS-SUS-16-039~\cite{Sirunyan:2017lae}\\(arxiv:1709.05406) }          &\tabincell{l}{
			$ \tilde{\chi}_2^0 \tilde{\chi}_1^{\pm} \to \ell \tilde{\nu} \ell \tilde{\ell}$\\
			$ \tilde{\chi}_2^0 \tilde{\chi}_1^{\pm} \to \tilde{\tau} \nu \tilde{\ell} \ell$\\
			$ \tilde{\chi}_2^0 \tilde{\chi}_1^{\pm} \to \tilde{\tau} \nu \tilde{\tau} \tau$\\
			$ \tilde{\chi}_2^0 \tilde{\chi}_1^{\pm} \to WZ \tilde{\chi}_1^0 \tilde{\chi}_1^0$\\
			$ \tilde{\chi}_2^0 \tilde{\chi}_1^{\pm} \to WH \tilde{\chi}_1^0 \tilde{\chi}_1^0$
			}
		& $n\ell(\geq 0)(\tau)+E_{\rm T}^{\rm miss}$
		& $35.9~{\rm fb}^{-1}$               \\  \addlinespace
	\tabincell{c}{\bf CMS-SUS-16-034~\cite{Sirunyan:2017qaj}\\ (arXiv:1709.08908)}
		& $\tilde{\chi}_2^0 \tilde{\chi}_1^{\pm} \to W\tilde{\chi}_1^0 Z(h) \tilde{\chi}_1^0$
		& ${\rm n}\ell (\geq 2) + {\rm n}j(\geq 1) E_{\rm T}^{\rm miss}$
		& $35.9~{\rm fb}^{-1}$               \\ \addlinespace
	\tabincell{c}{\bf CERN-EP-2017-303~\cite{Aaboud:2018jiw}\\ (arXiv:1803.02762)} &\tabincell{l}{
			$ \tilde{\chi}_2^0 \tilde{\chi}_1^{\pm} \to WZ \tilde{\chi}_1^0 \tilde{\chi}_1^0$ \\
			$ \tilde{\chi}_2^0 \tilde{\chi}_1^{\pm} \to \nu \tilde{\ell} \ell \tilde{\ell}$ \\
			$ \tilde{\chi}_1^{\pm} \tilde{\chi}_1^{\mp} \to \nu\tilde{\ell} / \ell\tilde{\nu} \to \ell \ell \nu \nu \tilde{\chi}_1^0 \tilde{\chi}_1^0 $ \\
			$ \tilde{\ell} \tilde{\ell} \to \ell \ell \tilde{\chi}_1^0 \tilde{\chi}_1^0$
			}
		& ${\rm n}\ell (\geq 2) + E_{\rm T}^{\rm miss}$
		& $35.9~{\rm fb}^{-1}$              \\ \addlinespace
	\tabincell{c}{\bf CERN-EP-2018-306~\cite{Aaboud:2018ngk}\\ (arXiv:1812.09432)}
		& $\tilde{\chi}_2^0 \tilde{\chi}_1^{\pm} \to Wh \tilde{\chi}_1^0\tilde{\chi}_1^0$
		& ${\rm n}\ell(\geq0) + {\rm n}j(\geq0) + {\rm n}b(\geq0) + {\rm n}\gamma(\geq0) + E_{\rm T}^{\rm miss}$
		& $35.9~{\rm fb}^{-1}$             	\\ \addlinespace
	\tabincell{c}{\bf CERN-EP-2018-113~\cite{Aaboud:2018sua}\\ (arXiv:1806.02293)}
		& $\tilde{\chi}_2^0 \tilde{\chi}_1^{\pm} \to WZ \tilde{\chi}_1^0 \tilde{\chi}_1^0$
		& ${\rm n}\ell(\geq2) + {\rm n}j(\geq0) + E_{\rm T}^{\rm miss}$
		& $35.9~{\rm fb}^{-1}$              \\ \addlinespace
	\tabincell{c}{\bf CERN-EP-2019-263~\cite{Aad:2019vvi}\\ (arXiv:1912.08479)}
		& $\tilde{\chi}_2^0 \tilde{\chi}_1^{\pm} \to W\tilde{\chi}_1^0 Z \tilde{\chi}_1^0 \to \ell\nu \ell \ell \tilde{\chi}_1^0 \tilde{\chi}_1^0 $
		& $3\ell + E_{\rm T}^{\rm miss}$
		& $139~{\rm fb}^{-1}$               \\ \addlinespace
	\tabincell{c}{\bf CERN-EP-2019-106~\cite{Aad:2019vnb}\\ (arXiv:1908.08215)}   &\tabincell{l}{
			$ \tilde{\ell} \tilde{\ell} \to \ell \ell \tilde{\chi}_1^0 \tilde{\chi}_1^0$ \\
			$ \tilde{\chi}_1^{\pm} \tilde{\chi}_1^{\mp} \to \nu \tilde{\ell} / \ell \tilde{\nu} \to \ell \ell \nu \nu \tilde{\chi}_1^0 \tilde{\chi}_1^0 $
			}
		& $2 \ell+ E_{\rm T}^{\rm miss}$
		& $139~{\rm fb}^{-1}$              \\ \addlinespace
	\tabincell{c}{\bf CERN-EP-2019-188~\cite{Aad:2019vvf}\\ (arXiv:1909.09226)}
		& $ \tilde{\chi}_{2}^0 \tilde{\chi}_1^{\pm} \to Wh \tilde{\chi}_1^0 \tilde{\chi}_1^0$
		& $1\ell + h(\to bb) + E_{\rm T}^{\rm miss}$
		& $139~{\rm fb}^{-1}$               \\  \addlinespace
\tabincell{c}{\bf CMS-SUS-20-001~\cite{CMS:2020bfa}\\ (arXiv:2012.08600)}  & \tabincell{l}{
			$ \tilde{\chi}_2^0 \tilde{\chi}_1^{\mp} \to Z \tilde{\chi}_1^0 W^{\mp}\tilde{\chi}_1^0 $\\
			$ \tilde{\ell} \tilde{\ell} \to \ell \ell \tilde{\chi}_1^0 \tilde{\chi}_1^0 $
		}
		& $2\ell + E_{\rm T}^{\rm miss}$
		& $137~{\rm fb}^{-1}$   \\ \hline \hline
	\end{tabular}}
	\caption{\label{tab:sms}Signal of final state for electroweakino pair-production processes considered in this work. Relevant experimental analyses were performed in simplified models by ATLAS and CMS collaborations, and their results have been encoded in \texttt{SmodelS-1.2.3}.}
\end{table}
\texttt{SModelS}~\cite{Kraml:2013mwa, Ambrogi:2017neo, Dutta:2018ioj, Heisig:2018kfq, Khosa:2020zar, Alguero:2020grj} enables the fast interpretation of the LHC data via simplified model results from ATLAS and CMS searches for SUSY particles. It decomposes all of the  signatures occurring in a given SUSY model into simplified model topologies\footnote{Simplified model topologies are also referred to as simplified model spectra. An event topology is defined by the vertex structure as well as the SM and BSM final states. In each topology, the intermediate $\mathbb{Z}_2$ odd BSM particles are characterized only by their masses, production rates, and decay modes. } via a generic procedure. In practice, the cross-section upper limits and efficiency maps are used in re-interpreting each topology result. Compared with Monte Carlo simulation, \texttt{SModelS} is much easier and faster. It not only allows for re-interpreting searches of the cut-and-count methodology, but it also allows for other searches, such as those relying on boosted decision tree (BDT) variables. The power of \texttt{SModelS} comes from its superfast speed and its large and continuously updated database. The applicability of \texttt{SModelS} is limited by the simplified model results available in the database. Moreover, when the tested spectra split into many different channels, as is often the case in a complex model, the derived results are generally conservative.
\par In this work, the samples are refined with the analyses at 13 TeV LHC in \texttt{SModelS}, which are summarized in Table~\ref{tab:sms}.

\bibliography{references}
\bibliographystyle{CitationStyle}
\end{document}